\newif\ifconfver
\confverfalse      %declaring conference version false
\ifconfver
    \documentclass[10pt,twocolumn,twoside]{IEEEtran}
\else
    \documentclass[11pt,draftcls,onecolumn]{IEEEtran}
\fi

\usepackage{cite}
\usepackage{graphicx}
\usepackage{psfrag}
\usepackage{subfigure}
\usepackage{url}
\usepackage{stfloats}
\usepackage{amsfonts,amssymb,amsmath,bm,paralist,theorem,cite,ifthen,color}
\usepackage{array}
\usepackage{caption}
\usepackage{calc}
\usepackage{subfig}
\usepackage{longtable}
\usepackage{stfloats}
\usepackage{multirow}
\usepackage{algorithmic,algorithm}
\usepackage{graphics,booktabs,epsfig,enumerate}
\usepackage{amsfonts,amssymb,amsmath,amsbsy,bm,paralist,theorem,cite,ifthen,color}

{\begin{list}{}{
    \settowidth{\labelwidth}{\mbox{\textnormal{#1}}}%
    \setlength{\leftmargin}{\labelwidth+\labelsep}}}%
{\end{list}}

    \makeatletter
    \def\multilimits@{\bgroup
  \Let@
  \restore@math@cr
  \default@tag
 \baselineskip\fontdimen10 \scriptfont\tw@
 \advance\baselineskip\fontdimen12 \scriptfont\tw@
 \lineskip\thr@@\fontdimen8 \scriptfont\thr@@
 \lineskiplimit\lineskip
 \vbox\bgroup\ialign\bgroup\hfil$\m@th\scriptstyle{##}$\hfil\crcr}
    \def\Sb{_\multilimits@}
    \def\endSb{\crcr\egroup\egroup\egroup}
\makeatother

\newtheorem{Proposition}{Proposition}
\newtheorem{Theorem}{Theorem}
\newtheorem{Definition}{Definition}

\newtheorem{Claim}{Claim}

\newtheorem{Remark}{Remark}
\theorembodyfont{\rmfamily}

\newcommand\wb{\ensuremath{{\bm w}}}

\newcommand\ub{\ensuremath{{\bm u}}}

\newcommand\hb{\ensuremath{{\bm h}}}

\newcommand\Qb{\ensuremath{{\bm Q}}}
\newcommand\qb{\ensuremath{{\bm q}}}

\newcommand\lambdab{\ensuremath{{\bm \lambda}}}

\newcommand\Wb{\ensuremath{{\bm W}}}

\newcommand\tr{\ensuremath{{\rm Tr}}}

\newcommand\rank{\ensuremath{{\rm rank}}}

\newcommand\Cplx{\ensuremath{{\mathbb{C}}}}

\newcommand\zerob{\ensuremath{{{\bf 0}}}}

\definecolor{orange}{RGB}{255,107,0}

\begin{document}

\markboth{Accepted by IEEE TRANSACTIONS ON Signal Processing, Nov.
2012}{Accepted by IEEE TRANSACTIONS ON Signal Processing, Nov. 2012}

%\title{A Convex Approximation Approach to Weighted Sum Rate Maximization of
%Multiuser MISO Interference Channel under Outage Constraints}
\title{Coordinated Beamforming for Multiuser MISO Interference Channel under Rate Outage Constraints$^\S$}
%\title{Coordinated Beamforming for Multiuser MISO Interference Channel under Outage Constraints}
%\title{Utility Maximization of Multiuser MISO Interference Channel under Outage Constraints}
\ifconfver \else {\linespread{1.1} \rm \fi

\author{\vspace{0.5cm} Wei-Chiang Li$^\star$, Tsung-Hui Chang$^\dag$, Che Lin$^\star$, and Chong-Yung Chi$^\star$
\thanks{Copyright (c) 2012 IEEE. Personal use of this material is permitted. However, permission to use this material for any other purposes must be obtained from the IEEE by sending a request to
pubs-permissions@ieee.org.}
\thanks{$^\S$
The work is supported by the National Science Council, R.O.C., under
Grant NSC 99-2221-E-007-052-MY3 and under Grant NSC
101-2218-E-011-043. Part of this work was presented at the IEEE
ICASSP, Prague, Czech, May 22-27, 2011 \cite{Li2011}.}
\thanks{$^\dag$
Tsung-Hui Chang is the corresponding author. Address: Department of
Electronic and Computer Engineering, National Taiwan University of
Science and Technology, Taipei, Taiwan 10607, R.O.C. E-mail:
tsunghui.chang@ieee.org.}
%Tel: +886-3-5715131X34033, Fax: +886-3-5751787.}
\thanks{$^\star$
Wei-Chiang Li, Che Lin and Chong-Yung Chi are with Institute of Communications Engineering \& Department of Electrical Engineering, National Tsing Hua University, Hsinchu, Taiwan 30013, R.O.C.
E-mail: weichiangli@gmail.com,~\{clin,
cychi\}@ee.nthu.edu.tw}
}

\maketitle

\vspace{-\baselineskip}

\begin{abstract}
This paper studies the coordinated beamforming design problem for
the multiple-input single-output (MISO) interference channel,
assuming only channel distribution information (CDI) at the
transmitters. Under a given requirement on the rate outage
probability for receivers, we aim to maximize the system utility
(e.g., the weighted sum rate, weighted geometric mean rate, and the
weighed harmonic mean rate) subject to the rate outage constraints
and individual power constraints. The outage constraints, however,
lead to a complicated, nonconvex structure for the considered
beamforming design problem and make the optimization problem
difficult to handle. {Although} this nonconvex optimization problem
can be solved in an exhaustive search manner, this brute-force
approach is only feasible when the number of transmitter-receiver
pairs is small. For a system with a large number of
transmitter-receiver pairs, computationally efficient alternatives
are necessary. The focus of this paper is hence on the design of
such efficient approximation methods. In particular, by employing
semidefinite relaxation (SDR) and first-order approximation
techniques, we propose an efficient successive convex approximation
(SCA) algorithm that provides high-quality approximate beamforming
solutions via solving a sequence of convex approximation problems.
The solution thus obtained is further shown to be a stationary point
for the SDR of the original outage constrained beamforming design
problem. {Furthermore}, we propose a distributed SCA algorithm where
each transmitter optimizes its own beamformer using local CDI and
information obtained from limited message exchange with the other
transmitters. Our simulation results demonstrate that the proposed
SCA algorithm and its distributed counterpart indeed converge, and
near-optimal performance can be achieved for all the considered
system utilities.
 \ifconfver \else
\\\\
\noindent {\bfseries EDICS}:   SAM-BEAM, MSP-APPL, MSP-CODR, SPC-APPL \fi
\end{abstract}

\ifconfver \else \IEEEpeerreviewmaketitle} \fi

%---------------------------------------------------------------------------
\ifconfver \else
\newpage
\fi

%--------------------------------------------------------------------
%%%%%%%%%%%%%%%%%%%%%%%%%%%%%%%%%%%%%%%%%%%%%%%%%%%%%%%%%%%%%%%%%%%%%
\section{Introduction}\label{sec: intro}
%%%%%%%%%%%%%%%%%%%%%%%%%%%%%%%%%%%%%%%%%%%%%%%%%%%%%%%%%%%%%%%%%%%%%

Inter-cell interference is known to be one of the main bottlenecks
that limit the system performance of a wireless cellular network
where all transmitters share a universal frequency band. The
performance degradation caused by such interference is severe
especially for the users at the cell edge and can only be alleviated
when {some sort of} cooperation is available between base stations
(BSs) \cite{Dai2008}.
%Coordinated transmission %, where multiple BSs collaborate on the data
%transmission,
%has thus been proposed as an effective approach to
%mitigating inter-cell interference and improving the network
%spectral efficiency \cite{Gesbert10JSAC}.
According to the level of cooperation, the coordinated transmission
can be roughly divided into two classes: Network multiple-input
multiple-output (MIMO) and interference coordination
\cite{Gesbert10JSAC}. In network MIMO, all BSs work as a single
virtual BS using all the available antennas for data transmission
and reception. Each of the BSs requires to know all the channel
state information (CSI) and data streams of users, demanding a large
amount of message exchange between BSs \cite{Bjornson11TSP}.
Interference coordination, by contrast, only needs CSI sharing
between BSs; based on the shared CSI, the BSs coordinate with each
other in the design of transmission strategies, e.g., coordinated
beamforming \cite{Dahrouj_2010,Venturino10TWC} or power allocation
\cite{Venturino09CM}. Our interest in this paper lies in the
coordinated beamforming design.
%the information signals required by all the users
%are jointly transmitted by all the base stations. In this scenario,
%the base stations can be viewed as a virtual multi-antenna base
%station. This kind of cooperation requires to exchange both the
%channel state information (CSI) and data streams through the
%backhaul signaling. However, the exchange of data streams demands a
%considerable amount of backhaul bandwidth. To relax the requirement
%on the backhaul bandwidth, an alternative is to consider coordinated
%beamforming where each base station only transmit the information
%signals required by the local users, but the transmission
%strategies, e.g., beamformer or power allocation, are jointly
%designed to mitigate the inter-cell interference, hence, only CSI
%exchange is required \cite{Gesbert10JSAC,Dahrouj_2010}.}
%For such coordinated transmission, it is necessary to exchange among
%BSs the channel state information (CSI) or even the data streams
%intended for users through the backhaul signaling. The exchange of
%these data streams, however, demands a considerable amount of
%backhaul bandwidth. A more viable alternative is to consider
%interference coordinated transmission designs where the BSs require
%only to exchange the CSI \cite{Gesbert10JSAC,Dahrouj_2010}.
To this end, we adopt the commonly used interference channel (IFC)
model
\cite{Carleial78,Shang_07,Annapureddy2011}. %, where multiple
%transmitters communicate with their respective receivers over a
%common frequency band, thus interfering with each other.
Under this model, a Pareto optimal transmission scheme is that the
rate tuple of receivers {resides} on the boundary of the
achievable rate
region \cite{Larsson_etal2009_mag}. %Such a transmission scheme is called Pareto optimal
%\cite{Larsson_etal2009_mag}.
It is always desirable to have a Pareto
optimal transmission scheme since, otherwise, the achievable rates
of some of the receivers can be further improved.

Consider a multiple-input single-output (MISO) IFC, where the
transmitters are equipped with multiple antennas while the
receivers, i.e., mobile users, have only single antenna. We assume
that the receivers employ single-user detection wherein the
cross-link interference is treated as noise. Under such
circumstance, analyses in \cite{Jorswieck08,Shang2011,Mochaourab_11}
have shown that the Pareto optimal transmission strategy is transmit
beamforming. While beamforming is a structurally simple transmission
strategy, finding the optimal transmit beamformers for the MISO IFC
is intrinsically difficult. More precisely, it has been proved
\cite{Liu_11} that finding the optimal beamformers that maximize
system utilities, such as the weighted sum rate, the geometric mean
rate, or the harmonic mean rate, is NP-hard in general. As a result,
{lots of} efforts have focused on characterizing the optimal
beamformer structures \cite{Jorswieck08,Mochaourab_11,Zakhour_09} in
order to reduce the search dimension for finding the optimal
beamforming vectors, or on investigating suboptimal but
computationally efficient beamforming algorithms
\cite{Zhang_Cui_2010,Zakhour_09,Liu_11}. Another approach to
studying these resource conflicts encountered in the IFC is to use
Game theory; see \cite{Larsson_etal2009_mag,Larsson_08,Schmidt_09}
for related works.

%Another approach to studying these resource conflicts encountered in
%the IFC is to use Game theory
%\cite{Larsson_etal2009_mag,Larsson_08,Schmidt_09}. In particular,
%considering the two-user MISO IFC and using game theoretic
%approaches, the authors in \cite{Larsson_08} revealed the
%fundamental need for base station cooperation, by showing that the
%Nash bargaining solution can potentially achieve a much better sum
%rate performance than that achieved by the non-cooperative Nash
%equilibrium solution. To avoid excessive information exchange for
%user cooperation, pricing based game approaches were also considered
%in \cite{Schmidt_09}, where the transmitters only exchange a small
%number of pricing weights that are used to regulate the cross-link
%interference of the network.
%
%
%each transmitter models the cross-link interference induced
%by itself as a price against its achievable rate so that each transmitter can optimize its own beamformer while taking into account the cross-link interference at the same time.}

The aforementioned beamforming designs all assume that the
transmitters have the complete knowledge of CSI. To provide the
transmitters with complete CSI, the receivers need to periodically
send the CSI (e.g., for frequency division duplexing systems) or
training signals (e.g., for time division duplexing systems) back to
the transmitters. In contrast to the CSI, channel distribution
information (CDI) can remain unchanged for a relatively long period
of time and thus the amount of feedback information can be
significantly reduced. With CDI at the transmitters, the ergodic
rate region of the $K$-user MISO IFC has been analyzed and the
structure of the Pareto optimal beamformers has been characterized
in \cite{Bjornson_10}. For a two-user case, an efficient algorithm
for finding the Pareto boundary of the ergodic rate region was
presented in \cite{Karipidis_09}. Unlike the ergodic achievable rate
where the packet delay is not taken into consideration, the outage
constrained achievable rate is more suitable for delay-sensitive
applications, such as those involving voice or video data
communications. For such outage constrained achievable rate region,
the authors of \cite{Lindblom09,Lindblom_11} presented a numerical
method for finding the Pareto boundary; however, the complexity of
this algorithm increases exponentially with the number of
transmitter-receiver pairs. Developing efficient beamforming design
algorithms that can approach the outage constrained Pareto boundary
is therefore important. While several efficient beamforming
algorithms can be found in \cite{Kandukuri02,Ghosh_10}, a different
power-minimization design criterion was considered, instead of rate
utility maximization.

In this paper, we investigate efficient coordinated beamforming
design algorithms for maximizing the system utility under rate
outage constraints and individual power constraints. Specifically,
we assume that the MISO channel between each transmitter and
receiver is composed of zero-mean circularly symmetric complex
Gaussian fading coefficients where the corresponding covariance
matrix is known to the transmitter. We formulate an outage
constrained coordinated beamforming design problem, aiming at
finding the Pareto optimal beamformers that maximize the system
utility (e.g., the weighted sum rate) subject to a pre-assigned rate
outage probability requirement and power constraints. However, due
to the complicated nonconvex outage constraints,
%the inherent problem structure is quite involved and is
%difficult to handle. As aforementioned, the numerical method
%presented in \cite{Lindblom09,Lindblom_11} demands an exponential
%complexity. The focus here is hence on efficient approximate
%beamforming design algorithms. More precisely,
we propose a
successive convex approximation (SCA) algorithm, where the original
problem is successively approximated by a convex problem and the
beamforming solution is refined in an iterative manner. The convex
approximation formulation is obtained by applying the convex
optimization based semidefinite relaxation (SDR) technique
\cite{Luo2010_SPM}, followed by a logarithmic change of variables
and first-order approximation techniques. We analytically show that the proposed
SCA algorithm can yield a beamforming solution that is a stationary
point for the SDR of the original problem.
%, provided that the approximation formulation admits
%a rank-one beamforming solution.
We further propose a round-robin-fashioned distributed SCA algorithm
where each transmitter optimizes only its beamformer using local CDI
with limited communication overhead of message exchange with the
other transmitters. It is shown by simulations that the two proposed
algorithms yield near-optimal performance with lower complexity
compared with those reported in \cite{Lindblom09,Lindblom_11}.
%It is shown
%that the proposed distributed SCA algorithm can also attain a
%stationary point of the rate outage constrained design problem.

The remaining part of this paper is organized as follows. The system
model and the outage constrained coordinated beamforming problem are
presented in Section \ref{sec:system model}. In Section
\ref{sec:PropAlgrthm}, we present the proposed SCA algorithm and
analyze its convergence property. In Section \ref{sec:Decen_Alg},
the distributed SCA algorithm is developed and analyzed.
Simulation results that demonstrate the efficacy of the proposed algorithms
%the near-optimal performance of
%the proposed SCA and distributed SCA algorithms
are presented in
Section \ref{sec:SimuRslt}. Finally, the conclusions are drawn in
Section \ref{sec:conclusions}.

\textbf\textit{Notation:} %The boldfaced lowercase letters are
%used to denote vectors and the boldfaced uppercase letters are used
%to denote matrices.
The $n$-dimensional complex vectors and complex Hermitian matrices
are denoted by $\mathbb{C}^n$ and $\mathbb{H}^n$, respectively. The
$n\times{n}$ identity matrix is denoted by $\mathbf{I}_n$. The
superscripts `$T$' and `$H$' represent the matrix transpose and
conjugate transpose,
respectively. We denote $\|\cdot\|$ %and $\|\cdot\|_F$
as the vector Euclidean norm. $\mathbf{A}\succeq\mathbf{0}$ and
$\mathbf{a}\succeq\mathbf{0}$ respectively mean that matrix
$\mathbf{A}$ is positive semidefinite (PSD) and vector $\mathbf{a}$
is elementwise nonnegative. The trace and rank of matrix
$\mathbf{A}$ are denoted as $\tr(\mathbf{A})$ and
$\mathrm{rank}(\mathbf{A})$, respectively. We use the expression
$\mathbf{x}\sim\mathcal{CN}(\bm{\mu},\mathbf{Q})$ if $\mathbf{x}$ is
circularly symmetric complex Gaussian distributed with mean
$\bm{\mu}$ and covariance matrix $\mathbf{Q}$. We denote
$\exp(\cdot)$ (or simply $e^{(\cdot)}$) as the exponential function,
while $\ln(\cdot)$ and $\mathrm{Pr}\{\cdot\}$ represent the natural
log function and the probability function, respectively. For a
variable $a_{ik}$, where $i,k\in \{1,\ldots,K\}$, $\{a_{ik}\}_k$
denotes the set $\{a_{i1},\dots,a_{iK}\}$, $\{a_{ik}\}_{k\ne{i}}$
denotes the set $\{a_{ik}\}_k$ excluding $a_{ii}$, and $\{a_{ik}\}$
is defined as the set containing all possible $a_{ik}$, i.e.,
$\{a_{11},\ldots,a_{1K},a_{21},\ldots,a_{KK}\}$.

\vspace{-0.2cm}
%%%%%%%%%%%%%%%%%%%%%%%%%%%%%%%%%%%%%%%%%%%%%%%%%%%%%%%%%%%%%%%%%%
\section{Signal Model and Problem Statement} \label{sec:system model}
%%%%%%%%%%%%%%%%%%%%%%%%%%%%%%%%%%%%%%%%%%%%%%%%%%%%%%%%%%%%%%%%%%

We consider the $K$-user MISO IFC where each {transmitter} is
equipped with $N_t$ antennas and {each receiver} with a single
antenna. It is assumed that transmitters employ transmit beamforming
to {communicate} {with} their respective receivers. Let $s_i(t)$
denote the information signal sent from transmitter $i$, and let
$\wb_i \in \Cplx^{N_t}$ be the {corresponding} beamforming vector.
The received signal at receiver $i$ is given by
\begin{align}\label{received signal}
x_i(t) = \hb_{ii}^H\wb_is_i(t)+\sum_{k=1,k\neq{i}}^K\hb_{ki}^H\wb_k
s_k(t)+n_i(t),
\end{align}
where $\hb_{ki}\in\mathbb{C}^{N_t}$ denotes the channel vector from
transmitter $k$ to receiver $i$, and
$n_i(t)\sim\mathcal{CN}(0,\sigma_i^2)$ is the additive white
Gaussian noise at receiver $i$ where $\sigma^2_i>0$ is the noise
variance. As can be seen from \eqref{received signal}, in addition
to the noise, each receiver suffers from the cross-link interference
$\sum_{k\ne{i}}\hb_{ki}^H\wb_ks_k(t)$. We assume that all receivers
employ single-user detection {where} the cross-link interference is
simply treated as background noise. Under Gaussian signaling, i.e.,
$s_i(t)\sim\mathcal{CN}(0,1)$, the instantaneous achievable rate of
the $i$th transmitter-receiver pair {is known to be}
{\small\begin{align*} r_i\left(\{\hb_{ki}\}_k,\{\wb_k\}\right)
=\log_2\left(1+\frac{\left|\hb_{ii}^H\wb_i\right|^2}{\sum_{k\neq{i}}\left|\hb_{ki}^H\wb_k\right|^2+\sigma^2_i}\right).
\end{align*}}\vspace{-0.2cm}

In this paper, we assume that the channel coefficients $\hb_{ki}$
are block-faded (i.e., quasi-static), and that the transmitters have
only the statistical information of the channels, i.e., the CDI. In
particular, it is assumed that
$\hb_{ki}\sim\mathcal{CN}(\mathbf{0},\Qb_{ki})$ for all
$k,i=1,\ldots,K$, where $\Qb_{ki} \succeq \zerob$ denotes the
channel covariance matrix and is known to all the transmitters.
Since the transmission rate $R_i$ cannot be adapted without CSI, the
communication would be in outage whenever the transmission rate
$R_i>0$ is higher than the instantaneous capacity that the channel
can support. For a given outage probability requirement
$(\epsilon_1,\dots,\epsilon_K)$, the beamforming vectors $\{\wb_i\}$
{must} satisfy
$\Pr\{r_i(\{\hb_{ki}\}_k,\{\wb_i\})<R_i\}\le\epsilon_i$. Following
\cite{Lindblom_11}, {we define the corresponding }
$\epsilon_i$-outage achievable rate region {as follows.}

\begin{Definition} {\rm \cite{Lindblom_11}}
Let $P_i>0$ denote the power constraint of transmitter $i$, for
$i=1,\ldots,K$. The rate tuple $(R_1,\dots,R_K)$ is said to be
achievable if
%\begin{align*}
 $\Pr\left\{r_i\!\left(\{\hb_{ki}\}_k,\{\wb_k\}\right)<
R_i\right\}\leq \epsilon_i,~i=1,\dots,K$,
%\end{align*}
for some $(\wb_1,\dots,\wb_K) \in \mathcal{W}_1 \times \cdots \times
\mathcal{W}_K$ where {$\epsilon_i\in(0,1)$ is the maximum tolerable
outage probability of receiver $i$ and}
$\mathcal{W}_i\triangleq\{\wb\in\mathbb{C}^{N_t}|~
\|\wb\|^2\le{P_i}\}$. The $\epsilon_i$-outage achievable rate region
is given by
{\small\[
\mathcal{R}=\bigcup_{\substack{\wb_i\in\mathcal{W}_i,\\i=1,\dots,K}}\left\{(R_1,\dots,R_K)|~\Pr\left\{r_i\left(\{\hb_{ki}\}_k,\{\wb_k\}\right)<R_i\right\}\le\epsilon_i,~i=1,\dots,K\right\}.
\]}
\end{Definition}

Given {an} outage {requirement} $(\epsilon_1,\ldots,\epsilon_K)$
{and an individual power constraint $(P_1,\dots,P_K)$,} our goal is
to optimize $\{\wb_k\}$ such that {the predefined system utility
function $U(R_1,\dots,R_K)$ is maximized.} {To this end,} we
consider the following outage constrained coordinated beamforming
design problem
%\begin{center}
%\fbox{\parbox[]{0.97 \linewidth}{\vspace{0.3cm}
%\hspace{0.3cm}\underline{\textbf{Outage Constrained Coordinated Beamforming Design:}}
\begin{subequations}\label{UMX}
\begin{align}
\max_{\substack{\wb_i \in \Cplx^{N_t},R_i\geq 0, \\ i=1,\ldots,K}}~&U(R_1,\dots,R_K)\label{UMX_a}\\
\text{s.t.}~&\Pr\left\{r_i(\{\hb_{ki}\}_k,\{\wb_k\})< R_i\right\}\leq \epsilon_i,\label{UMX_b}\\
&~\|\wb_i\|^2\leq P_i,~~i=1,\dots,K.\label{UMX_c}
\end{align}
\end{subequations}
%\vspace{-0.6cm}
%}}
%\end{center}
Note that, as each user would prefer a higher transmission rate, a
sensible system utility function should be strictly increasing with
respect to the individual rate $R_i$ for $i=1,\dots,K$, such that
the optimal $(R_1,\dots,R_K)$ of problem \eqref{UMX} would lie on
the so-called Pareto boundary of $\mathcal{R}$
\cite{Larsson_etal2009_mag}. In this paper, we consider the
following system utility function which captures a tradeoff between
the system throughput and user fairness \cite{Mo2000}
\begin{equation}\label{utility}
U_{\beta}(\{R_i\})=\begin{cases}
                         \sum_{i=1}^K\frac{\alpha_iR_i^{1-\beta}}{1-\beta},&0\le\beta<\infty,~\beta\ne1,\\
                         \sum_{i=1}^K\alpha_i\ln(R_i),&\beta=1,
                         \end{cases}
\end{equation}
%\begin{align}\label{utility}
%U_{\beta}(R_1,\dots,R_K)=\left\{
%                         \begin{array}{ll}
%                         \sum_{i=1}^K\frac{\alpha_iR_i^{1-\beta}}{1-\beta},&~0\le\beta<\infty,~\beta\ne1,\\
%                         \sum_{i=1}^K\alpha_i\ln(R_i),&~\beta=1,
%                         \end{array}\right.
%\end{align}
where the coefficients $\alpha_i\in[0,1]$ for $i=1,\dots,K$ with
$\sum_{i=1}^K\alpha_i=1$ represent the user priority, and the
parameter $\beta\in[0,\infty)$ reflects the user fairness. %,
%i.e., higher fairness among the users is achieved by the optimal
%solution of \eqref{UMX} with larger $\beta$.
For example, for $\beta$ being $0,1,2$ and infinity,
$U_{\beta}(\{R_i\})$ corresponds to the weighted sum rate, weighted
geometric mean rate, weighted harmonic mean rate and the weighted
minimal rate, respectively. Hence, for $\beta$ being $0,1,2$ and
infinity, maximizing $U_{\beta}(\{R_i\})$ is respectively equivalent
to achieving the maximal throughput, proportional fairness, minimal
potential delay, and the max-min
fairness of users\cite{Bonald2001}. It can be verified %by the
%second-order condition \cite{BK:BoydV04}
that $U_{\beta}(\cdot)$ is concave in $\{R_i\}$ for all $\beta$.
However, since the outage constraints
\eqref{UMX_b} %are not convex and
have a complicated structure as will be seen later, solving problem
\eqref{UMX} is still challenging.

%=========================================================================
%{In this case, the following three utilities are commonly
%considered for tradeoff between the system throughput and user
%fairness on the Pareto boundary \cite{Mo2000}:
%\begin{itemize}
%\item {\bf Weighted Sum Rate:}
%$U_S(R_1,\dots,R_K)={\sum_{i=1}^K}~\alpha_iR_i$,
%
%\item {\bf Weighted Proportional Fairness Rate:}
%$U_P(R_1,\dots,R_K)={\prod_{i=1}^K}~R_i^{\alpha_i}$,
%
%\item {\bf Weighted Harmonic Mean Rate:}
%$U_H(R_1,\dots,R_K)=1/({\sum_{i=1}^K}~\alpha_iR_i^{-1})$.
%\end{itemize}
%The coefficients $\alpha_1,\ldots,\alpha_K$ represent the user
%priority, which {\blue satisfies} $\alpha_i\in [0,1]$ for
%$i=1,\dots,K$ and $\sum_{i=1}^K\alpha_i=1$. It can be seen that
%$U_S$, ${U_P}$ and ${U_H}$ are all increasing with respect to $R_i$
%for all $i$; moreover, $U_S$ is a concave function, and maximizing
%$U_P$ and $U_H$ is equivalent to maximizing the concave functions
%$\sum_{i=1}^K\alpha_i\ln{R_i}$ and $-\sum_{i=1}^K\alpha_iR_i^{-1}$,
%respectively.
%
%Although the aforementioned objective functions or the equivalent
%objective functions are concave, solving problem \eqref{UMX} is
%still challenging {since}, as we will see, the outage constraints
%\eqref{UMX_b} are not convex and have a complicated structure.}
%============================================================================

One possible approach to solving such a nonconvex problem is {via
exhaustive search \cite{Lindblom09}}. {In} \cite{Lindblom09}, each
of the cross-link interference {is discretized} into $M$ levels,
{and} given a set of cross-link interference levels, the maximum
achievable rate for each receiver can be computed \cite{Lindblom09}.
Since there are {a total of} ${K(K-1)}$ cross-user links, one has to
exhaustively search over $M^{K(K-1)}$ rate tuples. The complexity of
this method thus increases exponentially with $K(K-1)$, {making}
this approach only viable when $K$ is small. For a simple example of
$K=3$ and $M=10$, this method requires searching over $10^6$ rate
tuples, which is computationally prohibitive in practice.

%%%%%%%%%%%%%%%%%%%%%%%%%%%%%%%%%%%%%%%%%%%%%%%%%%%%%%%%%%%%%%%%%%
\section{Proposed Convex Approximation Method}
\label{sec:PropAlgrthm}
%%%%%%%%%%%%%%%%%%%%%%%%%%%%%%%%%%%%%%%%%%%%%%%%%%%%%%%%%%%%%%%%%%

Our goal in this section is to develop an efficient approximation
algorithm that obtains near-optimal solutions of problem \eqref{UMX}
for any number of transmitter-receiver {pairs, $K$}. {To
begin with}, we note from \cite[Appendix I]{Kandukuri02} that the
{outage} probability function in \eqref{UMX_b} can actually be
expressed in closed form as
\begin{align}\label{outage prob_CLSFORM}
\Pr\left\{r_i(\{\hb_{ki}\}_k,\{\wb_k\})<R_i\right\}=1-\exp\left(\frac{-(2^{R_i}-1)\sigma_i^2}{\wb_i^H\Qb_{ii}\wb_i}\right)
\prod_{k\neq{i}}\frac{\wb_i^H\Qb_{ii}\wb_i}{\wb_i^H\Qb_{ii}\wb_i+(2^{R_i}-1)\wb_k^H\Qb_{ki}\wb_k}.
\end{align}
So, problem \eqref{UMX} can be {rewritten as}
\begin{subequations}\label{UMX_CLSFORM}
\begin{align}
\max_{\substack{\wb_i \in \Cplx^{N_t},R_i\geq 0, \\ i=1,\ldots,K}}~&U(R_1,\dots,R_K) \label{UMX_CLSFORM_a}\\
\text{s.t.}~&\rho_i~\exp\left(\frac{(2^{R_i}-1)\sigma_i^2}{\wb_i^H\Qb_{ii}\wb_i}\right)\prod_{k\ne{i}}\left(1+\frac{(2^{R_i}-1)\wb_k^H\Qb_{ki}\wb_k}{\wb_i^H\Qb_{ii}\wb_i}\right)\le1,\label{UMX_CLSFORM_b}\\
&\|\wb_i\|^2\leq P_i,~i=1,\dots,K, \label{UMX_CLSFORM_c}
\end{align}
\end{subequations}
where $\rho_i\triangleq 1-\epsilon_i$. {Although the outage
probability can now be expressed in closed form}, problem
\eqref{UMX_CLSFORM} is still difficult to {solve}, {since the}
constraints in \eqref{UMX_CLSFORM_b} {are still nonconvex and
complicated}. In the ensuing subsections, we present a convex
approximation method {to handle} problem \eqref{UMX_CLSFORM}
efficiently.

%%%%%%%%%%%%%%%%%%%%%%%%%%%%%%%%%%%%%%%%%%%%%%%%%%%%%%%%%%%%%%%%%%
\subsection{Convex Approximation Formulation}\label{sec:ConsrvApprx}
%%%%%%%%%%%%%%%%%%%%%%%%%%%%%%%%%%%%%%%%%%%%%%%%%%%%%%%%%%%%%%%%%%

{The} proposed convex approximation method starts with applying
semidefinite relaxation (SDR), a convex optimization based
approximation technique \cite{Luo2010_SPM}. Specifically, {through}
SDR, we approximate the quadratic terms
$\wb_k^H\Qb_{ki}\wb_k=\tr(\wb_k\wb_k^H\Qb_{ki})$ in
\eqref{UMX_CLSFORM_b} by the linear terms $\tr(\Wb_k\Qb_{ki})$,
{where} the rank-one matrices $\wb_k\wb_k^H$ {are replaced by} the
PSD matrices $\Wb_k$ {of arbitrary $\mathrm{rank}(\Wb_k){\le}N_t$}.
The approximated problem is thus given by
{\small\begin{subequations}\label{UMX_SDR}
\begin{align}
\max_{\substack{\Wb_i \in \mathbb{H}^{N_t},R_i\geq 0, \\ i=1,\ldots,K}}~&U(R_1,\dots,R_K) \label{UMX_SDR_a}\\
\text{s.t.}~&\rho_i~
\exp\left(\frac{(2^{R_i}-1)\sigma_i^2}{\tr\left(\Wb_i\Qb_{ii}\right)}\right)\prod_{k\ne{i}}
\left(1+\frac{(2^{R_i}-1)\tr\left(\Wb_k\Qb_{ki}\right)}{\tr\left(\Wb_i\Qb_{ii}\right)}\right)\le1, \label{UMX_SDR_b}\\
&\tr\left(\Wb_i\right)\leq P_i,\label{UMX_SDR_c}\\
&\Wb_i\succeq0,~i=1,\dots,K.\label{UMX_SDR_d}
\end{align}
\end{subequations}}\hspace{-0.19cm}
We should mention that SDR has been widely used in various
beamforming design problems (see \cite{Luo10} for a review), where,
in most cases, a convex semidefinite program (SDP) approximation
formulation can be directly obtained via SDR and thus can be
efficiently solved. Problem \eqref{UMX_SDR}, however, is still not
convex yet due to the constraints in \eqref{UMX_SDR_b}. Therefore,
further approximations are needed for problem \eqref{UMX_SDR}.

In contrast to SDR that essentially results in a larger problem
feasible set, the {second approximation} is \emph{restrictive}, in
the sense that the obtained solution must also be feasible to
problem \eqref{UMX_SDR}. To illustrate this {restrictive
approximation}, let us consider the following change of variables:
%\begin{subequations}\label{change of variables}
%\begin{align}
%  e^{x_{ki}}&\triangleq \tr(\Wb_k\Qb_{ki}), \\
%  e^{y_{i}}&\triangleq 2^{R_i}-1,\label{change of variables a}\\
%  z_i&\triangleq
%  \frac{2^{R_i}-1}{\tr(\Wb_i\Qb_{ii})}=e^{y_i-x_{ii}}, \label{change of variables b}
%\end{align}
%\end{subequations}
\begin{align}\label{change of variables}
  e^{x_{ki}}&\triangleq \tr(\Wb_k\Qb_{ki}), ~
  e^{y_{i}}\triangleq 2^{R_i}-1,~
  z_i\triangleq
  \frac{2^{R_i}-1}{\tr(\Wb_i\Qb_{ii})}=e^{y_i-x_{ii}},
\end{align}
for $i,k=1,\dots,K$, where $x_{ki},y_i,z_i$ are slack variables. By
substituting \eqref{change of variables} into \eqref{UMX_SDR}, one
can reformulate problem \eqref{UMX_SDR} as the following problem
{\small \begin{subequations}\label{UMX_ChVar}
\begin{align}
\max_{\substack{\{\Wb_i\}\in\mathcal{S},R_i\geq 0, \\ x_{ki},y_i,z_i \in \mathbb{R}, \\k,i=1,\ldots,K}}~&U(R_1,\dots,R_K),\label{UMX_ChVar_a}\\
\text{s.t.}~&~\rho_ie^{\sigma_i^2z_i}\prod_{k\ne{i}}\left(1+e^{-x_{ii}+x_{ki}+y_i}\right)\leq 1,\label{UMX_ChVar_b}\\
&~\tr(\Wb_k\Qb_{ki})\leq e^{x_{ki}},\label{UMX_ChVar_c}\\
&~\tr(\Wb_i\Qb_{ii})\geq e^{x_{ii}},\label{UMX_ChVar_c2}\\
&~R_i\le\log_2(1+e^{y_i}),\label{UMX_ChVar_d}\\
&~e^{y_i-x_{ii}}\le{z_i},~\forall{k\in\mathcal{K}^c_i},~i=1,\dots,K,\label{UMX_ChVar_e}
%&~\tr(\Wb_i)\le{P_i},~\Wb_i\succeq\zerob,~i=1,\dots,K,\label{UMX_ChVar_f}
\end{align}
\end{subequations}}\hspace{-0.19cm}
where $\mathcal{K}^c_i\triangleq \{1,\ldots,K\}\backslash\{i\}$, and
the set $\mathcal{S}$ is defined in \eqref{additional constraint}
below. Notice that we have replaced the equalities in \eqref{change
of variables} with inequalities as in \eqref{UMX_ChVar_c} to
\eqref{UMX_ChVar_e}. It can be verified by the monotonicity of the
objective function that all the inequalities in \eqref{UMX_ChVar_b}
to \eqref{UMX_ChVar_e} would hold with equalities at the {optimal
points}. We also note that, for example, if the optimal solution
satisfies $\tr(\Wb_i\Qb_{ii})=0$ in \eqref{UMX_ChVar_c2}, then the
optimal $x_{ii}$ has to be minus infinity which is not attainable.
Similar issues occur for $\tr(\Wb_k\Qb_{ki})$ and $x_{ki}$. In view
of this, in \eqref{UMX_ChVar} we have enforced $\Wb_1,\ldots, \Wb_K$
to lie in the subset
\begin{equation}\label{additional constraint}
\mathcal{S}\triangleq\{\Wb_1,\ldots,\Wb_K\succeq\mathbf{0}|~\tr(\Wb_i)\le{P_i},~\tr(\Wb_i\Qb_{ik})\ge\delta~\forall{i},k=1,\dots,K\},
\end{equation}
where $\delta >0$. As long as $\delta$ is set to a small number, the
rate loss due to \eqref{additional constraint} would be negligible.
%Besides, due to \eqref{additional constraint} and the monotonicity of $U(R_1,\dots,R_K)$, one can show that the optimal rates $R_1,\ldots,R_K$ yielded by \eqref{UMX_ChVar} must be strictly greater than zero.

%Therefore, {\red if $\delta=0$}, problem \eqref{UMX_ChVar} is
%equivalent to problem \eqref{UMX_SDR}. {\red However, due to the
%exponential change of variables \eqref{change of variables}, the
%optimal value of \eqref{UMX_ChVar} may be unattainable when
%$\delta=0$. Hence, we set $\delta$ to be a small positive value.
%With the constraint $\{\Wb_i\}\in\mathcal{S}$, it can be verified by
%the monotonicity of the constraint functions in
%\eqref{UMX_ChVar_b}-\eqref{UMX_ChVar_e} that the optimal solution of
%\eqref{UMX_ChVar} is attainable since the optimal rates $\{R_i\}$
%should be strictly greater than zero. It will be demonstrated latter
%by numerical simulation that the performance degradation due to this
%constraint is ignorable when $\delta$ is small.}

It is interesting to see that constraint
\eqref{UMX_ChVar_b} is now convex; constraints
\eqref{UMX_ChVar_c2} and \eqref{UMX_ChVar_e} are also convex.
Constraints \eqref{UMX_ChVar_c} and \eqref{UMX_ChVar_d} are not
convex; nevertheless, they are relatively easy to {handle} compared
{with} the original \eqref{UMX_SDR_b}. Let
$(\bar{\wb}_1\bar{\wb}_1^H,\ldots, \bar{\wb}_K\bar{\wb}_K^H,\bar{R}_1,\ldots,\bar{R}_K)$ be a feasible point of problem \eqref{UMX_ChVar}. Define
%\begin{subequations}\label{feasible point}
%\begin{align}
%   \bar{x}_{ki}&\triangleq\ln(\bar{\wb}_k^H\Qb_{ki}\bar{\wb}_k),~k=1,\dots,K,\label{feasible point_a}\\
%   \bar{y}_i&\triangleq\ln(2^{\bar{R}_i}-1),\label{feasible point_b}\\
%   \bar{z}_i&\triangleq e^{\bar{y}_i -\bar{x}_{ii}},\label{feasible_point_c}
%\end{align}
%\end{subequations}
%\begin{subequations}\label{feasible point}
%\begin{align}
%   \bar{x}_{ki}&\triangleq\ln(\bar{\wb}_k^H\Qb_{ki}\bar{\wb}_k),~k=1,\dots,K,\\
%   \bar{y}_i&\triangleq\ln(2^{\bar{R}_i}-1),~
%   \bar{z}_i\triangleq e^{\bar{y}_i -\bar{x}_{ii}},
%\end{align}\end{subequations}
\begin{equation}\label{feasible point}
\bar{x}_{ki}\triangleq\ln(\bar{\wb}_k^H\Qb_{ki}\bar{\wb}_k),~\bar{y}_i\triangleq\ln(2^{\bar{R}_i}-1),~\bar{z}_i\triangleq{e}^{\bar{y}_i
-\bar{x}_{ii}},
\end{equation}
for $i,k=1,\ldots,K$. Then, $\{\bar{x}_{ki}\}$, $\{\bar{y}_i\}$,
$\{\bar{z}_i\}$ together with $\{\bar{R}_i\}$ and
$\bar{\Wb}_i\triangleq \bar{\wb}_i\bar{\wb}_i^H$, $i=1,\ldots,K$,
are feasible to problem \eqref{UMX_ChVar}. Here we conservatively
approximate \eqref{UMX_ChVar_c} and \eqref{UMX_ChVar_d} at the point
$(\{\bar{x}_{ki}\}_{k,i\neq k},\{\bar{y}_i\})$. Since both of
$e^{x_{ki}}$ and $\log_2(1+e^{y_i})$ are convex, their first-order
lower bounds at $\bar{x}_{ki}$ and $\bar{y}_i$ are respectively
given by
\begin{equation}\label{eq:linear_apprx}
e^{\bar{x}_{ki}}(x_{ki}-\bar{x}_{ki}+1)~~\text{and}~~\log_2(1+e^{\bar{y}_i})+\frac{e^{\bar{y}_i}(y_i-\bar{y}_i)}{\ln2\cdot(1+e^{\bar{y}_i})}.
\end{equation}
Consequently, restrictive approximations for \eqref{UMX_ChVar_c} and
\eqref{UMX_ChVar_d} are given by
\begin{subequations}\label{eq:Linrz}
\begin{align}
\tr(\Wb_k\Qb_{ki})&\le{e}^{\bar{x}_{ki}}(x_{ki}-\bar{x}_{ki}+1),~k\in\mathcal{K}^c_{i},\label{eq:Linrz_a}\\
R_i&\le\log_2(1+e^{\bar{y}_i})+\frac{e^{\bar{y}_i}(y_i-\bar{y}_i)}{\ln2\cdot(1+e^{\bar{y}_i})}.\label{eq:Linrz_b}
\end{align}
\end{subequations}
%\begin{subequations}\label{eq:Linrz}
%\begin{align}
%\tr(\Wb_k\Qb_{ki})&\le{e}^{\bar{x}_{ki}}(x_{ki}-\bar{x}_{ki}+1),~k\in\mathcal{K}^c_{i},\label{eq:Linrz_a}\\
%R_i&\le\frac{1}{\ln2}\left(\ln(1+e^{\bar{y}_i})+\frac{e^{\bar{y}_i}}{1+e^{\bar{y}_i}}(y_i-\bar{y}_i)\right).\label{eq:Linrz_b}
%\end{align}
%\end{subequations}
By replacing \eqref{UMX_ChVar_c} and
\eqref{UMX_ChVar_d} with \eqref{eq:Linrz_a} and \eqref{eq:Linrz_b}, we obtain the following
approximation for problem \eqref{UMX_CLSFORM}:
%\begin{center}
%\fbox{\parbox[]{0.97 \linewidth}{\vspace{-0.3cm}
{\small \begin{subequations}\label{UMX_ChVar2}
\begin{align}
\max_{\substack{\{\Wb_i\}\in\mathcal{S},R_i\ge0, \\ x_{ki},y_i,z_i \in \mathbb{R}, \\k,i=1,\ldots,K}}~&U(R_1,\dots,R_K), \label{UMX_ChVar2_a}\\
\text{s.t.}~&\rho_ie^{\sigma_i^2z_i}\prod_{k\ne{i}}\left(1+e^{-x_{ii}+x_{ki}+y_i}\right)\leq 1,\label{UMX_ChVar2_b}\\
&\tr(\Wb_k\Qb_{ki})\le{e}^{\bar{x}_{ki}}(x_{ki}-\bar{x}_{ki}+1),\label{UMX_ChVar2_c}\\
&\tr(\Wb_i\Qb_{ii})\geq e^{x_{ii}},\label{UMX_ChVar2_c2}\\
&R_i\le\log_2(1+e^{\bar{y}_i})+\frac{e^{\bar{y}_i}(y_i-\bar{y}_i)}{\ln2\cdot(1+e^{\bar{y}_i})},\label{UMX_ChVar2_d}\\
&e^{y_i-x_{ii}}\le{z_i},~\forall{k\in\mathcal{K}^c_i,}~i=1,\dots,K.\label{UMX_ChVar2_e}
%&~\tr(\Wb_i)\leq{P_i},~i=1,\dots,K, \label{UMX_ChVar2_f}
\end{align}
\end{subequations}}
%\vspace{-0.6cm}}}
%\end{center}
%\end{subequations}
%where
%%\begin{equation}
%$\Theta(\bar{y}_i)\triangleq(\theta_1(\bar{y}_i))^{\theta_1(\bar{y}_i)}(\theta_2(\bar{y}_i))^{\theta_2(\bar{y}_i)}.
%$ %\end{equation}
Problem \eqref{UMX_ChVar2} is a convex optimization problem; it can
be efficiently solved by standard convex solvers such as
\texttt{CVX} \cite{cvx}.

%The idea of removing the nonconvex rank-one constraints of
%$\{\Wb_i\}_{i=1}^K$ in \eqref{WSR_ChVar2} is known as semidefinite
%relaxation (SDR) in convex optimization theory \cite{Luo2010_SPM}.
Let $(\hat \Wb_1,\ldots, \hat \Wb_K)$ and $(\hat R_1,\ldots,\hat
R_K)$ denote the optimal beamforming matrices and achievable rates
yielded by the approximation problem \eqref{UMX_ChVar2}. Since the
lower bounds in \eqref{eq:linear_apprx} may not be exactly tight, it
may hold, for $(\hat \Wb_1,\ldots, \hat \Wb_K)$ and $(\hat
R_1,\ldots,\hat R_K)$ and for some $i=1,\ldots,K,$ that
{\small\begin{align}
\rho_i~\exp\left(\frac{(2^{\hat{R}_i}-1)\sigma_i^2}{\tr(\hat{\Wb}_i\Qb_{ii})}\right)\prod_{k\ne{i}}
\left(1+\frac{(2^{\hat{R}_i}-1)\tr(\hat{\Wb}_k\Qb_{ki})}{\tr(\hat{\Wb}_i\Qb_{ii})}\right)<1,
\end{align}}
\hspace{-0.19cm}i.e., the SINR outage probability is strictly less
than $\epsilon_i$ and thus the outage constraint is over satisfied.
Alternatively, one can obtain a tight rate tuple $(\tilde
R_1,\ldots,\tilde R_K)$, where $\tilde R_i \geq \hat R_i$ for all
$i=1,\ldots,K$, by solving the equations
{\small\begin{align}\label{bar R}
\rho_i~\exp\left(\frac{(2^{R_i}-1)\sigma_i^2}{\tr(\hat{\Wb}_i\Qb_{ii})}\right)\prod_{k\ne{i}}
\left(1+\frac{(2^{R_i}-1)\tr(\hat{\Wb}_k\Qb_{ki})}{\tr(\hat{\Wb}_i\Qb_{ii})}\right)=1,
\end{align}}
\hspace{-0.19cm}for $i=1,\ldots,K$. Note that each equation in
\eqref{bar R} can be efficiently solved by simple line search. The
obtained $(\hat \Wb_1,\ldots, \hat \Wb_K)$ and $(\tilde
R_1,\ldots,\tilde R_K)$ then serve as an approximate solution for
problem \eqref{UMX_SDR}.

In summary, the reformulation above consists of two approximation
steps: {a) the} rank relaxation of $\wb_k\wb_k^H$ to
$\Wb_k\succeq\mathbf{0}$ by SDR, and {b)} constraint restrictions of
\eqref{UMX_ChVar_c} and \eqref{UMX_ChVar_d} by \eqref{UMX_ChVar2_c}
and \eqref{UMX_ChVar2_d}. Note that if problem \eqref{UMX_ChVar2}
yields a rank-one optimal $(\Wb_1,\ldots,\Wb_K)$, a rank-one beamforming
solution {can be readily obtained} by rank-one decomposition of
$\Wb_i=\wb_i\wb_i^H$ for all $i=1,\dots,K$. It is then
straightforward to verify by the restrictiveness of
\eqref{UMX_ChVar2_c} and \eqref{UMX_ChVar2_d} that this rank-one
beamforming solution $(\wb_1,\ldots,\wb_K)$ is also feasible to the original problem
\eqref{UMX_CLSFORM} [i.e., problem \eqref{UMX}], thereby satisfying
the desired rate outage requirement. In view of this, it is
important to investigate the conditions under which problem
\eqref{UMX_ChVar2} can yield rank-one optimal $(\Wb_1,\ldots,\Wb_K)$. The
following proposition provides one such condition:

\begin{Proposition}\label{proposition_sufficient_condition_for_rank1}
Assume that \eqref{UMX_ChVar2} is feasible. Then there exists an optimal
$(\Wb_1,\dots,\Wb_K)$ satisfying
$
\mathrm{rank}\left(\Wb_i\right)\le1,~i=1,\dots,K,
$
if the number of users is no larger than three, i.e., $K\le3$.
\end{Proposition}

\noindent\emph{Proof:} Let
$(\{\hat\Wb_i\},\{\hat R_i\},\{\hat x_{ik}\},\{\hat y_i\},\{\hat z_i\})$
denote an optimal solution of problem \eqref{UMX_ChVar2}. Consider
\begin{subequations}\label{W_LP}
\begin{align}
\max_{\Wb_i\succeq\mathbf{0}}~&~\tr(\Wb_i\Qb_{ii})\\
\text{s.t.}~&~\delta \leq \tr(\Wb_i\Qb_{ik})\le{e}^{\bar{x}_{ik}}
\left(\hat x_{ik}-\bar{x}_{ik}+1\right),~k\in\mathcal{K}_i^c \label{W_LP b}\\
%&~\tr(\Wb_i\Qb_{ik})\ge\delta,~k\in\mathcal{K}_i^c \label{W_LP c}\\
&~\tr(\Wb_i\Qb_{ii})\ge\delta,~\tr(\Wb_i)\le{P}_i, \label{W_LP d}
\end{align}
\end{subequations}
for all $i=1,\dots,K$. By \eqref{additional constraint} and
\eqref{UMX_ChVar2_c}, $\hat \Wb_i$ is {also} feasible to the above
problem \eqref{W_LP}. Moreover, by \eqref{UMX_ChVar2_b},
\eqref{UMX_ChVar2_c2}, \eqref{UMX_ChVar2_d}, \eqref{UMX_ChVar2_e}
and by the monotonicity of $U(R_1,\ldots,R_K)$, one can show, by
contradiction, that $\hat \Wb_i$ is actually optimal to problem
\eqref{W_LP}, for all $i=1,\dots,K$. Let $\Wb_i'$ be an optimal
solution to \eqref{W_LP}, for $i=1,\dots,K$. Then, one can also
verify that $(\Wb_1',\dots,\Wb_K')$ is optimal to problem
\eqref{UMX_ChVar2}. We hence focus on problem \eqref{W_LP}. Firstly,
since problem \eqref{W_LP} is assumed to be feasible and the
objective is to maximize $\tr(\Wb_i\Qb_{ii})$, the constraint
$\tr(\Wb_i\Qb_{ii})\ge\delta$ in \eqref{W_LP d} is actually
irrelevant and can be dropped without affecting the optimal
solution. Secondly, it is easy to observe that, for each
$k\in\mathcal{K}_i^c$, it is either $\delta
<\tr(\Wb_i\Qb_{ik})={e}^{\bar{x}_{ik}} \left(\hat
x_{ik}-\bar{x}_{ik}+1\right)$ or
$\tr(\Wb_i\Qb_{ik})={e}^{\bar{x}_{ik}}\left(\hat
x_{ik}-\bar{x}_{ik}+1\right)=\delta$ at the optimum; that is, for
either case, it is equivalent to having one equality constraint in
\eqref{W_LP b} at the optimum for each $k\in\mathcal{K}_i^c$. As a
result, problem \eqref{W_LP} equivalently has only $K$ constraints.
According to \cite[Theorem 3.2]{Huang_Palomar2010}, {there always
exists an optimal solution $\Wb_i$ of} problem \eqref{W_LP}
satisfying
\begin{align}\label{rank constraint}
\rank(\Wb_i)\le\sqrt{K}~\text{for}~i=1,\dots,K.
\end{align}
%If the obtained optimal $\Wb_i$ does not satisfy \eqref{rank constraint}, a rank reduction procedure \cite{Huang_Palomar2010} can be employed
%to obtain another optimal $\Wb_i$ that satisfies \eqref{rank constraint}.
Therefore, if $K\le3$, {there always exists a rank-one optimal
$(\Wb_1,\ldots,\Wb_K)$ for problem \eqref{UMX_ChVar2}}.
\hfill{$\blacksquare$}

We should mention that $K\le3$ is only a sufficient condition but
not a necessary condition. For $K>3$, there may exist other
conditions under which a rank-one {optimal} solution exists for
problem \eqref{UMX_ChVar2}. If the optimal $(\Wb_1,\ldots,\Wb_K)$ is
not of rank one, then one may resort to the rank-one approximation
procedures such as Gaussian randomization \cite{Luo2010_SPM,Luo10}
to obtain an approximate solution to \eqref{UMX}. Note that, in that
case, the utility achieved by the randomized solution would be no
larger than $U(\tilde R_1,\ldots,\tilde R_K)$.
Surprisingly, in our computer simulations, we found that problem
\eqref{UMX_ChVar2} is always solved with rank-one optimal $\{\Wb_i\}$. %, which
%implies that, for the tested problem instances, we can simply
%perform rank-one decomposition, i.e., $\Wb_i=\wb_i\wb_i^H$, to
%obtain feasible beamforming solutions {for} the original problem
%\eqref{UMX}.
Some insightful analyses, which explain why problem \eqref{W_LP} is
often solved with rank-one optimal $\Wb_i$ for randomly generated
problem instances, can be found in \cite{Zhang_Cui_2010}.

%Since SDR is in general an approximation, the optimal
%$\{\Wb_i\}_{i=1}^K$ of problem \eqref{UMX_ChVar2} may not be of rank
%one. Surprisingly, it is found that, for the problem instances we
%tested in simulations, problem \eqref{UMX_ChVar2} always yields
%rank-one optimal $\{\Wb_i\}_{i=1}^K$, i.e., $\Wb_i=\wb_i(\wb_i)^H$
%for all $i$, provided that $\Wb_i \neq \zerob$. This implies that an
%approximate beamforming solution to problem \eqref{UMX_CLSFORM} can
%be directly obtained by decomposing the optimal $\{\Wb_i\}_{i=1}^K$
%of problem \eqref{UMX_ChVar2}.

\vspace{-0.4cm}
%%%%%%%%%%%%%%%%%%%%%%%%%%%%%%%%%%%%%%%%%%%%%%%%%%%%%%%%%%%%%%%%%%
\subsection{Successive Convex Approximation (SCA)}\label{sec:Successive_CVX_Aprx}
%%%%%%%%%%%%%%%%%%%%%%%%%%%%%%%%%%%%%%%%%%%%%%%%%%%%%%%%%%%%%%%%%%

Formulation \eqref{UMX_ChVar2} is obtained by approximating problem
\eqref{UMX_ChVar} at the given feasible point
$(\{\bar{\wb}_i\bar{\wb}_i^H\},\{\bar{R}_i\})$, {as described in
\eqref{feasible point}}. This approximation can be further improved
by successively approximating problem \eqref{UMX_ChVar} based on the
optimal solution $(\{{\Wb}_i\},\{{R}_i\})$ obtained by solving
\eqref{UMX_ChVar2} in the previous approximation. Specifically, let
$(\hat{\Wb}_1[n-1],\ldots,\hat{\Wb}_K[n-1])$ be the optimal
beamforming matrices obtained in the $(n-1)$th iteration, and,
similar to \eqref{bar R}, let
$(\tilde{R}_i[n-1],\ldots,\tilde{R}_i[n-1])$ be the corresponding
achievable rate tuple obtained by solving the following $K$
equations {\small\begin{align}\label{bar R2}
\rho_i~\exp\left(\frac{(2^{R_i}-1)\sigma_i^2}{\tr(\hat{\Wb}_i[n-1]\Qb_{ii})}\right)\prod_{k\ne{i}}
\left(1+\frac{(2^{R_i}-1)\tr(\hat{\Wb}_k[n-1]\Qb_{ki})}{\tr(\hat{\Wb}_i[n-1]\Qb_{ii})}\right)=1,~i=1,\ldots,K.
\end{align}}\hspace{-.19cm}
Moreover, let
\begin{subequations}\label{feasible point_n}
  \begin{align}
  \bar{x}_{ki}[n-1]&=\ln({\tr(\hat{\Wb}_k[n-1]\Qb_{ki})}),\\
  \bar{y}_i[n-1]&=\ln(2^{{\tilde{R}_i[n-1]}}-1),~i,k=1,\ldots,K.
  \end{align}
\end{subequations}
By replacing $\bar{x}_{ki}$ and $\bar{y}_i$ in \eqref{UMX_ChVar2}
with $\bar{x}_{ki}[n-1]$ and $\bar{y}_i[n-1]$ in \eqref{feasible
point_n} for $k\in\mathcal{K}_i^c$, $i=1,\dots,K$, we solve, in the
$n$th iteration, the following convex {optimization} problem
{\small\begin{subequations}\label{eq:central_optsol_n_ite}
\begin{align}
(\{\hat{\Wb}_i[n]\},&\{\hat{R}_i[n]\},\{\hat{x}_{ik}[n]\},\{\hat{y}_i[n]\},\{\hat{z}_i[n]\})=\nonumber\\
&\mathrm{arg}\max_{\substack{\{\Wb_i\}\in\mathcal{S},R_i\ge0\\x_{ik},y_i,z_i\in\mathbb{R}\\i,k=1,\dots,K}}~U(R_1,\dots,R_K)\label{eq:central_optsol_n_ite_a}\\
&~~~~~~~~~~\text{s.t.}~~~~\rho_ie^{\sigma_i^2z_i}\prod_{k\ne{i}}\left(1+e^{-x_{ii}+x_{ki}+y_i}\right)\le1,\label{eq:central_optsol_n_ite_b}\\
&~~~~~~~~~~~~~~~~~\tr(\Wb_k\Qb_{ki})\le{e}^{\bar{x}_{ki}[n-1]}\left(x_{ki}-\bar{x}_{ki}[n-1]+1\right),\label{eq:central_optsol_n_ite_c}\\
&~~~~~~~~~~~~~~~~~\tr(\Wb_i\Qb_{ii})\ge{e}^{x_{ii}},\label{eq:central_optsol_n_ite_d}\\
&~~~~~~~~~~~~~~~~~R_i\le\log_2(1+e^{\bar{y}_i[n-1]})+\frac{e^{\bar{y}_i[n-1]}(y_i-\bar{y}_i[n-1])}{\ln2\cdot(1+e^{\bar{y}_i[n-1]})},\label{eq:central_optsol_n_ite_e}\\
&~~~~~~~~~~~~~~~~~e^{y_i-x_{ii}}\le{z_i},~\forall{k\in\mathcal{K}_i^c},~i=1,\dots,K.\label{eq:central_optsol_n_ite_f}
%&~~~~~~~~~~~~~~~~~\tr(\Wb_i)\le{P_i},~i=1,\dots,K.\label{eq:central_optsol_n_ite_g}
\end{align}
\end{subequations}}
\hspace{-0.3cm} Note that the rate $\tilde{R}_i[n-1]$ obtained by \eqref{bar R2} is no less than $\hat {R}_i[n-1]$ for all $i=1,\ldots,K$, and thus the former is used to compute $\{\bar y_i[n-1]\}$ as the point for successive approximation.
In fact, successive approximation ensures monotonic improvement of the utility ${U}(\tilde{R}_1[n],\dots,\tilde{R}_K[n])$. Let us define
\begin{equation}\label{def:zbar}
\bar{z}_i[n-1]\triangleq{e}^{\bar{y}_i[n-1]-\bar{x}_{ii}[n-1]},~i=1,\dots,K.
\end{equation}
Then, by \eqref{bar R2}, \eqref{feasible point_n} and
\eqref{def:zbar}, one can show that
$(\{\hat{\Wb}_i[n-1]\},\{\tilde{R}_i[n-1]\},\{\bar{x}_{ik}[n-1]\},
\{\bar{y}_i[n-1]\},\{\bar{z}_i[n-1]\})$ is a feasible point of
\eqref{eq:central_optsol_n_ite}. As a result, we have
\begin{align}\label{rate increase}
U(\tilde{R}_1[n],\dots,\tilde{R}_K[n]) \geq
{U}(\hat{R}_1[n],\dots,\hat{R}_K[n])\ge{U}(\tilde{R}_1[n-1],\dots,\tilde{R}_K[n-1]),~\forall{n\ge1}.
\end{align}
The proposed \emph{successive convex approximation}
(SCA) algorithm is summarized in Algorithm 1. %in the following Algorithm
%\ref{alg:centralized}.

%\begin{algorithm}[h]
%  \caption{Proposed successive convex approximation algorithm for solving problem \eqref{UMX_CLSFORM}}
%\begin{algorithmic}[1]
%  \STATE {\bf Input} a feasible point
%  $(\{\bar{\wb}_i\}_{i=1}^K,\{\bar{R}_i\}_{i=1}^K)$ of problem
%  \eqref{UMX_CLSFORM}, and a solution accuracy $\delta>0$.
%  \STATE {Obtain} $\{\{\bar{x}_{ki}\}_{k\neq i},\bar{y}_i\}_{i=1}^K$ by \eqref{feasible point} and obtain
%        $\theta_1(\bar{y}_i)={e^{\bar{y}_i}}/({e^{\bar{y}_i}+1})$
%        and $\theta_2(\bar{y}_i)={1}/({e^{\bar{y}_i}+1})$ for
%        $i=1,\ldots,K.$
%  \STATE {Solve} problem \eqref{UMX_ChVar2} to obtain the
%  optimal beamforming matrices $\{\Wb_i^\star\}_{i=1}^K$ and rates
%  $\{R_i^\star\}_{i=1}^K$.
%
%  \STATE Obtain $\wb_i^\star$ by decomposition of
%  $\Wb_i^\star=\wb_i^\star(\wb_i^\star)^H$ for $i=1,\ldots,K$.
%
%  \STATE {\bf Output} the approximate beamforming solution
%  $(\wb_1^\star,\ldots,\wb_K^\star)$ and achievable rate tuple $(R_1^\star,\ldots,R_K^\star)$
%  if $|U(R_1^\star,\dots,R_K^\star)-U(\bar{R}_1,\dots,\bar{R}_K)|/U(\bar{R}_1,\dots,\bar{R}_K)<\delta$; otherwise update $\bar{\wb}_i:=\wb_i^\star$ and
%  $\bar{R}_i:=R_i^\star$ for all $i$, and go to Step 2.
%\end{algorithmic}
%\end{algorithm}
%\vspace{-0.2cm}

\begin{algorithm}[t]
\caption{{SCA} algorithm for solving problem \eqref{UMX}}
\begin{algorithmic}[1]\label{alg:centralized}
\STATE {\bf Given}
$(\bar{\wb}_1\bar{\wb}_1^H,\ldots,\bar{\wb}_K\bar{\wb}_K^H)\in\mathcal{S}$
and $(\bar{R}_1,\ldots,\bar{R}_K)$ that are feasible to
\eqref{UMX_SDR}. \STATE Set
{$\hat{\Wb}_i[0]=\bar{\wb}_i\bar{\wb}^H_i$ and
$\tilde{R}_i[0]=\bar{R}_i$} for all $i=1,\ldots,K,$ and set $n=0$.

\REPEAT

\STATE $n:=n+1$.

\STATE {Obtain} $\{\bar{x}_{ki}[n-1]\}$ and
       $\{\bar{y}_i[n-1]\}$ by
       \eqref{feasible point_n}, and solve problem \eqref{eq:central_optsol_n_ite} to obtain the
       optimal solution
       $\hat \ub[n]\triangleq (\{\hat{\Wb}_i[n]\},\{\hat{R}_i[n]\},\{\hat{x}_{ik}[n]\},\{\hat{y}_i[n]\},\{\hat{z}_i[n]\})$.

%\STATE {Solve} problem \eqref{eq:central_optsol_n_ite} to obtain the
%         optimal solution $(\{\hat{\Wb}_i[n]\},\{\hat{R}_i[n]\},\{\hat{x}_{ik}[n]\},\{\hat{y}_i[n]\},\{\hat{z}_i[n]\})$.

\STATE {{Compute} $(\tilde{R}_1[n],\ldots,\tilde{R}_K[n])$ by solving \eqref{bar R2}}.

\UNTIL the stopping criterion is met. % (e.g.,
%$U({\tilde{R}_1[n],\dots,\tilde{R}_K[n]})=U({\tilde{R}_1[n-1],\dots,\tilde{R}_K[n-1]})$).

\STATE Obtain $\wb_i^\star$ by decomposition of
$\hat\Wb_i[n]=\wb_i^\star(\wb_i^\star)^H$ for all $i$, if
$\hat\Wb_i[n]$ are all of rank one; otherwise perform Gaussian
randomization \cite{Luo10} to obtain a rank-one approximate
solution of {\eqref{UMX}}.

%\STATE {\bf Output}
%  {\blue $\{\hat{\wb}_i\}$ and the associated rate tuple
%  {\red$\{\bar{R}_i\}$}, which is obtained by solving
%  \eqref{max_achievable_rate}, as the approximate solution of problem
%  \eqref{UMX}}.
\end{algorithmic}
\end{algorithm}
\vspace{-0.3cm}

\vspace{-0.2cm}
%%%%%%%%%%%%%%%%%%%%%%%%%%%%%%%%%%%%%%%%%%%%%%%%%%%%%%%%%%%%%%%%%%%
\subsection{Convergence Analysis} \label{sec:convergence_analysis}
%%%%%%%%%%%%%%%%%%%%%%%%%%%%%%%%%%%%%%%%%%%%%%%%%%%%%%%%%%%%%%%%%%%

%For Algorithm \ref{alg:centralized}, it is clear that the utility
%function $U(\hat{R}_1[n],\dots,\hat{R}_K[n])$ is nondecreasing from
%iteration to iteration, {\blue and} will converge due to the finite
%power constraints, i.e., \eqref{eq:central_optsol_n_ite_g}. In fact,
%We show here that the optimal $(\Wb_1,\ldots,\Wb_K,R_1,\ldots,R_K)$ obtained by solving
%\eqref{eq:central_optsol_n_ite} and \eqref{bar R2} converges to a stationary point of
%problem \eqref{UMX_SDR}.
Convergence properties of Algorithm 1 is given below.
\vspace{-0.3cm}
%==================================================================
\begin{Theorem}\label{thm:centralized_convergence} Suppose
that the utility $U(R_1,\dots,R_K)$ is differentiable and strictly
increasing with respect to $R_i$, for $i=1,\dots,K$. Then, the
sequence,
$\{U({\tilde{R}_1[n],\dots,\tilde{R}_K[n]})\}_{n=1}^\infty$
generated by Algorithm 1, converges, and any limit point of the
sequence $\{(\hat\Wb_1[n],\ldots,\hat\Wb_K[n]), (\tilde
R_1[n],\ldots, \tilde R_K[n]) \}_{n=1}^\infty$ is a stationary point
of problem \eqref{UMX_ChVar} as well as a stationary point of
problem \eqref{UMX_SDR} with extra constraints
$\tr(\Wb_i\Qb_{ik})\ge\delta$ for $i,k=1,\dots,K$ $\mathrm{(see
~\eqref{additional constraint})}$.
\end{Theorem}\vspace{-0.1cm}
%==================================================================
\emph{Proof of Theorem \ref{thm:centralized_convergence}:} As
discussed earlier, the utility
$U(\tilde{R}_1[n],\dots,\tilde{R}_K[n])$ is nondecreasing with $n$.
Moreover, due to the individual power constraints, the sequence
$\{U(\tilde{R}_1[n],\dots,\tilde{R}_K[n])\}_{n=1}^\infty$ is
bounded, which implies the convergence of
$U(\tilde{R}_1[n],\dots,\tilde{R}_K[n])$.

%Because $\bar \ub[n-1]\triangleq (\{\hat{\Wb}_i[n-1]\},\{\tilde{R}_i[n-1]\},\{\bar{x}_{ik}[n-1]\},
%\{\bar{y}_i[n-1]\},\{\bar{z}_i[n-1]\})$ is feasible to \eqref{eq:central_optsol_n_ite}, by
%\eqref{rate increase} and by the convergence of $U(\tilde{R}_1[n],\dots,\tilde{R}_K[n])$, $\tilde \ub[n-1]$ is asymptotically optimal to \eqref{eq:central_optsol_n_ite}; that is
%\begin{align}\label{good neighbors}
% \lim_{n\rightarrow \infty} \hat\ub[n] - \bar \ub[n-1] =\zerob,
%\end{align}
%where $\hat\ub[n]\triangleq
%(\{\hat{\Wb}_i[n]\},\{\hat{R}_i[n]\},\{\hat{x}_{ik}[n]\},\{\hat{y}_i[n]\},\{\hat{z}_i[n]\})$ denotes the optimal solution of \eqref{eq:central_optsol_n_ite}.

Let $\hat\ub[n]\triangleq
(\{\hat{\Wb}_i[n]\},\{\hat{R}_i[n]\},\{\hat{x}_{ik}[n]\},\{\hat{y}_i[n]\},\{\hat{z}_i[n]\})$,
denote the optimal solution of \eqref{eq:central_optsol_n_ite}. To
prove that any limit point of $\hat{\ub}[n]$ is a stationary point
of \eqref{UMX_ChVar}, two key observations are needed. Firstly, we
note that the proposed SCA algorithm is in fact an inner
approximation algorithm in the nonconvex optimization literature
\cite{Marks1978}. In particular, the nonconvex constraints in
\eqref{UMX_ChVar_c} and \eqref{UMX_ChVar_d}, i.e.,
\begin{align*}
\Psi_{ki}(\Wb_k,x_{ki})\triangleq&\tr(\Wb_k\Qb_{ki})-e^{x_{ki}}\le0,~k\in\mathcal{K}_i^c,\\
\Phi_i(R_i,y_i)\triangleq&{R_i}-\log_2(1+e^{y_i})\le0,~i=1,\dots,K,
\end{align*}
are respectively replaced by
\begin{align}
&\bar{\Psi}_{ki}(\Wb_k,x_{ki}|~\bar{x}_{ki}[n-1])
\triangleq\tr(\Wb_k\Qb_{ki})-e^{\bar{x}_{ki}[n-1]}(x_{ki}-
\bar{x}_{ki}[n-1]+1)\le0,\label{psi}\\
&\bar{\Phi}_{i}(R_i,y_i|~\bar{y}_i[n-1])\triangleq{R_i}-\log_2(1+e^{\bar{y}_i[n-1]})+\frac{e^{\bar{y}_i[n-1]}(y_i-\bar{y}_i[n-1])}{\ln2\cdot(1+e^{\bar{y}_i[n-1]})}\le0,\label{phi}
\end{align}
for all $k\in\mathcal{K}_i^c$, $i=1,\dots,K$. One can verify that
$\bar{\Psi}_{ki}(\Wb_k,x_{ki}|\bar{x}_{ki}[n\!-\!1])$ and
$\bar{\Phi}_{i}(R_i,y_i|\bar{y}_i[n\!-\!1])$ satisfy
%\begin{subequations}\label{eq:inner_approximation}
\begin{align}
%\Psi_{ki}(\Wb_k,x_{ki})&\le\bar{\Psi}_{ki}(\Wb_k,x_{ki}|~\bar{x}_{ki}[n-1])\label{eq:inner_approximation_a}\\
\Psi_{ki}(\hat{\Wb}_k[n-1],\bar{x}_{ki}[n-1])&=\bar{\Psi}_{ki}(\hat{\Wb}_k[n-1],
\bar{x}_{ki}[n-1]|~\bar{x}_{ki}[n-1])=0\label{eq:inner_approximation_b}\\
\frac{\partial{\Psi}_{ki}(\Wb_k,x_{ki})}{\partial\Wb_k}&=\frac{\partial
\bar{\Psi}_{ki}(\Wb_k,x_{ki}|~\bar{x}_{ki}[n-1])}{\partial\Wb_k}\label{eq:inner_approximation_c}
\\
\left.\frac{\partial{\Psi}_{ki}(\Wb_k,x_{ki})}{\partial{x}_{ki}}\right|_{x_{ki}
=\bar{x}_{ki}[n-1]}&=\left.\frac{\partial\bar{\Psi}_{ki}(\Wb_k,x_{ki}|~\bar{x}_{ki}[n-1])}{\partial{x}_{ki}}\right|_{x_{ki}=\bar{x}_{ki}[n-1]}\label{eq:inner_approximation_d}
\end{align}
%\end{subequations}
%\begin{subequations}\label{eq:inner_approximation2}
\begin{align}
%\Phi_i(R_i,y_i)&\le\bar{\Phi}_i(R_i,y_i|~\bar{y}_i[n-1])\label{eq:inner_approximation_e}\\
\Phi_i(\hat{R}_i[n-1],\bar{y}_i[n-1])&=\bar{\Phi}_i(\hat{R}_i[n-1],
\bar{y}_i[n-1]|~\bar{y}_i[n-1])=0\label{eq:inner_approximation_f}\\
\frac{\partial{\Phi}_i(R_i,y_i)}{\partial{R}_i}&=
\frac{\partial\bar{\Phi}_i(R_i,y_i|~\bar{y}_i[n-1])}
{\partial{R}_i}\label{eq:inner_approximation_g}
%\\
\end{align}
\begin{align}
\left.\frac{\partial{\Phi}_i(R_i,y_i)}{\partial{y}_i}
\right|_{y_i=\bar{y}_i[n-1]}&=\left.\frac{\partial
\bar{\Phi}_i(R_i,y_i|~\bar{y}_i[n-1])}{\partial{y}_i}
\right|_{y_i=\bar{y}_i[n-1]}\label{eq:inner_approximation_h},
\end{align}
%\end{subequations}
for all $k\in\mathcal{K}_i^c$ and $i=1,\dots,K$.

Secondly, the restrictive approximations made in
\eqref{eq:central_optsol_n_ite_c} and
\eqref{eq:central_optsol_n_ite_e} are asymptotically tight as
$n\to\infty$:\vspace{-0.1cm}
%==================================================================
\begin{Claim}\label{claim:x_converge}
It holds true that\vspace{-0.1cm}
\begin{align}
&\lim_{n\to\infty}|\hat{x}_{ik}[n]-\bar{x}_{ik}[n-1]|=0,\label{claim 1 a}\\
&\lim_{n\to\infty}|\hat{y}_i[n]-\bar{y}_i[n-1]|=0,~\lim_{n\to\infty}(\tilde{R}_i[n]-\hat{R}_i[n])=0,\label{claim
1 b}
\end{align}
for all $k\in\mathcal{K}_i^c$, $i=1,\dots,K$.
\end{Claim}
%==================================================================
Claim \ref{claim:x_converge} is proved in Appendix \ref{appendix:
proof of claim12}. Moreover, by the monotonicity of
$U(R_1,\ldots,R_K)$ and due to \eqref{additional constraint}, it is
not difficult to verify that:
%==================================================================
\begin{Claim}\label{claim:bounded}
The sequence $\{\hat \ub[n]\}_{n=0}^\infty$ generated by Algorithm 1 is bounded.
\end{Claim}
%==================================================================

Now let us consider the KKT conditions of
\eqref{eq:central_optsol_n_ite}. Denote
$\mathcal{L}(\hat{\ub}[n],\bm{\lambda}[n]|\{\{\bar{x}_{ki}[n-1]\}_{k\neq{i}}\}_i,\{\bar{y}_i[n-1]\})$
as the Lagrangian of \eqref{eq:central_optsol_n_ite}. For ease of
explanation, let
$\Theta_i(\{x_{ki}\}_k,y_i,z_i)\triangleq\rho_ie^{\sigma_i^2z_i}\prod_{k\ne{i}}(1+e^{-x_{ii}+x_{ki}+y_i})-1$
denote the constraint function in \eqref{eq:central_optsol_n_ite_b},
and consider the following Lagrangian-stationarity condition:{\small
\begin{align}\label{KKT one}
&\frac{\partial\mathcal{L}(\hat{\ub}[n],\bm{\lambda}[n]|\{\{\bar{x}_{ki}[n-1]\}_{k\neq{i}}\}_i,\{\bar{y}_i[n-1]\})}{\partial{x_{ki}}}\nonumber\\
&=\lambda_i^{\mathrm{b}}[n]\frac{\partial\Theta_i(\{\hat{x}_{ji}[n]\}_j,\hat{y}_i[n],\hat{z}_i[n])}{\partial{x}_{ki}}+\lambda_{ki}^{\mathrm{c}}[n]\frac{\partial\bar{\Psi}_{ki}(\hat{\Wb}_k[n],\hat{x}_{ki}[n]|~\bar{x}_{ki}[n-1])}{\partial{x_{ki}}}=0,~\forall{k}\ne{i},
%  &\frac{\partial \mathcal{L}(\hat{\ub}[n],{\red\bm{\lambda}[n]}| \{\bar x_{ki}[n-1]\}_{k\neq i},\{\bar y_i[n-1]\})}{\partial \Wb_i}\notag =\lambda_i^P[n] \Ib_{N_t}- \\
%  &~~~~~~(\lambda_i^d[n]+\lambda_{ii}^{\mathcal{S}}[n])\Qb_{ii}
%  + \sum_{k\neq i}\left(\lambda_{ik}^c[n]
%  \frac{\partial\bar{\Psi}_{ik}(\hat{\Wb}_i[n],
%   \hat{x}_{ik}[n]|~\bar{x}_{ik}[n-1])}{\partial\Wb_i} - \lambda_{ik}^{\mathcal{S}}[n]\Qb_{ik}
%  \right)\succeq \zerob,
\end{align}}
\!\!where
$\lambdab[n]\triangleq(\{\lambda_i^{\mathrm{b}}[n]\},\{\{\lambda_{ik}^{\mathrm{c}}[n]\}_{k\ne{i}}\}_i,\{\lambda_i^{\mathrm{d}}[n]\},\{\lambda_i^{\mathrm{e}}[n]\},\{\lambda_i^{\mathrm{f}}[n]\},\{\lambda_i^{P}[n]\},\{\lambda_{ik}^{\delta}[n]\})$
are dual variables associated with constraints
\eqref{eq:central_optsol_n_ite_b}-\eqref{eq:central_optsol_n_ite_f},
the transmit power constraint and $\tr(\Wb_i\Qb_{ik})\ge\delta$.
Since problem \eqref{eq:central_optsol_n_ite} satisfies the Slater's
condition, the dual variables are bounded
\cite{BK:Bertsekas2003_analysis}. Moreover, $\hat \ub[n]$ is bounded
as well by Claim \ref{claim:bounded}. Therefore, there exists a
subsequence $\{n_1,\dots,n_\ell,\dots\}\subseteq\{1,\dots,n,\dots\}$
and a primal-dual limit point, denoted by
$\hat\ub^\star\triangleq(\{\hat{\Wb}_i^\star\},\{\hat{R}_i^\star\},\{\hat{x}_{ik}^\star\},\{\hat{y}_i^\star\},\{\hat{z}_i^\star\})$
and
$\lambdab^\star\triangleq(\{\lambda_i^{\mathrm{b}\star}\},\{\{\lambda_{ik}^{\mathrm{c}\star}\}_{k\ne{i}}\}_i,\{\lambda_i^{\mathrm{d}\star}\},$
$\{\lambda_i^{\mathrm{e}\star}\},\{\lambda_i^{\mathrm{f}\star}\},\{\lambda_i^{P\star}\},\{\lambda_{ik}^{\delta\star}\})$
such that{\small
\begin{align}\label{limit point}
\lim_{\ell\to\infty}\hat{\ub}[n_\ell]=
\hat{\ub}^\star,~\lim_{\ell\to\infty}\lambdab[n_\ell]=\lambdab^\star,
\end{align}}
\!\! where $(\hat{\ub}^\star,\lambdab^\star)$ is primal-dual
feasible to \eqref{eq:central_optsol_n_ite}. Consider \eqref{KKT
one} over the subsequence $\{n_1,\dots,n_\ell,\dots\}$. By taking
$\ell\rightarrow \infty$ in \eqref{KKT one}, and by
\eqref{eq:inner_approximation_d}, \eqref{claim 1 a} and \eqref{limit
point}, we obtain
\begin{align*}
\lambda_i^{\mathrm{b}\star}\frac{\partial\Theta_i(\{\hat{x}_{ji}^\star\}_j,\hat{y}_i^\star,\hat{z}_i^\star)}{\partial{x}_{ki}}+\lambda_{ki}^{\mathrm{c}\star}\frac{\partial\Psi_{ki}(\hat{\Wb}_k^\star,\hat{x}_{ki}^\star)}{\partial{x_{ki}}}=0,
%\lambda_i^{P\star} \Ib_{N_t}- (\lambda_i^{\mathrm{d}\star}+\lambda_{ii}^{\mathcal{S}\star})\Qb_{ii}
%  + \sum_{k\neq i}\left(\lambda_{ik}^{\mathrm{c}\star}[n]
%  \frac{\partial{\Psi}_{ik}(\hat{\Wb}_i^\star,
%   \hat{x}_{ik}^\star)}{\partial\Wb_i} - \lambda_{ik}^{\mathcal{S}\star}\Qb_{ik}
%  \right)\succeq \zerob,
\end{align*}
\!\!which is the Lagrangian-stationarity condition of
\eqref{UMX_ChVar} corresponding to $x_{ki}$. By applying similar
arguments above to all the other KKT conditions of
\eqref{eq:central_optsol_n_ite} and by Claims \ref{claim:x_converge}
and 2, we end up with the conclusion that $\hat\ub^\star$ satisfies
the KKT conditions of problem \eqref{UMX_ChVar} and thus is a
stationary point.

What remains is to show that any stationary point of \eqref{UMX_ChVar} is also a stationary point of \eqref{UMX_SDR} if the constraint set \eqref{additional constraint} is added in \eqref{UMX_SDR}. This can be proved by carefully examining the equivalence of the KKT conditions of the two problems. %The details are presented in Appendix \ref{}
Due to the space limitation, the detailed derivations are omitted
here. \hfill{$\blacksquare$}

As the SCA algorithm only guarantees to provide a stationary point,
the approximation accuracy depends on the initial point
$(\{\hat{\Wb}_i[0]\},\{\tilde{R}_i[0]\})$. A {possible choice is} to
initialize Algorithm 1 {via} some heuristic transmission strategies.
For example, one can obtain {an initial} point
$(\{\bar{\wb}_i\},\{\bar{R}_i\})$ of problem \eqref{UMX_CLSFORM}
through the simple maximum-ratio transmission (MRT) strategy. In
this strategy, the beamforming vectors $\{\bar{\wb}_i\}$ are simply
set to $\bar{\wb}_i = \sqrt{P_i} \qb_i$ where $\qb_i \in
\Cplx^{N_t}$ is the principal eigenvector of $\Qb_{ii}$ for
$i=1,\ldots,K$, with $\|\qb_i\|=1$. The corresponding rate
$\tilde{R}_i$ can be obtained by solving \eqref{bar R} with
$\{\hat{\Wb}_i\}=\{\bar{\wb}_i\bar{\wb}_i^H\}$.
%For the $i$th transmitter-receiver pair, the associated
%$\epsilon_i$-outage achievable rate of MRT is given by the maximum
%$\bar{R}_i$ that satisfies the following inequality [see
%\eqref{UMX_CLSFORM}]
%\begin{equation}\label{max_achievable_rate}
%\rho_i~
%\exp\left(\frac{(2^{\bar{R}_i}-1)\sigma_i^2}{\bar{\wb}_i^H\Qb_{ii}\bar{\wb}_i}\right)\prod_{k\ne{i}}
%\left(1+\frac{(2^{\bar{R}_i}-1)\bar{\wb}_k^H\Qb_{ki}\bar{\wb}_k}{\bar{\wb}_i^H\Qb_{ii}\bar{\wb}_i}\right)\le1.
%\end{equation}
Analogously, one can also obtain an initial point by the zero-forcing (ZF) transmission strategy,
provided that the column space of $\Qb_{ii}$ is not subsumed by the
column space of $\sum_{k\ne{i}}^K\Qb_{ik}$, for all $i=1,\dots,K$.

%\begin{align}
%\max_{\Wb_i\succeq\mathbf{0}}~&~\tr(\Wb_i\Qb_{ii})\\
%\text{s.t.}~&~\tr(\Wb_i\Qb_{ik})\le{e}^{\bar{x}_{ik}}\left(x_{ik}^\star-\bar{x}_{ik}+1\right),~k\in\mathcal{K}_i^c\\
%&~\tr(\Wb_i)\le{P}_i,
%\end{align}

%%%%%%%%%%%%%%%%%%%%%%%%%%%%%%%%%%%%%%%%%%%%%%%%%%%%%%%%%%%%%%%%%%
\section{Distributed Implementation}\label{sec:Decen_Alg}
%%%%%%%%%%%%%%%%%%%%%%%%%%%%%%%%%%%%%%%%%%%%%%%%%%%%%%%%%%%%%%%%%%

For Algorithm \ref{alg:centralized}, we have implicitly assumed that
there exists a control center in the network, collecting all the CDI
of users and {computing} the beamforming solution in a centralized
manner. In this section, we propose a distributed version for
Algorithm 1, where each transmitter $i$ only needs to optimize its own
beamformer, using only its local CDI, i.e., $\{\Qb_{ik}\}_k$,
and some information obtained from the other transmitters.
Since each of the subproblems involved has a much smaller problem size than the original problem \eqref{UMX_ChVar}, even for a centralized implementation, the proposed distributed optimization algorithm can be used to reduce the computation overhead of the control center.

The idea of the proposed distributed algorithm is to solve problem
\eqref{UMX_ChVar} from one transmitter to another, in a round-robin
fashion (i.e., the Gauss-Seidel fashion). Suppose that transmitter 1
optimizes its beamformer first, followed by transmitter 2 and so on,
and let $n$ denote the index of the current round. Then, in the
$n$th round, transmitter $i$ solves the following problem {
\begin{subequations}\label{UMX_subprob_CLSFORM}
\begin{align}
&(\hat{\wb}_i[n],\hat{R}_1[n,i],\dots,\hat{R}_K[n,i])=\mathrm{arg}\max_{\substack{\|\wb_i\|^2\le{P}_i\\R_1,\dots,R_K\ge0}}~U(R_1,\dots,R_K)\label{UMX_subprob_CLSFORM_a}\\
&~~~~~~\text{s.t.}~~~~\rho_i~\exp\left(\frac{(2^{R_i}-1)\sigma_i^2}{\wb_i^H\Qb_{ii}\wb_i}\right)
\prod_{k\ne{i}}\left(1+\frac{\left(2^{R_i}-1\right)e^{\bar{x}_{ki}
[n-u_{ki}]}}{\wb_i^H\Qb_{ii}\wb_i}\right)\le1,\label{UMX_subprob_CLSFORM_b}\\
&~~~~~~~~~~~~~\rho_j~\exp\left(\frac{(2^{R_j}-1)\sigma_j^2}
{e^{\bar{x}_{jj}[n-u_{ji}]}}\right)\left(1+\frac{\left(2^{R_j}-1\right)
\wb_i^H\Qb_{ij}\wb_i}{e^{\bar{x}_{jj}[n-u_{ji}]}}\right)\nonumber\\
&~~~~~~~~~~~~~~~~~~~\times\prod_{\substack{k\ne{j}\\k\ne{i}}}
\left(1+\frac{\left(2^{R_j}-1\right)e^{\bar{x}_{kj}[n-u_{ki}]}}
{e^{\bar{x}_{jj}[n-u_{ji}]}}\right)\le1,~j\in\mathcal{K}_i^c,\label{UMX_subprob_CLSFORM_c}
%\\
%&~~~~~~~~~~~~~~~~~\|\wb_i\|^2\le{P}_i,\label{UMX_subprob_CLSFORM_d}
\end{align}
\end{subequations}}\hspace{-0.2cm}
where
$\bar{x}_{kj}[n-u_{ki}]\!=\!\ln(\hat{\wb}_k^H[n-u_{ki}]\Qb_{kj}\hat{\wb}_k[n-u_{ki}])$,
and $u_{ki}$ is equal to one if $k>i$ and zero otherwise.
%where
%\begin{equation}
%\bar{x}_{kj}[n-u_{kj}]=\ln\left(\hat{\wb}_k^H[n-u_{kj}]\Qb_{kj}\hat{\wb}_k[n-u_{kj}]\right),~k,j=1,\dots,K,\label{distributed_xbar}
%\end{equation}
%and $u_{kj}$ is equal to one if {\red$k\ge{j}$} and zero otherwise.
%
%and $u_{kj}$ is defined as
%\[
%u_{kj}=\left\{\begin{array}{ll}
%            1,&~\text{for}~k-j\ge0\\
%            0,&~\text{otherwise}.
%            \end{array}\right.
%\]

Note that for \eqref{UMX_subprob_CLSFORM}, only $\hat\wb_i[n]$ is
optimized while $\{\bar{x}_{kj}[n-u_{kj}]\}_{k\neq i,j}$ are given.
{Once the beamforming solution of \eqref{UMX_subprob_CLSFORM} is
obtained, $\{\bar{x}_{ik}[n]\}_k$ are updated according to the
optimal $\hat \wb_i[n]$ and then passed to all the other
transmitters for their subsequent beamforming
optimization\footnote{In this paper, we assume that the
communication between transmitters for message exchange is
error-free.}. There are two interesting points to note {here}.
Firstly, as {can be} seen {from} \eqref{UMX_subprob_CLSFORM_b} and
\eqref{UMX_subprob_CLSFORM_c}, transmitter $i$ not only optimizes
its rate $R_i$, but also {takes into account} the rate outage
constraints for all the other users. The constraints in
\eqref{UMX_subprob_CLSFORM_c} {indicate} that transmitter $i$ {needs
to} regulate its own transmission in order not to violate the outage
{requirement} of the other users. Secondly, {to solve}
\eqref{UMX_subprob_CLSFORM}, transmitter $i$ only needs the local
CDI,
i.e., $\{\Qb_{ik}\}_k$. %, in contrast to Algorithm
%\ref{alg:centralized}.

Similar difficulties arise here as in problem \eqref{UMX_CLSFORM}
since problem \eqref{UMX_subprob_CLSFORM} is not convex. We {hence}
apply the same approximation techniques in Section
\ref{sec:ConsrvApprx} to {approximate} \eqref{UMX_subprob_CLSFORM}.
In particular, we first apply SDR, followed by the reformulation as
described by \eqref{change of variables}, and {the} first-order
approximations in \eqref{eq:Linrz}. {The resulting convex
optimization problem reads}
\begin{subequations}\label{UMX_subprob_aprx}
\begin{align}
&(\hat{\Wb}_i[n],\{\hat{R}_k[n,i]\},\{\hat{x}_{ik}[n]\}_k,\{\hat{y}_k[n,i]\},\{\hat{z}_k[n,i]\})={\mathrm{arg}}\max_{\substack{\Wb_i\in\mathcal{S}_i,R_k,x_{ik},y_k,z_k\\k=1,\dots,K}}~U(R_1,\dots,R_K)\label{UMX_subprob_aprx_a}\\
&~~~~\text{s.t.}~~\rho_ie^{\sigma_i^2z_i}\prod_{k\ne{i}}\left(1+e^{-x_{ii}+\bar{x}_{ki}[n-u_{ki}]+y_i}\right)\le1,\label{UMX_subprob_aprx_b}\\
&~~~~~~~~~\rho_je^{\sigma_j^2z_j}\left(1+e^{-\bar{x}_{jj}[n-u_{ji}]+x_{ij}+y_j}\right)\!\!\prod_{k\ne{j},k\ne{i}}\!\!\left(1+e^{-\bar{x}_{jj}[n-u_{ji}]+\bar{x}_{kj}[n-u_{ki}]+y_j}\right)\le1,~j\in\mathcal{K}_i^c,\label{UMX_subprob_aprx_c}\\
&~~~~~~~~~\tr\left(\Wb_i\Qb_{ii}\right)\ge{e}^{x_{ii}},\label{UMX_subprob_aprx_d}\\
&~~~~~~~~~\tr\left(\Wb_i\Qb_{ik}\right)\le{e}^{\bar{x}_{ik}[n-1]}\left(x_{ik}-\bar{x}_{ik}[n-1]+1\right),~k\in\mathcal{K}_i^c,\label{UMX_subprob_aprx_e}\\
&~~~~~~~~~{R_j}\le\frac{1}{\ln2}\left(\ln(1+e^{\bar{y}_j[n,i-1]})+\frac{e^{\bar{y}_j[n,i-1]}}{1+e^{\bar{y}_j[n,i-1]}}(y_j-\bar{y}_j[n,i-1])\right),~j=1,\dots,K,\label{UMX_subprob_aprx_f}\\
&~~~~~~~~~e^{y_i-x_{ii}}\le{z}_i,~e^{y_j-\bar{x}_{jj}[n-u_{ji}]}\le{z}_j,~j\in\mathcal{K}_i^c,\label{UMX_subprob_aprx_h}
%&~~~~~~~~~~~~~~~~~\tr\left(\Wb_i\right)\le{P}_i,\label{UMX_subprob_aprx_i}
\end{align}
\end{subequations}
where
$\mathcal{S}_i\triangleq\{\Wb_i\succeq\mathbf{0}|~\tr(\Wb_i)\le{P_i},~\tr(\Wb_i\Qb_{ik})\ge\delta,k=1,\ldots,K\}$,
\begin{align}
\bar{x}_{kj}[n-u_{ki}]&=\ln\left(\tr(\hat{\Wb}_k[n-u_{ki}]\Qb_{kj})\right),\label{distributed_xbar}\\
\bar{y}_j[n,i-1]&=\ln\left(2^{\tilde{R}_j[n,i-1]}-1\right),\label{distributed_ybar}
\end{align}
for $j,k=1,\dots,K$, and, similar to \eqref{bar R2},
$\tilde{R}_j[n,i]\geq \hat {R}_j[n,i]$ is obtained by solving the
following equations
\begin{equation}\label{bar R3}
\rho_j~\exp\left(\frac{(2^{R_j}-1)\sigma_j^2}{e^{\bar{x}_{jj}[n-u_{ji}]}}\right)\prod_{k\ne{j}}\left(1+(2^{R_j}-1)e^{\bar{x}_{kj}[n-u_{ki}]-\bar{x}_{jj}[n-u_{ji}]}\right)=1,
\end{equation}
for $j=1,\ldots,K.$ It is worth {mentioning} that {problem
\eqref{UMX_subprob_aprx} is only solved once and successive
approximation is not performed as in Algorithm
\ref{alg:centralized}. As long as problem \eqref{UMX_subprob_aprx}
is solved by transmitter $i$, the algorithm directly goes to the
next optimization problem performed by transmitter $i+1$.}
Successive approximation is {now} performed implicitly from one
transmitter to another in a round-robin fashion. We summarize the
proposed distributed SCA algorithm in Algorithm
\ref{alg:distributed_modified}.

{\footnotesize
\begin{algorithm}[h]
  \caption{Distributed SCA algorithm for solving problem \eqref{UMX}}
\begin{algorithmic}[1]\label{alg:distributed_modified}
  \STATE {\bf Given} an initial beamforming matrix $\hat{\Wb}_i[0]$ at transmitter $i$, for $i=1,\ldots,K$.
  \STATE For all $i=1,\dots,K$, transmitter $i$ computes $\{\bar{x}_{ik}[0]\}_k$ by
  \eqref{distributed_xbar}, and pass them to the other
  transmitters.
  \STATE Set $n=0$
  \REPEAT
    \STATE $n=n+1$
    \FOR{$i=1,\dots,K$}
    \STATE User $i$ solves \eqref{bar R3} to obtain
      $\{\tilde{R}_j[n,i-1]\}_j$ and compute $\{\bar{y}_j[n,i-1]\}_j$, followed by solving \eqref{UMX_subprob_aprx} to
      obtain the solution
      $(\hat{\Wb}_i[n],\{\hat{R}_k[n,i]\},\{\hat{x}_{ik}[n]\}_k,\{\hat{y}_k[n,i]\},
      \{\hat{z}_k[n,i]\})$.
    \STATE User $i$ computes $\{\bar{x}_{ik}[n]\}_k$ by
      \eqref{distributed_xbar} and {passes them to all} the other
      transmitters.
    \ENDFOR
  \UNTIL the predefined stopping criterion is met.
  \STATE For $i=1,\ldots,K$, each transmitter $i$ decomposes $\hat{\Wb}_i[n]=\hat{\wb}_i\hat{\wb}_i^H$, if $\hat{\Wb}_i[n]$ is of rank one; otherwise perform Gaussian randomization to obtain a rank-one approximate solution. %The Gaussian randomization for transmitter $i$ is similar to that in Table I, except that only $\hat\Wb_i[n]$ is considered and \eqref{max_achievable_rate} is replaced by \eqref{bar R3}.
  %\STATE {\bf Output} {\blue $\{\hat{\wb}_i\}$ and the associated rate tuple ${\red\{\tilde{R}_i\}}$, which is obtained by solving \eqref{max_achievable_rate}, as the approximate solution of problem
%\eqref{UMX}}.
\end{algorithmic}
\end{algorithm}}

Analogous to Algorithm \ref{alg:centralized}, we can show that Algorithm
\ref{alg:distributed_modified} generates a stationary point of
problem \eqref{UMX_ChVar} as stated in the following theorem.

\begin{Theorem}\label{thm:distributed_convergence} Suppose
that $U(R_1,\dots,R_K)$ is differentiable and is strictly increasing
with respect to $R_i$, for $i=1,\dots,K$. Then, the sequence
$\{U({\tilde{R}_1[n,i],\dots,\tilde{R}_K[n,i]})\}_{n=1}^\infty$
generated by Algorithm \ref{alg:distributed_modified} converges to a
common value for all $i=1,\dots,K$. Moreover, for all $i$, any limit
point of the sequence
$\{(\hat{\Wb}_1[n],\ldots,\hat{\Wb}_k[n],\tilde{R}_1[n,i],\ldots,\tilde{R}_1[n,i])\}_{n=1}^\infty$
is a stationary point of problem \eqref{UMX_ChVar} as well as a
stationary point of problem \eqref{UMX_SDR}
$\mathrm{(with~the~extra~constraints~in~\eqref{additional
constraint})}$.
\end{Theorem}

Different from the proof for Theorem 1, the proof for Theorem 2 is
more involved, since the beamforming vectors of transmitters are not
simultaneously optimized as in Algorithm 1 but are individually
optimized in a round-robin manner. The detailed proof of Theorem 2
is presented in Appendix \ref{sec:proof_converge_distributed}.

\begin{Remark}{\rm An important issue concerning
distributed optimization algorithms is the communication overhead
introduced by message exchange between transmitters. To address
this, we compare the communication overhead of the proposed
Algorithm \ref{alg:distributed_modified} with the following two
schemes. Scheme 1: All the transmitters directly exchange their CDI
so that the design problem \eqref{UMX} can be handled independently
by each transmitter. Scheme 2: A control center gathers the CDI from
all transmitters, optimizes the beamforming vectors, and distributes
the beamforming solutions to the transmitters. We consider a
cellular system where all the transmitters (i.e., BSs) are connected
by dedicated backhaul links (e.g., optical fibers) and the BSs
exchange messages in a point-to-point fashion. Since, in Algorithm
\ref{alg:distributed_modified}, transmitter $i$ needs to inform
$\{\bar{x}_{ik}[n]\}_k$ ($K$ real values) to all the other $K-1$
transmitters in each round, the communication overhead due to
transmitter $i$ is quantified by the amount of $K(K-1)$ real values.
Hence, the total communication overhead of Algorithm
\ref{alg:distributed_modified} is $N\times K\times K(K-1)=
K^2(K-1)N$ real values, where $N$ is the number of rounds run by
Algorithm \ref{alg:distributed_modified}. For scheme 1, each
transmitter needs to send $K$ covariance matrices (which contain
$KN_t^2$ real values) to all the other $K-1$ transmitters.
Therefore, the associated total communication overhead is given by
$K\times(K-1)\times KN_t^2=K^2(K-1)N_t^2$ real values. Therefore,
for scheme 1, if $N<N_t^2$, then the proposed Algorithm
\ref{alg:distributed_modified} has a smaller amount of communication
overhead. For scheme 2, there are $K^2$ covariance matrices sent
from the transmitters to the control center, and the optimal
solution $\{\wb_i^\star,R_i^\star\}$ passed from the control center
to transmitter $i$ for $i=1,\dots,K$, respectively. Hence, the
communication overhead is $K^2N_t^2+K(2N_t+1)$ real values.
Therefore, for scheme 2, the proposed Algorithm
\ref{alg:distributed_modified} has a smaller amount of communication
overhead if $N<N_t^2/(K-1)+(2N_t+1)/(K^2-K)$.
%From the above
%analysis, we can see that scheme 2 is the most efficient one in
%terms of the communication overhead under point-to-point message
%exchange when the ratio $N_t^2/K$ is small, i.e., when the number of
%transmitter-receiver pairs increases or only a small number of
%transmit antennas is available; otherwise, the proposed Algorithm
%\ref{alg:distributed_modified} should has the least communication
%overhead
As we show in the simulation section, Algorithm
\ref{alg:distributed_modified} in general can converge in less than
15 rounds for a system with $K\le6$ and $N_t=8$.

We should mention here that, while in general the proposed
distributed algorithm is more efficient in terms of computation and
communication overhead, it may result in larger transmission delays
(due to the iterative optimizations between transmitters) compared
with the centralized schemes.}
\end{Remark}
\vspace{-0.2cm}
\begin{Remark}{\rm
We should emphasize that the proposed beamforming design is based on
the users' statistical channel information, which usually changes
{much more} slowly compared to the instantaneous CSI, so beamforming
optimization need not be performed frequently. As a result, the
throughput loss induced by the round-robin optimization in Algorithm
2 {should} not be a serious concern.}
\end{Remark}

%According to Theorem \ref{thm:distributed_convergence}, if the
%$\{\hat{\Wb}_i[n]\}$ obtained by \eqref{UMX_subprob_aprx} are of
%rank one, then $\{\hat{\wb}_i\}$ is a stationary point of the
%original outage constrained problem {\red\eqref{UMX_CLSFORM} with
%the extra constraint $\wb_i^H\Qb_{ii}\wb_i$}.
%In the next section,
%we present some simulation results to demonstrate the near-optimal
%performance of the proposed algorithms.

\vspace{-0.4cm}
%%%%%%%%%%%%%%%%%%%%%%%%%%%%%%%%%%%%%%%%%%%%%%%%%%%%%%%%%%%%%%%%%%
\section{Simulation Results}\label{sec:SimuRslt}
%%%%%%%%%%%%%%%%%%%%%%%%%%%%%%%%%%%%%%%%%%%%%%%%%%%%%%%%%%%%%%%%%%

In the section, we demonstrate the performance of the proposed
Algorithm \ref{alg:centralized} and Algorithm
\ref{alg:distributed_modified} for solving the outage constrained
coordinated beamforming problem in \eqref{UMX}. In the simulations,
we consider $\beta=0$, $\beta=1$, and $\beta=2$ for the objective
function $U_{\beta}(R_1,\dots,R_K)$, corresponding to maximization
of the weighted sum rate, the weighted geometric mean rate, and the
weighted harmonic mean rate, respectively. All receivers are assumed
to have the same noise power, i.e., $\sigma_1^2=\cdots=\sigma_K^2
\triangleq \sigma^2$, and all power constraints are set to one,
i.e., $P_1=\cdots=P_K=1$. The parameter $\delta$ in
\eqref{additional constraint} is set to $10^{-5}$. The channel
covariance matrices $\Qb_{ki}$ are randomly generated. We normalize
the maximum eigenvalue of $\Qb_{ii}$, i.e.,
$\lambda_{\max}(\Qb_{ii})$, to one for all $i$, and normalize
$\lambda_{\max}(\Qb_{ki})$ to a value $\eta  \in(0,1]$ for all $k\in
 \mathcal{K}^c_i$, $i=1,\ldots,K$. The parameter $\eta$, thereby, represents the relative cross-link interference
level. If not mentioned specifically, all $\Qb_{ki}$ are of full
rank, and the outage probability {requirements are} set to {the same
value, i.e.,} $\epsilon_1=\cdots=\epsilon_K=0.1$, indicating a
$10\%$ outage probability. The stopping criterion
of Algorithm 1 is {\small\[
\frac{|U(\tilde{R}_1[n],\dots,\tilde{R}_K[n])-U(\tilde{R}_1[n-1],\dots,\tilde{R}_K[n-1])|}{U(\tilde{R}_1[n-1],\dots,\tilde{R}_K[n-1])}<0.01.
\]}\hspace{-.19cm}
That is, Algorithm \ref{alg:centralized} stops if the improvement in
system utility is less than $1\%$ of the system utility achieved in
the previous iteration. The simple MRT solution is used to
initialize {both} Algorithm 1 and Algorithm 2. The convex solver
\texttt{CVX} \cite{cvx} is used to solve {the convex problems
\eqref{eq:central_optsol_n_ite} and \eqref{UMX_subprob_aprx}.}

%%%%%%%%%%%%%%%%%%%%%%%%%%%%%%%%%%%%%%%%%%%%%%%%%%%%%%%%%%%%%%%%%%%%%%%%%%%%%%%%%%%%%%%%%%%
\begin{figure}[t]
\begin{center}
\subfigure[][]{\resizebox{.49\textwidth}{!}{\includegraphics{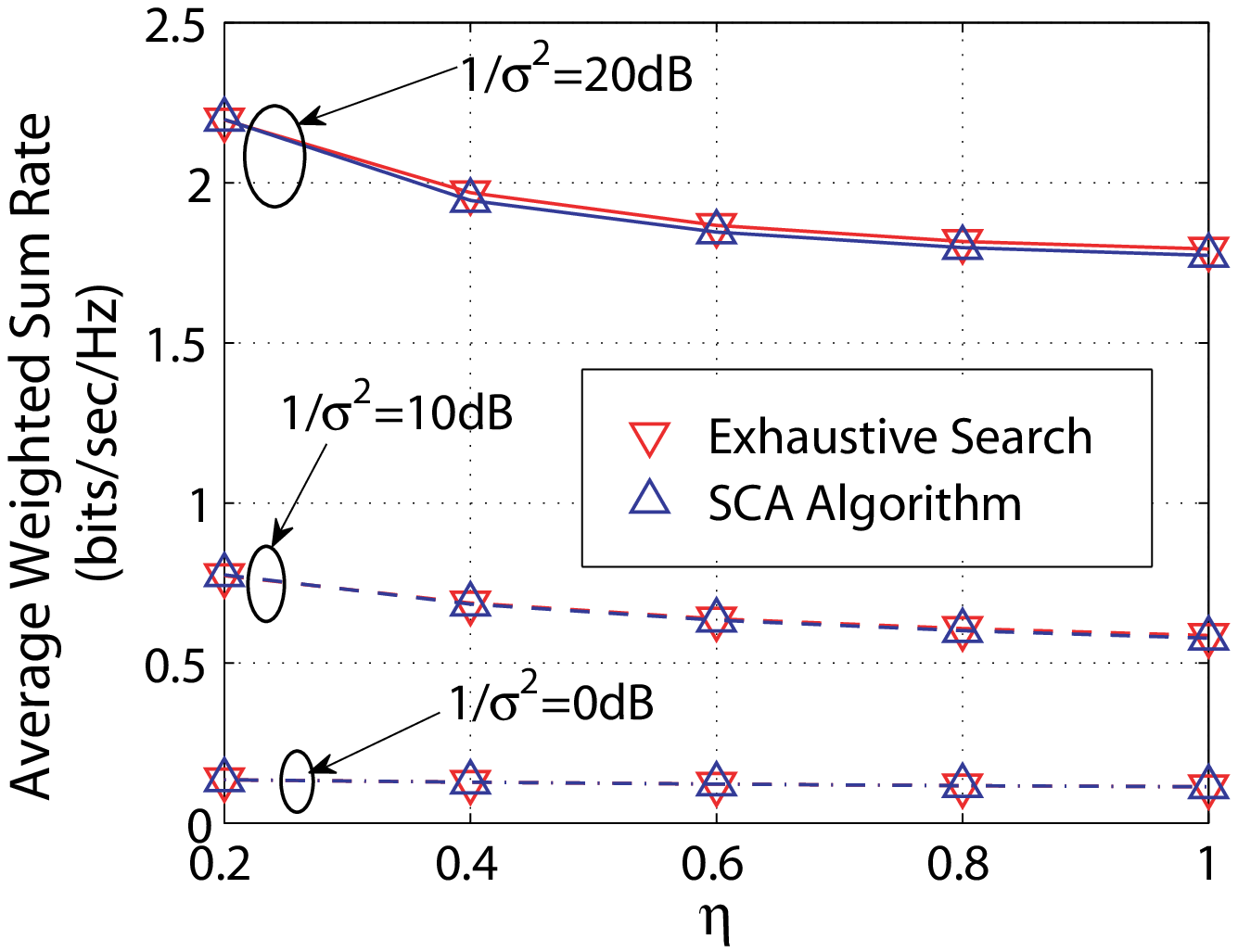}}}\label{fig:fig1_a}
\subfigure[][]{\resizebox{.49\textwidth}{!}{\includegraphics{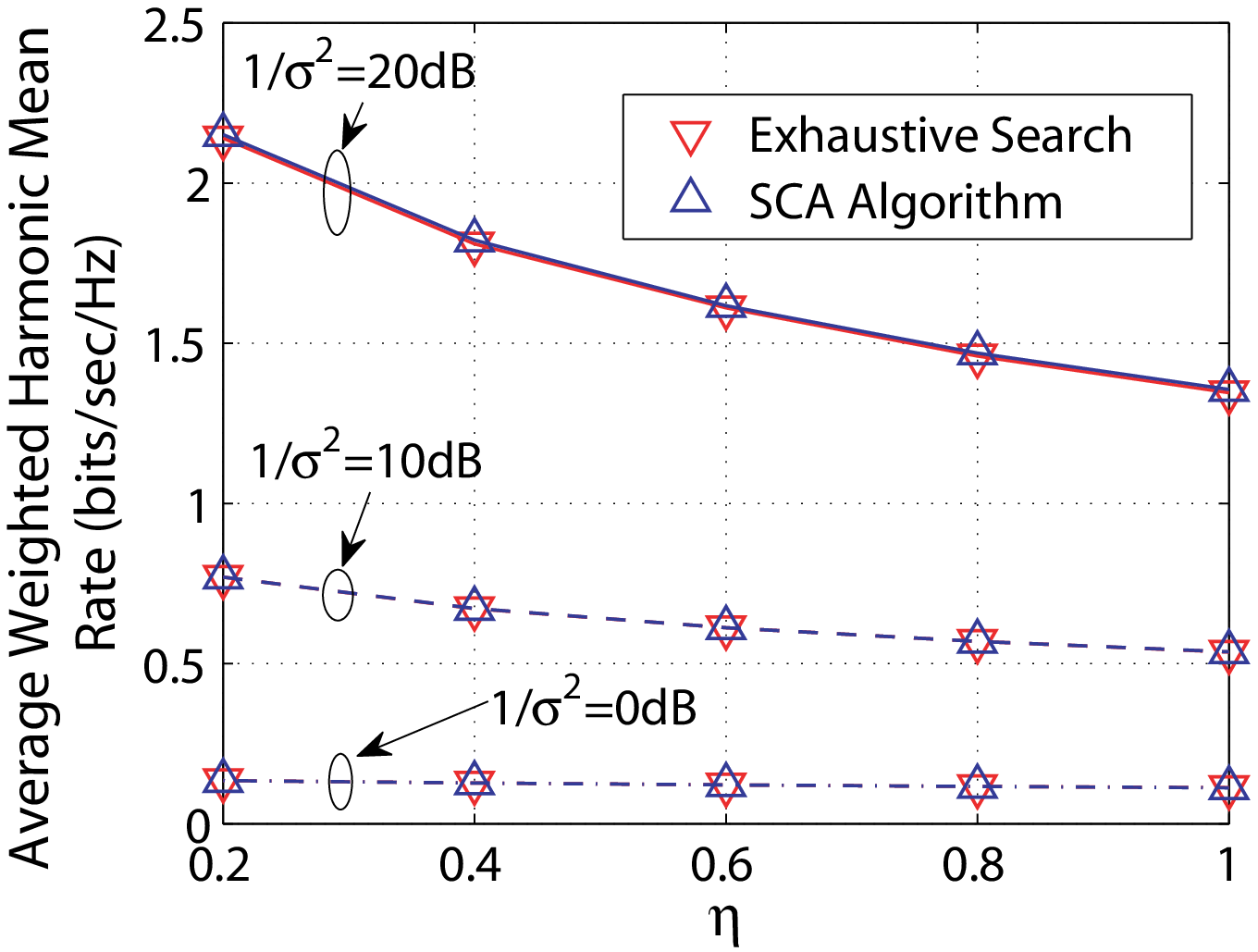}}}\label{fig:fig1_b}
\end{center}\vspace{-0.7cm}
\caption{Simulation results of the proposed SCA algorithm (Algorithm 1), for $K=2$, $N_t=4$, and
$(\alpha_1,\alpha_2)=(\frac{1}{2},\frac{1}{2})$; (a) weighted sum rate versus $\eta$, (b) weighted harmonic mean rate versus $\eta$. Each of the results is obtained by averaging over 500 realizations of $\{\Qb_{ki}\}$.} \label{fig:fig1}
\vspace{-0.6cm}
\end{figure}
%%%%%%%%%%%%%%%%%%%%%%%%%%%%%%%%%%%%%%%%%%%%%%%%%%%%%%%%%%%%%%%%%%%%%%%%%%%%%%%%%%%%%%%%%%%

{\bf Example 1:} We first examine the approximation performance of
the proposed SCA algorithm, by comparing it with the exhaustive
search method in \cite{Lindblom09}. In view of the {tremendous}
complexity overheads of this exhaustive search method, we consider a
simple case {where only} two transmitter-receiver pairs {are
present}, i.e. $K=2$, and set $N_t=4$. Figure \ref{fig:fig1}(a)
shows the simulation results {for the comparison of the achievable}
weighted sum rate {between the proposed SCA algorithm and the
exhaustive search method against the cross-link interference level
$\eta$, where the weights are given by
$(\alpha_1,\alpha_2)=(\frac{1}{2},\frac{1}{2})$}. Each {simulation
curve} is obtained by averaging over 500 realizations of {randomly
generated} $\{\Qb_{ki}\}$. {From this figure, we can observe} that,
for $1/\sigma^2=0$ dB and $1/\sigma^2=10$ dB, the proposed SCA
algorithm can attain almost the same average sum rate performance as
the exhaustive search method, {indicating} that the proposed SCA
algorithm yields near-optimal solutions {for} the outage constrained
beamforming design problem \eqref{UMX}. For $1/\sigma^2=20$ dB, it
can be observed that there is a small gap between the rate achieved
by the proposed SCA algorithm and that by the exhaustive search
method. Nonetheless, this gap is relatively small and is within
$2\%$ of the sum rate achieved by the exhaustive search method.
Figure \ref{fig:fig1}(b) displays the simulation results under the
same setting as in Figure \ref{fig:fig1}(a) except that the
objective function is now the average harmonic mean rate. As the
mean rate performance of SCA algorithm is almost the same as that of
the exhaustive search method, its solution is nearly optimal for
problem \eqref{UMX}.
%It can be seen that the SCA algorithm yields
%almost the same average harmonic mean rate performance as the
%exhaustive search method, so is nearly optimal for problem
%\eqref{UMX}.

%It can be seen that the proposed SCA
%algorithm yields almost the same average harmonic mean rate
%performance {compared} with the exhaustive search method, and is
%nearly optimal {for} problem \eqref{UMX} as well.

%%%%%%%%%%%%%%%%%%%%%%%%%%%%%%%%%%%%%%%%%%%%%%%%%%%%%%%%%%%%%%%%%%%%%%%%%%%%%%%%%%%%%%%%%%%
\begin{figure}[t]
\begin{center}
\subfigure[][]{\resizebox{.49\textwidth}{!}{\includegraphics{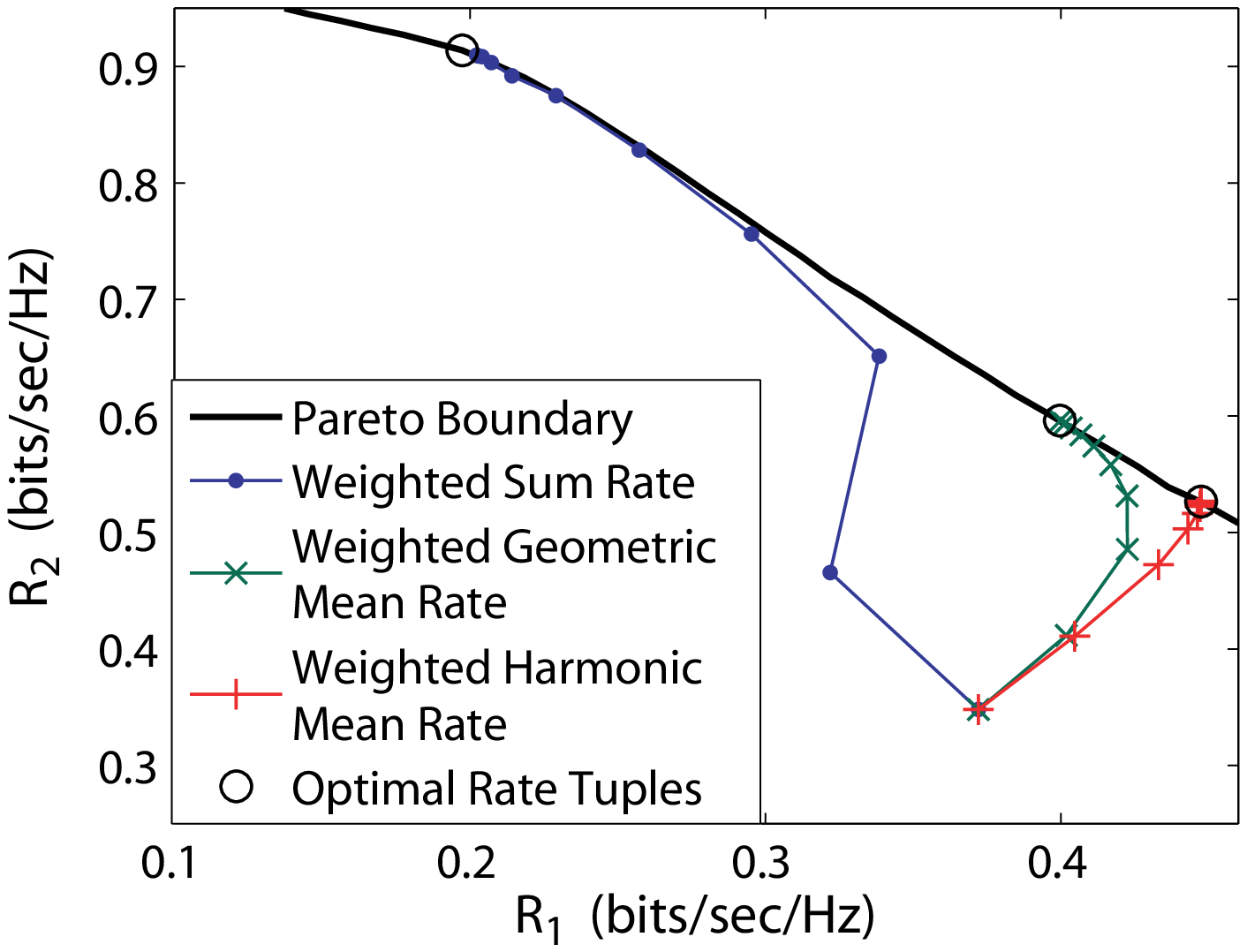}}}\label{fig:fig2_a}
\subfigure[][]{\resizebox{.49\textwidth}{!}{\includegraphics{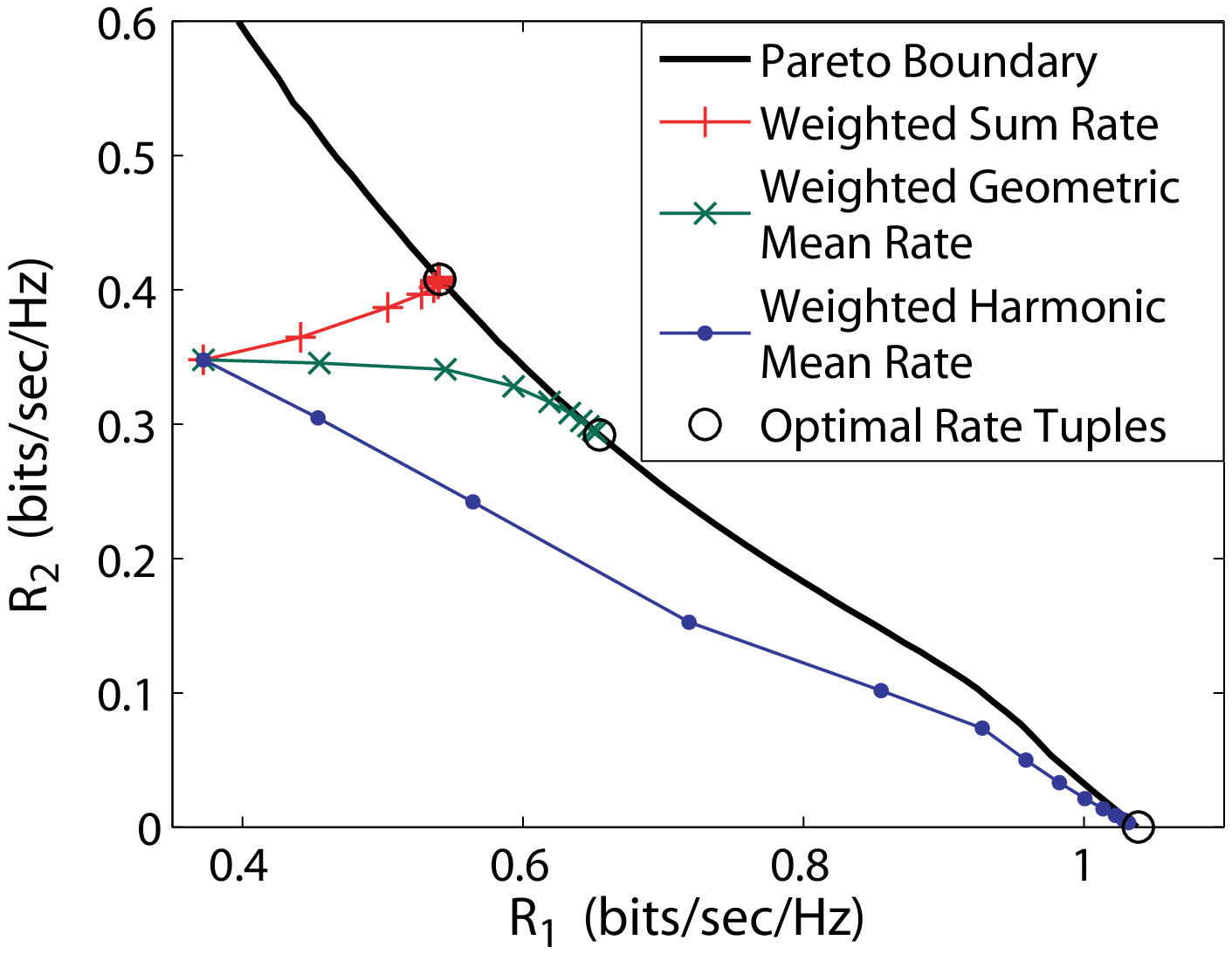}}}\label{fig:fig2_b}
\end{center}\vspace{-0.7cm}
\caption{Converge trajectories of the proposed SCA algorithm. $K=2$,
$N_t=4$, $\eta=0.4$; (a)
$(\alpha_1,\alpha_2)=(\frac{1}{2},\frac{1}{2})$, (b)
$(\alpha_1,\alpha_2)=(\frac{2}{3},\frac{1}{3})$. The results are
obtained using a typical set of randomly generated $\{\Qb_{ki}\}$. }
\label{fig:fig2}\vspace{-0.6cm}
\end{figure}
%%%%%%%%%%%%%%%%%%%%%%%%%%%%%%%%%%%%%%%%%%%%%%%%%%%%%%%%%%%%%%%%%%%%%%%%%%%%%%%%%%%%%%%%%%%

To {examine} how the proposed SCA algorithm converges, we illustrate
in Figure \ref{fig:fig2}(a) the trajectories of the optimal rate
tuple of problem \eqref{eq:central_optsol_n_ite} in each iteration
of Algorithm 1, {where} the weighted sum rate, the geometric mean
rate, and the harmonic mean rate are all considered. The user
priority {weights are} set to
$(\alpha_1,\alpha_2)=(\frac{1}{2},\frac{1}{2})$, and the Pareto
boundary is obtained by the exhaustive search method in
\cite{Lindblom09}. One can see from this figure that, for {all} rate
utility functions, the proposed SCA algorithm first approaches the
Pareto boundary and then converges to the corresponding optimal rate
tuple {along the boundary}. In Figure \ref{fig:fig2}(b), we display
similar results with {an} asymmetric user priority, {i.e.,}
$(\alpha_1,\alpha_2)=(\frac{2}{3},\frac{1}{3})$. It can be observed
that the SCA algorithm still converges to the optimal rate tuples
{in a similar fashion}.

%As seen in figure \ref{fig:fig1}(a), the average performance of
%centralized SCA algorithm is very close to the global optimal
%solution. In fact, It is observed in the numerical simulation that,
%for $K=2$, the SCA algorithm converges to the global optimal of
%problem \eqref{UMX} in most of the problem instances we tested. To
%show this, we plot the convergence trajectory of SCA algorithm,
%i.e., the rate tuples achieved by the solution of
%\eqref{eq:central_optsol_n_ite} in each iteration for different
%system utilities, and show that it converges to the optimal solution
%obtained by the exhaustive search method which samples 100 points on
%the Pareto boundary. In figure \ref{fig:fig3}(a), we set the user
%priority weights to $(\alpha_1,\alpha_2)=(\frac{1}{2},\frac{1}{2})$,
%and it can be seen that the solutions obtained by SCA algorithm
%{\color{red} almost overlap the solution obtained by exhaustive
%search method. The slight difference between them is due to the
%limited precision of the exhaustive search method.} In figure
%\ref{fig:fig3}(b), we show the convergence of SCA algorithm for an
%asymmetric weight, $(\alpha_1,\alpha_2)=(\frac{2}{3},\frac{1}{3})$.
%It is observed that SCA algorithm converges to the correct position
%in the outage constrained achievable rate region.
%
%

%%%%%%%%%%%%%%%%%%%%%%%%%%%%%%%%%%%%%%%%%%%%%%%%%%%%%%%%%%%%%%%%%%%%%%%%%%%%%%%%%%%%%%%%%%%
\begin{figure}[t]
\begin{center}
\subfigure[][]{\resizebox{.49\textwidth}{!}{\includegraphics{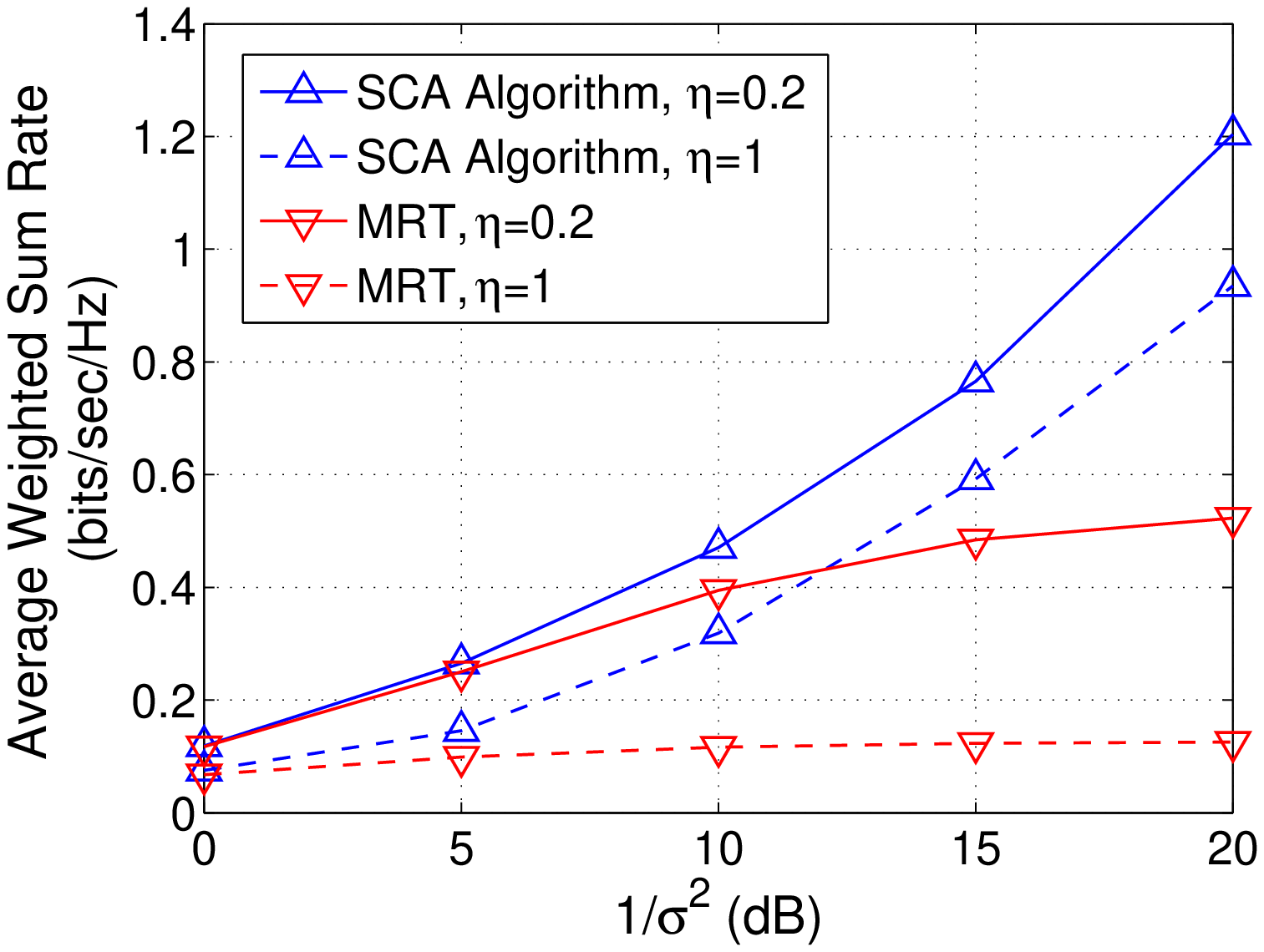}}}\label{fig:fig3_a}
\subfigure[][]{\resizebox{.49\textwidth}{!}{\includegraphics{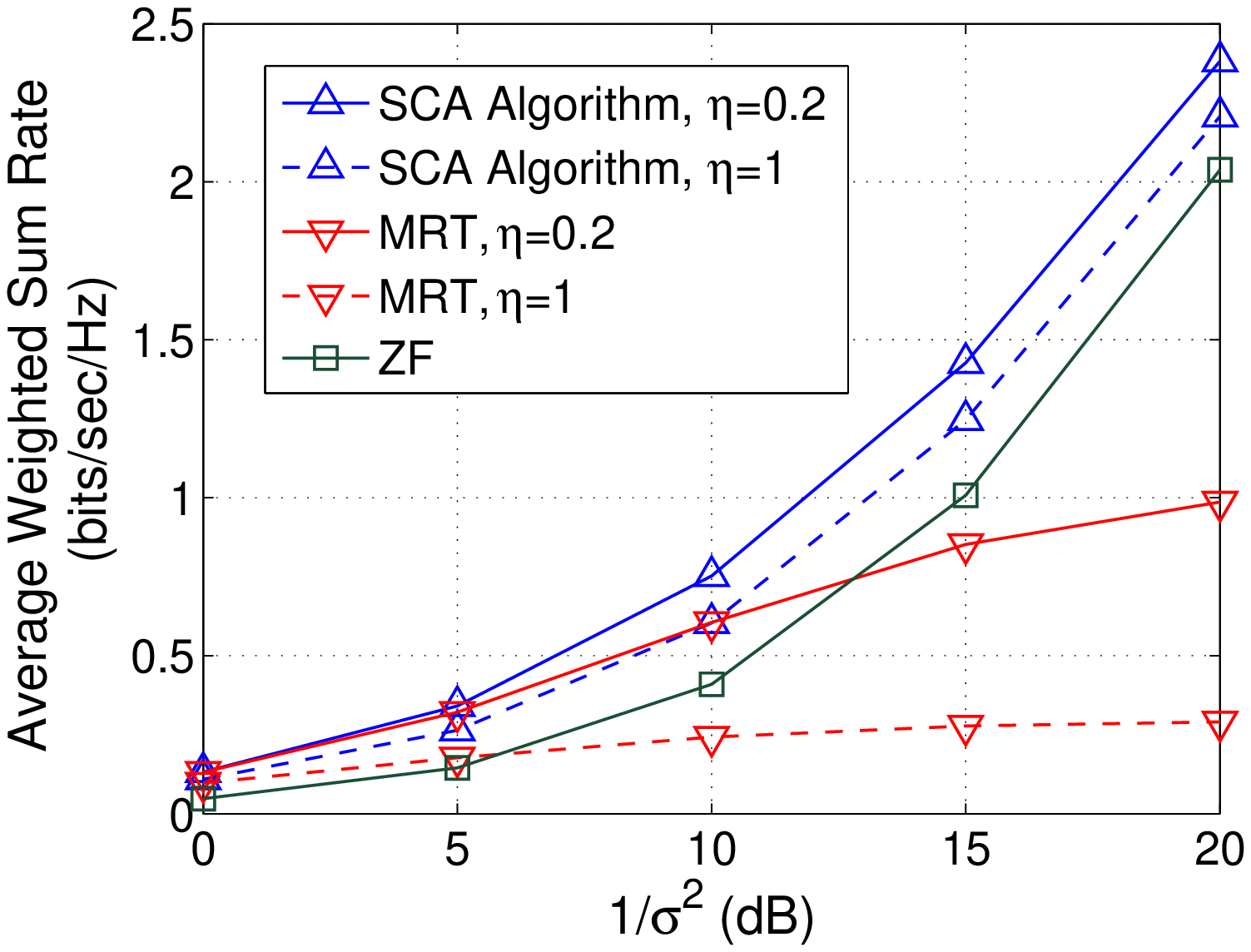}}}\label{fig:fig3_b}
\end{center}\vspace{-0.7cm}
\caption{Simulation results of average achievable sum rate versus
$1/\sigma^2$; (a) $K=N_t=4$, and full rank $\{\Qb_{ki}\}$, (b)
$K=4$, $N_t=8$ and $\rank(\Qb_{ki})=2$ for all $k,i$. The priority
weights are set to
$(\alpha_1,\alpha_2,\alpha_3,\alpha_4)=(\frac{1}{4},\frac{1}{4},\frac{1}{4},\frac{1}{4})$.
The results are obtained by averaging over 500 realizations of
$\{\Qb_{ki}\}$.} \label{fig:fig3}\vspace{-0.6cm}
\end{figure}
%%%%%%%%%%%%%%%%%%%%%%%%%%%%%%%%%%%%%%%%%%%%%%%%%%%%%%%%%%%%%%%%%%%%%%%%%%%%%%%%%%%%%%%%%%%

{\bf Example 2:}  To further {demonstrate} the {effectiveness} of
the proposed SCA algorithm, we {evaluate the performance of} the SCA
algorithm for the case of $K=N_t=4$ in this example. (Since under
this setting, the exhaustive search method in \cite{Lindblom09} is
too complex to implement, and, to the best of our knowledge, there
is no existing method for comparison, we {can} only compare
the proposed SCA algorithm with the heuristic MRT and ZF schemes.)
Figure \ref{fig:fig3}(a) shows the simulation results of {the}
average achievable sum rate versus $1/\sigma^2$. From this figure,
{one can observe} that the proposed SCA algorithm yields
better sum rate performance than the MRT scheme, especially when
$1/\sigma^2> 5$ dB.
For $1/\sigma^2\leq 5$ dB, the two methods exhibit comparable performance. %Since MRT is known to be asymptotically optimal in low SNR, the results in
%Figure \ref{fig:fig3}(a) for $1/\sigma^2\leq 5$ dB somehow implies the near-optimality of the proposed SCA algorithm.
In Figure \ref{fig:fig3}(b), we have shown the simulation results
for $K=4$, $N_t=8$ and $\rank(\Qb_{ki})=2$ for all $k,i$. Under this
setting, the ZF scheme is feasible and its average sum rate
performance is {also} shown in Figure \ref{fig:fig3}(b). It can be
observed from this figure that the ZF scheme outperforms the MRT
scheme for high $1/\sigma^2$ or when the cross-link interference is
strong ($\eta=1$). Nevertheless, as {can be} seen from Figure
\ref{fig:fig3}(b), the proposed SCA algorithm still outperforms both
the MRT and {the} ZF schemes.

%%%%%%%%%%%%%%%%%%%%%%%%%%%%%%%%%%%%%%%%%%%%%%%%%%%%%%%%%%%%%%%%%%%%%%%%%%%%%%%%%%%%%%%%%%%
\begin{figure}[t]
\begin{center}
\subfigure[][]{\resizebox{.49\textwidth}{!}{\includegraphics{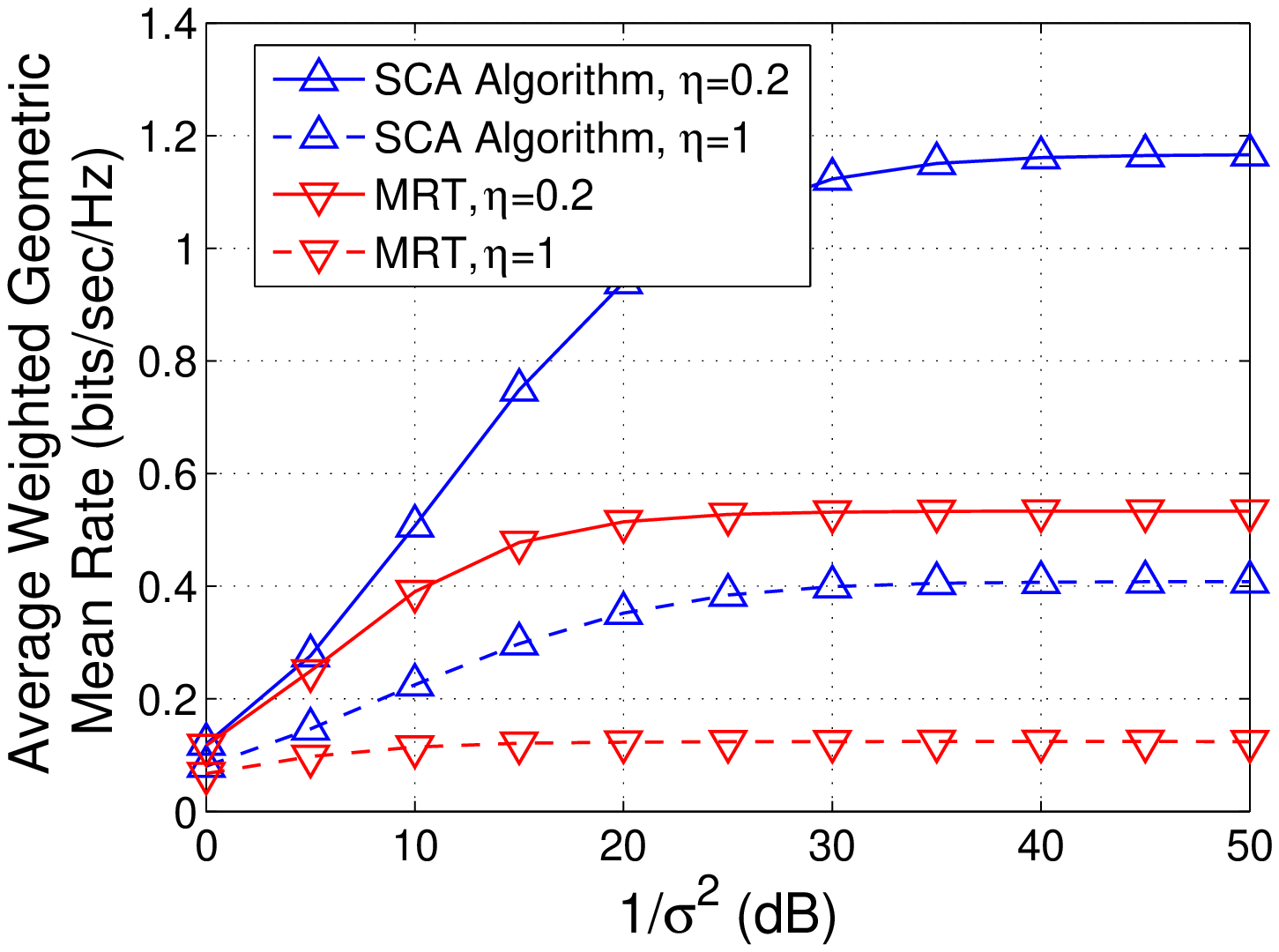}}}\label{fig:fig4_a}
\subfigure[][]{\resizebox{.49\textwidth}{!}{\includegraphics{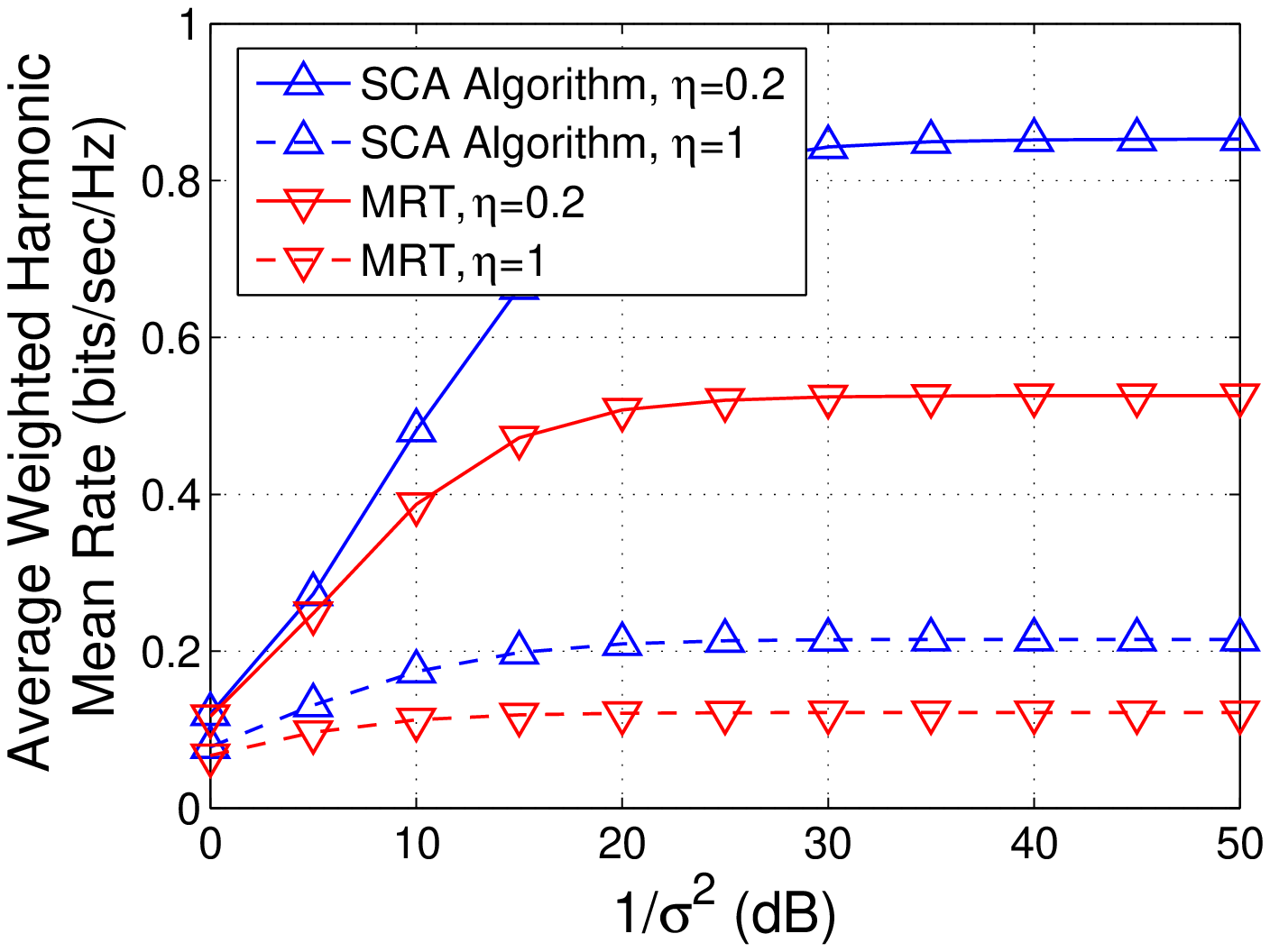}}}\label{fig:fig4_b}
\end{center}\vspace{-0.7cm}
\caption{Simulation results of the proposed SCA algorithm (Algorithm
1), for $K=N_t=4$ and
$(\alpha_1,\alpha_2,\alpha_3,\alpha_4)=(\frac{1}{8},\frac{1}{8},\frac{1}{4},\frac{1}{2})$;
(a) weighted geometric mean rate versus $1/\sigma^2$, (b) weighted
harmonic mean rate versus $1/\sigma^2$. Each of the results is
obtained by averaging over 500 realizations of $\{\Qb_{ki}\}$.}
\label{fig:fig4} \vspace{-0.6cm}\end{figure}
%%%%%%%%%%%%%%%%%%%%%%%%%%%%%%%%%%%%%%%%%%%%%%%%%%%%%%%%%%%%%%%%%%%%%%%%%%%%%%%%%%%%%%%%%%%

Figure \ref{fig:fig4} {demonstrates} the simulation results {for}
the weighted geometric mean rate and the weighted harmonic mean
rate, for $K=N_t=4$ and for an asymmetric weighting
$(\alpha_1,\alpha_2,\alpha_3,\alpha_4)=(\frac{1}{8},\frac{1}{8},\frac{1}{4},\frac{1}{2})$.
Performance comparison results similar to those in Figure
\ref{fig:fig3} can {also} be observed in this figure. In addition,
it is interesting to note from Figure \ref{fig:fig4} that, in
contrast to the sum rate performance as shown in Figure
\ref{fig:fig3}, the weighted geometric mean rates and weighted
harmonic mean rates achieved by the proposed SCA algorithm in Figure
\ref{fig:fig4}(a) and Figure \ref{fig:fig4}(b) saturate for high
$1/\sigma^2$. {These phenomena might result from the fact that user
fairness plays a more {prominent} role in the geometric mean rate
and the harmonic mean rate;} and thereby in the interference
dominated region (i.e., when $1/\sigma^2$ or $\eta$ is large), the
geometric mean rate and the harmonic mean rate cannot increase as
fast as the weighted sum rate.

%In figure \ref{fig:fig2}, we examine the performance of the
%centralized SCA algorithm for different system utility functions,
%i.e., weighted proportional fairness and weighted harmonic mean
%rate, respectively. In these two examples, the user priorities are
%assigned asymmetrically
%$(\alpha_1,\alpha_2,\alpha_3,\alpha_4)=(\frac{1}{8},\frac{1}{8},\frac{1}{4},\frac{1}{2})$.
%It can be seen that the SCA algorithm outperforms MRT in this two
%figures. It is worthwhile to note that, in contrast to the weighted
%sum rate, the average weighted proportional fairness and harmonic
%mean rate achieved by the SCA algorithm become saturated at high
%SNR, especially when the interference is strong ($\eta=1$). This is
%because that these two utilities are more restrictive on the user
%fairness such that the achievable rate of all users should be
%comparable for yielding higher system utility, hence, the achievable
%rates are limited by the inter user interference.

{\bf Example 3:} In this example, we examine the performance of the
proposed distributed SCA algorithm (Algorithm 2). Figure
\ref{fig:fig5}(a) shows the convergence behaviors (the evolution of
sum rate at each round) of the distributed SCA algorithm for
$N_t=8$, $K=4,6$, and for $N_t=12$, $K=6$, where $1/\sigma^2=10$ dB,
$\eta=0.4$. Each curve in Figure \ref{fig:fig5}(a) is obtained by
averaging over 500 sets of randomly generated $\{\Qb_{ki}\}$. It can
be observed from Figure \ref{fig:fig5}(a) that the sum rate
performance of the distributed SCA algorithm is almost the same as
its centralized counterpart for $N_t=8$, $K=4$; whereas there is a
gap between the sum rates achieved by the centralized and
distributed SCA algorithms for $N_t=8$, $K=6$. One explanation for
this gap is that, when the system is nearly fully loaded (i.e., when
$K$ is close to $N_t$), the distributed SCA algorithm, which updates
only the variables associated with one transmitter at a time, is
more likely to get stuck at a stationary point that is not as good
as that achieved by the centralized SCA algorithm which optimizes
all the variables in each iteration. As also shown in Fig. 5(a),
when we increase $N_t$ to 12, the decentralized algorithm again
converges to the centralized solution.
%{One}
%can see from this figure that the distributed SCA algorithm
%{converges} well within 10 rounds for {the considered} problem
%instances. Also note that, since the channel covariance matrices are
%randomly generated, there is no guarantee that the sum rate under
%$K=6$ is better than that under $K=4$ and $K=2$.
Figure \ref{fig:fig5}(b) shows that, for $N_t=8$, $K=4$ the
distributed SCA algorithm yields performance similar to that
achieved by its centralized counterpart for almost all of the 30
tested problem instances within 10 round-robin iterations.

%We should note that the (centralized) SCA
%algorithm does not necessarily perform better than the distributed
%SCA algorithm. As observed in Figure \ref{fig:fig5}(b), for the 1st,
%5th, and 16th problem instances the distributed SCA algorithm in
%fact outperforms its centralized counterpart, even though both of
%them are initialized by the same MRT scheme.

%%%%%%%%%%%%%%%%%%%%%%%%%%%%%%%%%%%%%%%%%%%%%%%%%%%%%%%%%%%%%%%%%%%%%%%%%%%%%%%%%%%%%%%%%%%
\begin{figure}[t]
\begin{center}
\subfigure[][]{\resizebox{.49\textwidth}{!}{\includegraphics{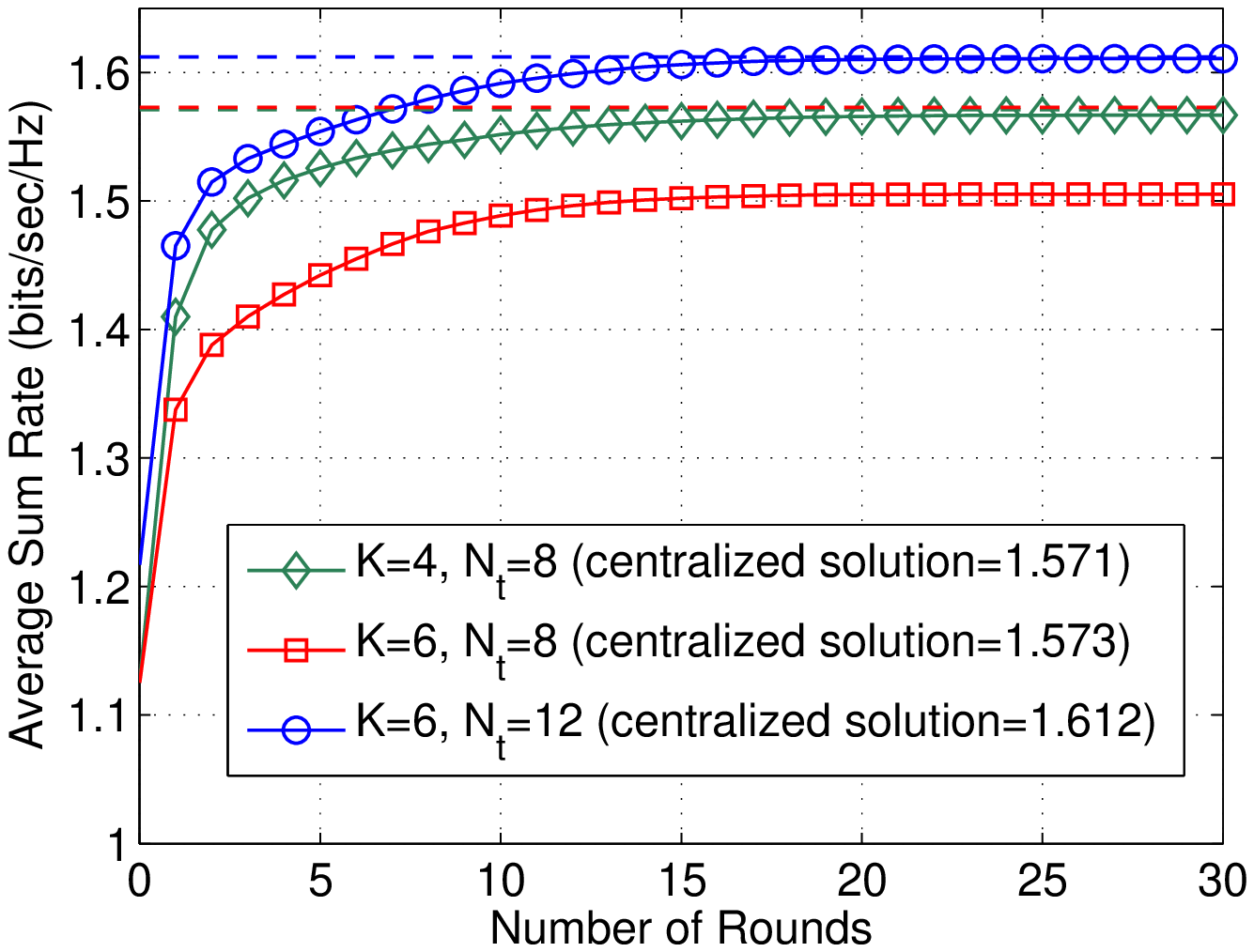}}}\label{fig:fig5_a}
\subfigure[][]{\resizebox{.49\textwidth}{!}{\includegraphics{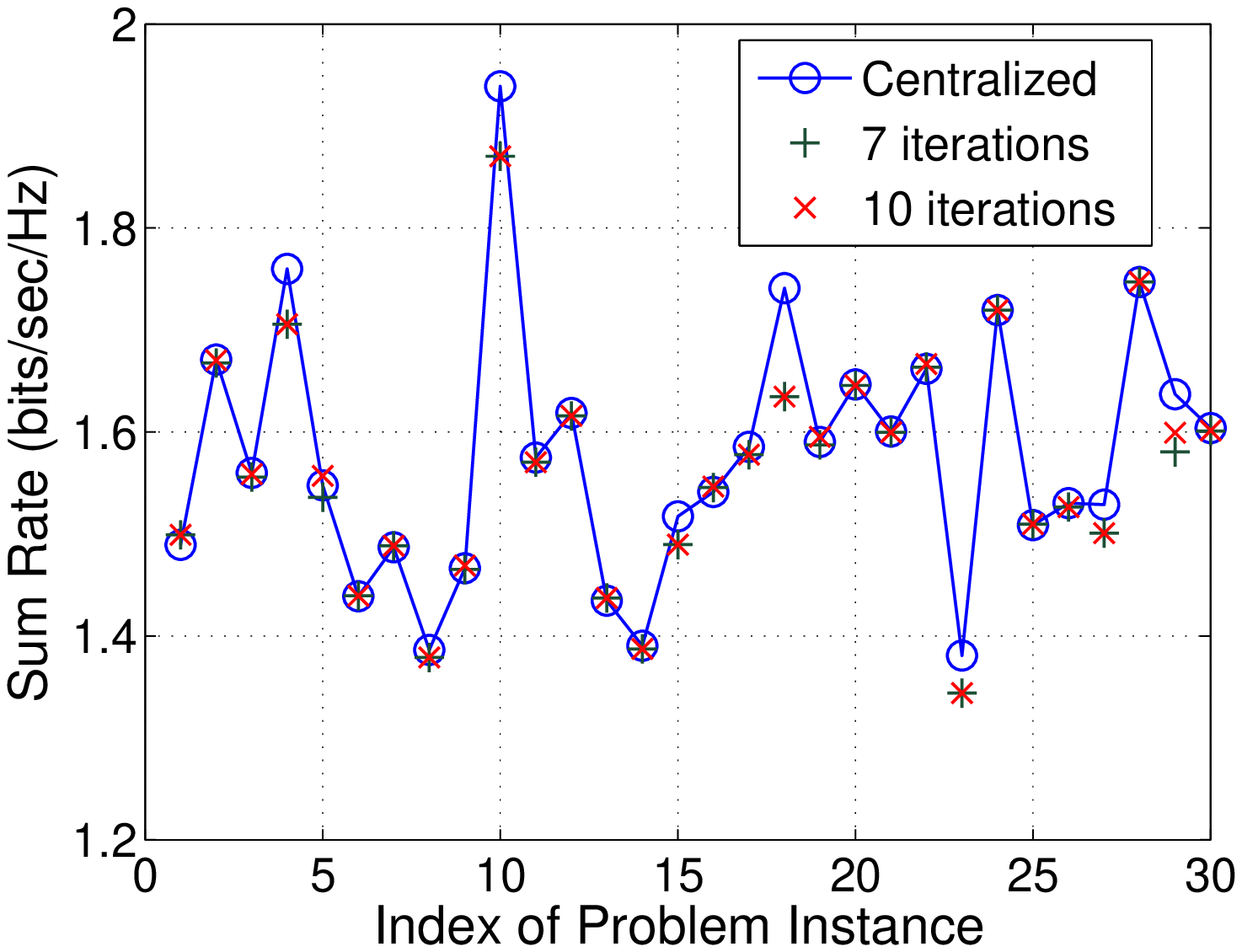}}}\label{fig:fig5_b}
\end{center}\vspace{-0.7cm}
\caption{Performance of Algorithm \ref{alg:distributed_modified},
for $1/\sigma^2=10$ dB and $\eta=0.4$; (a) convergence curves versus
round number for $N_t=8$, $K=4,6$, and for $N_t=12$, $K=6$, averaged
over 500 sets of randomly generated $\{\Qb_{ki}\}$, (b) comparison
with Algorithm \ref{alg:centralized} for $N_t=8$, $K=4$ over 30 sets
of randomly generated $\{\Qb_{ki}\}$.}
\label{fig:fig5}\vspace{-0.6cm}
\end{figure}
%%%%%%%%%%%%%%%%%%%%%%%%%%%%%%%%%%%%%%%%%%%%%%%%%%%%%%%%%%%%%%%%%%%%%%%%%%%%%%%%%%%%%%%%%%%

\vspace{-0.3cm}

%%%%%%%%%%%%%%%%%%%%%%%%%%%%%%%%%%%%%%%%%%%%%%%%%%%%%%%%%%%%%%%%%%
\section{Conclusions}\label{sec:conclusions}
%%%%%%%%%%%%%%%%%%%%%%%%%%%%%%%%%%%%%%%%%%%%%%%%%%%%%%%%%%%%%%%%%%

In this paper, we have presented two efficient approximation
algorithms for solving the rate outage constrained coordinated
beamforming design problem in \eqref{UMX}. In view of the fact that
the {original} design problem involves complicated nonconvex
constraints, we {first} {presented} an efficient SCA algorithm
(Algorithm 1) based on SDR and first-order approximation techniques.
We have shown that the proposed SCA algorithm, which involves
solving convex problem \eqref{eq:central_optsol_n_ite} iteratively,
can yield a stationary point of the outage constrained beamforming
design problem, provided that problem
\eqref{eq:central_optsol_n_ite} can yield a rank-one beamforming
solution. We further {presented} a distributed SCA algorithm
(Algorithm 2) that can {yield} approximate beamforming solutions of
problem \eqref{UMX} in a distributed, round-robin fashion, using
only local CDI and a small amount of messages exchanged among the
transmitters. The distributed SCA algorithm was also shown to
{provide} a stationary point of \eqref{UMX} provided that problem
\eqref{UMX_subprob_aprx} can yield a rank-one beamforming solution.
Finally, our simulation results demonstrated that the proposed SCA
algorithm yields near-optimal performance for $K=2$, and
significantly outperforms the heuristic MRT and ZF schemes.
Furthermore, the distributed SCA algorithm was also shown to exhibit
performance comparable to its centralized counterpart within 10
rounds of round-robin iterations for most of the problem instances.

%\vspace{-1cm}
%%%%%%%%%%%%%%%%%%%%%%%%%%%%%%%%%%%%%%%%%%%%%%%%%%%%%%%%%%%%%%%%%%
\appendices {\setcounter{equation}{0}
\renewcommand{\theequation}{A.\arabic{equation}}
%%%%%%%%%%%%%%%%%%%%%%%%%%%%%%%%%%%%%%%%%%%%%%%%%%%%%%%%%%%%%%%%%%

\section{Proof of Claim
\ref{claim:x_converge}}\label{appendix: proof of claim12}

Since constraint \eqref{eq:central_optsol_n_ite_e} holds with
equality at the optimal point, we have{\small
\begin{align}
\hat{R}_i[n]&=\log_2(1+e^{\bar{y}_i[n-1]})+\frac{e^{\bar{y}_i[n-1]}(\hat{y}_i[n]-\bar{y}_i[n-1])}{\ln2\cdot(1+e^{\bar{y}_i[n-1]})}
\le\log_2(1+e^{\hat{y}_i[n]});\label{eq:y_converge_b}
\end{align}}
\hspace{-0.19cm}similarly, from \eqref{eq:central_optsol_n_ite_c},
we have\vspace{-0.2cm}
%\begin{subequations}\label{eq:x_converge}
\begin{align}
e^{\bar{x}_{ik}[n]}&=\tr(\hat{\Wb}_i[n]\Qb_{ik})=
e^{\bar{x}_{ik}[n-1]}(\hat{x}_{ik}[n]-\bar{x}_{ik}[n-1]+1)
\le{e}^{\hat{x}_{ik}[n]},\label{eq:x_converge_b}
\end{align}
%\end{subequations}
\vspace{-0.1cm}for all $k\in\mathcal{K}_i^c$, $i=1,\dots,K$. We also
note from \eqref{eq:central_optsol_n_ite_d} and \eqref{feasible
point_n} that $\hat{x}_{ii}[n]=\bar{x}_{ii}[n]$ for all $i,n$. On
the other hand, by \eqref{bar R2}, the definition of
$\bar{x}_{ik}[n]$, $\bar{y}_i[n]$ in \eqref{feasible point_n}, and
the fact that \eqref{eq:central_optsol_n_ite_b},
\eqref{eq:central_optsol_n_ite_f} hold with equality at the optimum,
we can obtain\vspace{-0.2cm}
\begin{align}
1&=\rho_i\exp(\sigma_i^2e^{\bar{y}_i[n]-\bar{x}_{ii}[n]})\prod_{k\ne{i}}(1+e^{-\bar{x}_{ii}[n]+\bar{x}_{ki}[n]+\bar{y}_i[n]})\notag\\
&=\rho_i\exp(\sigma_i^2e^{\hat{y}_i[n]-\hat{x}_{ii}[n]})\prod_{k\ne{i}}(1+e^{-\hat{x}_{ii}[n]+\hat{x}_{ki}[n]+\hat{y}_i[n]}).\label{eq
equality}
\end{align}
Combining the above observations, i.e., \eqref{eq:x_converge_b},
\eqref{eq equality} and $\hat{x}_{ii}[n]=\bar{x}_{ii}[n]$, and by
the monotonicity of the exponential function, we obtain that
$\bar{y}_i[n]\ge\hat{y}_i[n]$, which implies
{\small\begin{equation}\label{eq:sandwich}
\hat{R}_i[n]\le\frac{1}{\ln2}\ln(1+e^{\hat{y}_i[n]})
\le\frac{1}{\ln2}\ln(1+e^{\bar{y}_i[n]})=\tilde{R}_i[n]~\forall{i,n}.
\end{equation}}

\vspace{-0.7cm}Suppose that
$e^{\hat{x}_{ki}[n]}-e^{\bar{x}_{ki}[n]}$ does not converge to zero
for some $i$ and $k\in\mathcal{K}_i^c$, then there exists an
$\epsilon>0$ such that, for all $N\ge1$,
$e^{\hat{x}_{ki}[n]}>e^{\bar{x}_{ki}[n]}+\epsilon$ for some
$n\ge{N}$. From \eqref{eq:y_converge_b} to \eqref{eq:sandwich}, we
must have $e^{\bar{y}_i[n]}>e^{\hat{y}_i[n]} + \epsilon'$ and thus
$\tilde{R}_i[n]>\hat{R}_i[n]+\epsilon''$, where $\epsilon',
\epsilon''>0$, which, together with \eqref{rate increase}, implies
that the utility $U(\tilde{R}_1[n],\ldots,\tilde{R}_k[n])$ diverges
as $n$ goes to infinity however. Therefore, we must have{\small
\begin{align}
&\lim_{n\to\infty}(e^{\hat{x}_{ik}[n]}-e^{\bar{x}_{ik}[n]})=0~\forall{i,k},\label{stop 1}\\
&\lim_{n\to\infty}(\tilde{R}_i[n]-\hat{R}_i[n])=0~\forall{i}.\label{stop
2}
\end{align}}
\hspace{-0.19cm}Now we use \eqref{stop 1} to prove \eqref{claim 1
a}. It follows from \eqref{eq:x_converge_b} and \eqref{stop 1} that
{\small\begin{align}\label{stop 3}
\lim_{n\to\infty}(e^{\hat{x}_{ik}[n]}-e^{\bar{x}_{ik}[n-1]}(\hat{x}_{ik}[n]-\bar{x}_{ik}[n-1]+1))=0
\end{align}}
\hspace{-0.19cm}for all $i$ and $k\in\mathcal{K}_i^c$. Consider the
2nd-order Taylor series expansion \cite{Gamboa2002} of
$e^{\hat{x}_{ik}[n]}$ at $\bar{x}_{ik}[n-1]$, i.e.,
\[
e^{\hat{x}_{ik}[n]}=e^{\bar{x}_{ik}[n-1]}(\hat{x}_{ik}[n]-\bar{x}_{ik}[n-1]+1)+e^{\theta[n]\hat{x}_{ik}[n]+(1-\theta[n])\bar{x}_{ik}[n-1]}(\hat{x}_{ik}[n]-\bar{x}_{ik}[n-1])^2,
\]
where $0\le\theta[n]\le1$ for all $n\ge1$. Substituting it into \eqref{stop 3} gives rise to
\[
\lim_{n\to\infty}e^{\theta[n]\hat{x}_{ik}[n]+(1-\theta[n])\bar{x}_{ik}[n-1]}(\hat{x}_{ik}[n]-\bar{x}_{ik}[n-1])^2=0.
\]
Since both $\bar{x}_{ik}[n]$ and $\hat{x}_{ik}[n]$ are bounded by Claim \ref{claim:bounded}, we conclude that \eqref{claim 1 a} is true.
%\[
%\lim_{n\to\infty}|\hat{x}_{ik}[n]-\bar{x}_{ik}[n-1]|=0,~k\in\mathcal{K}_i^c,~i=1,\dots,K,
%\]
%which completes the proof.

To show \eqref{claim 1 b}, we note from \eqref{eq:y_converge_b},
\eqref{eq:sandwich} and \eqref{stop 2} that
{\small\begin{align}\label{stop 4}
\lim_{n\to\infty}\left(\ln(1+\exp(\hat{y}_i[n]))-\ln(1+\exp(\bar{y}_i[n-1]))-\frac{\exp(\bar{y}_i[n-1])}{1+\exp(\bar{y}_i[n-1])}(\hat{y}_i[n]-\bar{y}_i[n-1])\right)=0.
\end{align}}
Analogously, by considering the 2nd-order Taylor series expansion of
$\ln(1+e^{\hat{y}_i[n]})$ at $\bar{y}_i[n-1]$, i.e.,
{\small\begin{align*}
\ln(1+e^{\hat{y}_i[n]})=\ln(1+e^{\bar{y}_i[n-1]})&+
\frac{\exp(\bar{y}_i[n-1])}{1+\exp(\bar{y}_i[n-1])}
(\hat{y}_i[n]-\bar{y}_i[n-1])
\notag \\
&+\frac{\exp(\theta[n]\hat{y}_i[n]+(1-\theta[n])\bar{y}_i[n-1])}{(1+\exp(\theta[n]\hat{y}_i[n]+(1-\theta[n])\bar{y}_i[n-1]))^2}(\hat{y}_i[n]-\bar{y}_i[n-1])^2,
\end{align*}}
where $0\le\theta[n]\le1$ for all $n\ge1$, and substituting it into
\eqref{stop 4}, we obtain {\small\begin{align*}
\lim_{n\to\infty}\frac{\exp(\theta[n]\hat{y}_i[n]+(1-\theta[n])\bar{y}_i[n-1])(\hat{y}_i[n]-\bar{y}_i[n-1])^2}{\left(1+\exp(\theta[n]\hat{y}_i[n]+(1-\theta[n])\bar{y}_i[n-1])\right)^2}=0.
\end{align*}}
Again, since $\bar{y}_i[n]$ and $\hat{y}_i[n]$ are bounded by Claim \ref{claim:bounded}, we obtain \eqref{claim 1 b}.
\hfill{$\blacksquare$}

\section{Proof of Theorem
\ref{thm:distributed_convergence}}\label{sec:proof_converge_distributed}
%%%%%%%%%%%%%%%%%%%%%%%%%%%%%%%%%%%%%%%%%%%%%%%%%%%%%%%%%%%%%%%%%%

Define
$\bar{z}_k[n,i-1]=e^{\bar{y}_k[n,i-1]-\bar{x}_{kk}[n-u_{k(i-1)}]}$
for all $k=1,\ldots,K$. Then it can be shown that\vspace{-0.1cm}
\[
\bar{\ub}[n-1,i]\triangleq\left(\hat{\Wb}_i[n-1],\{\tilde{R}_k[n,i-1]\}_k,
\{\bar{x}_{ik}[n-1]\}_k,\{\bar{y}_k[n,i-1]\}_k,
\{\bar{z}_k[n,i-1]\}_k\right),
\]
\vspace{-0.1cm}is a feasible point of \eqref{UMX_subprob_aprx}.
Hence,
%\begin{equation}\label{eq:U1}
${U}(\hat{R}_1[n,i],\dots,\hat{R}_K[n,i]) \geq U({\tilde{R}_1[n,i-1],\dots,\tilde{R}_K[n,i-1]})$ for all $i=1,\dots,K.$
%\end{equation}
In addition, analogous to \eqref{bar R}, we have
$\tilde{R}_j[n,i]\ge\hat{R}_j[n,i]$ for $i,j,n$, and thus
%\begin{equation}\label{eq:U2}
${U}(\tilde{R}_1[n,i],\dots,\tilde{R}_K[n,i])\geq U(\tilde{R}_1[n,i-1],\dots,\tilde{R}_K[n,i-1]),~i=1,\dots,K,$
%\end{equation}
%The above inequalities
which implies that the sequence
$\{U({\tilde{R}_1[1,1],\dots,\tilde{R}_K[1,1]}),\dots,U({\tilde{R}_1[1,K],
\dots,\tilde{R}_K[1,K]}),$
$U({\tilde{R}_1[2,1],\dots,\tilde{R}_K[2,1]}),\dots\}$ is
nondecreasing. Since it is also bounded,
$U({\tilde{R}_1[n,i],\dots,\tilde{R}_K[n,i]}),$ $i=1,\dots,K$, converge as
$n\rightarrow\infty$.

Now let us look at the KKT conditions of problem
\eqref{UMX_subprob_aprx}. Recall the definitions of
$\bar{\Psi}_{ki}(\cdot)$ and $\bar{\Phi}_{j}(\cdot)$ in \eqref{psi}
and \eqref{phi} and their inner approximation properties in
\eqref{eq:inner_approximation_b} to
\eqref{eq:inner_approximation_h}. Let
\begin{align}
\Theta_i^{[i]}(x_{ii},y_i,z_i,\{\bar x_{ki}[n-u_{ki}]\}_{k\neq{i}})&\triangleq\rho_ie^{\sigma_i^2z_i}\prod_{k\ne{i}}\left(1+e^{-x_{ii}+\bar{x}_{ki}[n-u_{ki}]+y_i}\right)-1,\\
\Theta_j^{[i]}(x_{ij},y_j,z_j,\{\bar x_{kj}[n-u_{ki}]\}_{k\neq{i}})&\triangleq\rho_je^{\sigma_j^2z_j}\left(1+e^{-\bar{x}_{jj}[n-u_{ji}]+x_{ij}+y_j}\right)\notag\\
&\times\prod_{k\ne{j},k\ne{i}}\left(1+e^{-\bar{x}_{jj}[n-u_{ji}]+\bar{x}_{kj}[n-u_{ki}]+y_j}\right)-1,~j\in\mathcal{K}_i^c.
%\Theta_i^{[i]}(x_{ii},y_i,z_i,\{\bar x_{ki}[n-u_{ki}]\}_{k\neq{i}})&\triangleq\rho_i\exp(\sigma_i^2z_i)\prod_{k\ne{i}}\left(1+\exp(-x_{ii}+\bar{x}_{ki}[n-u_{ki}]+y_i)\right)-1,\\
%\Theta_j^{[i]}(x_{ij},y_j,z_j,\{\bar x_{kj}[n-u_{ki}]\}_{k\neq{i}})&\triangleq\rho_j\exp(\sigma_j^2z_j)\left(1+\exp(-\bar{x}_{jj}[n-u_{ji}]+x_{ij}+y_j)\right)\notag\\
%&\times\prod_{\substack{k\ne{j}\\k\ne{i}}}\left(1+\exp(-\bar{x}_{jj}[n-u_{ji}]+\bar{x}_{kj}[n-u_{ki}]+y_j)\right)-1,~j\in\mathcal{K}_i^c.
\end{align}
\vspace{-0.2cm}Moreover, let
\[
\hat{\ub}[n,i]\triangleq(\hat{\Wb}_i[n],\{\hat{R}_k[n,i]\},\{\hat{x}_{ik}[n]\}_k,\{\hat{y}_k[n,i]\},\{\hat{z}_k[n,i]\})
\]
be the optimal solution of \eqref{UMX_subprob_aprx}, and let
\begin{align*}
&{\bm{\lambda}}[n,i]\triangleq({\lambda}^{\mathrm{b}}_i[n,i],
\{{\lambda}_k^{\mathrm{b}}[n,i]\}_{k\ne{i}},{\lambda}^{\mathrm{d}}[n,i],
\{{\lambda}_k^{\mathrm{e}}[n,i]\}_{k\ne{i}},\{{\lambda}_k^{\mathrm{f}}[n,i]\}_k,\\
&~~~~~~~~~~~~~~~~~~~~~~~~~~~~~~~~~{\lambda}^{\mathrm{g}}_i[n,i],
\{{\lambda}_k^{\mathrm{g}}[n,i]\}_{k\ne{i}},{\lambda}^P[n,i],
\{{\lambda}^{\delta}_k[n,i]\}_k)\succeq\mathbf{0},
\end{align*}
where
${\lambda}^{\mathrm{b}}_i[n,i]$,
$\{{\lambda}_k^{\mathrm{b}}[n,i]\}_{k\ne{i}}$,
${\lambda}^{\mathrm{d}}[n,i]$,
$\{{\lambda}_k^{\mathrm{e}}[n,i]\}_{k\ne{i}}$,
$\{{\lambda}_k^{\mathrm{f}}[n,i]\}_k$,
${\lambda}^{\mathrm{g}}_i[n,i]$,
$\{{\lambda}_k^{\mathrm{g}}[n,i]\}_{k\ne{i}}$ denote the dual
variables associated with constraints in \eqref{UMX_subprob_aprx_b}
to \eqref{UMX_subprob_aprx_h}, and $\lambda^P[n,i]$,
$\lambda_k^{\delta}[n,i]$, denote the dual variables associated with
constraint $\tr(\Wb_i)\le{P_i}$ and $\tr(\Wb_i\Qb_{ik})\ge\delta$,
respectively. Let
$\mathcal{L}^{[i]}(\hat{\ub}[n,i],{\bm{\lambda}}[n,i])$ be the
Lagrangian function. We can write the KKT conditions of
\eqref{UMX_subprob_aprx} as follows:
{\small\begin{subequations}\label{UMX_subprob_aprx_KKT_1}
\begin{align}
\!\!\!\!\!\!\!\!\!\!\frac{\partial\mathcal{L}^{[i]}(\hat{\ub}[n,i],
{\bm{\lambda}}[n,i])}{\partial\Wb_i}&={\lambda}^P[n,i]\mathbf{I}_{N_t}-({\lambda}^{\mathrm{d}}[n,i]+{\lambda}_i^{\delta}[n,i])\Qb_{ii}\notag \\
&+{\displaystyle\sum_{k\ne{i}}} \left({\lambda}_k^{\mathrm{e}}[n,i]
\frac{\partial\bar{\Psi}_{ik}(\hat{\Wb}_i[n],
   \hat{x}_{ik}[n]|~\bar{x}_{ik}[n-1])}{\partial\Wb_i}
-{\lambda}_k^{\delta}[n,i]\Qb_{ik}\right)\succeq\mathbf{0},\label{UMX_subprob_aprx_KKT_1_a}
\\
\frac{\partial\mathcal{L}^{[i]}(\hat{\ub}[n,i],{\bm{\lambda}}[n,i])}{\partial{R_j}}
&=-\frac{\partial{U}(\hat{R}_1[n,i],\ldots,\hat{R}_K[n,i])}
{\partial{R_j}}\notag \\
&~~~~~~~~~~~~~~~~~~~~~~~~+{\lambda}_{\mathrm{j}}^{\mathrm{f}} [n,i]
\frac{\partial\bar{\Phi}_{j}(\hat{R}_j[n,i],\hat{y}_j[n,i]|~\bar{y}_j[n,i-1])}
{\partial{R_j}}\ge0~\forall{j},\label{UMX_subprob_aprx_KKT_1_b}
\\
\frac{\partial\mathcal{L}^{[i]}(\hat{\ub}[n,i],{\bm{\lambda}}[n,i])}{\partial{x_{ii}}}&=
{\lambda}^{\mathrm{b}}_i[n,i]\frac{\partial \Theta_i^{[i]}(\hat{x}_{ii}[n],\hat{y}_i[n,i],\hat{z}_i[n,i],\{\bar x_{ki}[n-u_{ki}]\}_{k\neq i})  }{\partial{x_{ii}}} \notag \\
&~~~~~~~~~~~~~~~~~~~~~~~~+{\lambda}^{\mathrm{d}}[n,i]e^{\hat{x}_{ii}[n]}-
{\lambda}^{\mathrm{g}}_i[n,i]e^{\hat{y}_i[n,i]-\hat{x}_{ii}[n,i]}=0,
\label{UMX_subprob_aprx_KKT_1_c}
\\
\frac{\partial\mathcal{L}^{[i]}(\hat{\ub}[n,i],{\bm{\lambda}}[n,i])}{\partial{x_{ij}}}
&={\lambda}_j^{\mathrm{b}}[n,i]\frac{\partial \Theta_j^{[i]}(\hat{x}_{ij}[n],\hat{y}_j[n,i],\hat{z}_j[n,i],\{\bar x_{kj}[n-u_{ki}]\}_{k\neq i})}
{\partial{x_{ij}}}\notag
\\
&~~~~~~~~~~~~~~~~~~~~~~~~+{\lambda}_j^{\mathrm{e}}[n,i]\frac{\partial\bar{\Psi}_{ij}(\hat{\Wb}_i[n],
   \hat{x}_{ij}[n]|~\bar{x}_{ij}[n-1])}{\partial x_{ij}}=0~\forall j\in \mathcal{K}_i^c,\label{UMX_subprob_aprx_KKT_1_d}\\
\frac{\partial\mathcal{L}^{[i]}(\hat{\ub}[n,i],{\bm{\lambda}}[n,i])}{\partial{y_i}}
&={\lambda}^{\mathrm{b}}_i[n,i]\frac{\partial \Theta_i^{[i]}(\hat{x}_{ii}[n],\hat{y}_i[n,i],\hat{z}_i[n,i],\{\bar x_{ki}[n-u_{ki}]\}_{k\neq i})  }{\partial{y_i}}\notag \\
&~~~~~~~+{\lambda}_i^{\mathrm{f}}[n,i]
\frac{\partial\bar{\Phi}_{i}(\hat{R}_i[n,i],\hat{y}_i[n,i]|~\bar{y}_i[n,i-1])}
{\partial{y_i}}
+{\lambda}^{\mathrm{g}}_i[n,i]e^{\hat{y}_i[n,i]-\hat{x}_{ii}[n]}=0,
\label{UMX_subprob_aprx_KKT_1_e}
\\
\frac{\partial\mathcal{L}^{[i]}(\hat{\ub}[n,i],{\bm{\lambda}}[n,i])}{\partial{y_j}}
&={\lambda}_j^{\mathrm{b}}[n,i]\frac{\partial \Theta_j^{[i]}(\hat{x}_{ij}[n],\hat{y}_j[n,i],\hat{z}_j[n,i],\{\bar x_{kj}[n-u_{ki}]\}_{k\neq i})}
{\partial{y_j}}\notag \\
&+{\lambda}_j^{\mathrm{f}}[n,i]
\frac{\partial\bar{\Phi}_{j}(\hat{R}_j[n,i],\hat{y}_j[n,i]|~\bar{y}_j[n,i-1])}{\partial{y_j}}+
{\lambda}_j^{\mathrm{g}}[n,i]e^{\hat{y}_j[n,i]-\bar{x}_{jj}[n-u_{ji}]}=0~\forall
j\in\mathcal{K}_i^c, \label{UMX_subprob_aprx_KKT_1_f}
\\
\frac{\partial\mathcal{L}^{(i)}({\ub}[n,i],{\bm{\lambda}}[n,i])}{\partial{z_i}}
&={\lambda}^{\rm b}_i[n,i]\frac{\partial
\Theta_i^{[i]}(\hat{x}_{ii}[n],\hat{y}_i[n,i],\hat{z}_i[n,i],\{\bar
x_{ki}[n-u_{ki}]\}_{k\neq i})}{\partial{z_i}}-{\lambda}^{\rm
g}[n,i]=0,\label{UMX_subprob_aprx_KKT_1_g}
\\
\frac{\partial\mathcal{L}^{[i]}(\hat{\ub}[n,i],{\bm{\lambda}}[n,i])}{\partial{z_j}}
&={\lambda}_j^{\mathrm{b}}[n,i]\frac{\partial
\Theta_j^{[i]}(\hat{x}_{ij}[n],\hat{y}_j[n,i],\hat{z}_j[n,i],\{\bar
x_{kj}[n-u_{ki}]\}_{k\neq
i})}{\partial{z_j}}-{\lambda}_j^{\mathrm{g}}[n,i]=0,~j\in\mathcal{K}_i^c,
\label{UMX_subprob_aprx_KKT_1_h}
\end{align}
\end{subequations}}
and {\small
\begin{subequations}\label{UMX_subprob_aprx_KKT_2}
\begin{align}
{\lambda}^P[n,i]\cdot(\tr(\hat{\Wb}_i[n])-P_i)&=0,~
\frac{\partial\mathcal{L}^{[i]}(\hat{\ub}[n,i],{\bm{\lambda}}[n,i])}
{\partial{R_j}}\hat{R}_j[n,i]=0~\forall{j},\label{UMX_subprob_aprx_KKT_2_b}\\
\frac{\partial\mathcal{L}^{[i]}(\hat{\ub}[n,i],
{\bm{\lambda}}[n,i])}{\partial\Wb_i}\cdot \hat{\Wb}_i[n]&=\zerob,~
{\lambda}_j^{\delta}[n,i]\cdot(\delta-\tr(\hat{\Wb}_i[n]\Qb_{ik}))
=0~\forall{j}.\label{UMX_subprob_aprx_KKT_2_d}
\end{align}
\end{subequations}}
\hspace{-0.19cm}Note that we have omitted the
complementary slackness conditions for constraints
\eqref{UMX_subprob_aprx_b}-\eqref{UMX_subprob_aprx_h} since they are trivially
satisfied at $\hat{\ub}[n,i]$.

To show the desired results, we also need the following two claims:
%=============================================================================
\begin{Claim}\label{claim:xyR_converge_decentral}
It holds true that
\begin{subequations}
\begin{align}
&\lim_{n\to\infty}|\hat{x}_{ik}[n]-\bar{x}_{ik}[n-1]|=0~\forall i,k,\label{eq:x_converge_decentral}\\
&\lim_{n\to\infty}|\hat{x}_{ik}[n]-\bar{x}_{ik}[n]|=0~\forall{i,k},\label{eq:x2_converge_decentral}\\
&\lim_{n\to\infty}|\hat{y}_k[n,1]-\bar{y}_k[n-1,K]|=0,~\lim_{n\to\infty}|\hat{y}_k[n,i]-\bar{y}_k[n,i-1]|=0~\forall{i,k},\label{eq:y_converge_decentral}\\
&\lim_{n\to\infty}|\hat{y}_k[n,i]-\bar{y}_k[n,i]|=0~\forall{i,k},\label{eq:y2_converge_decentral}\\
&\lim_{n\to\infty}|\tilde{R}_k[n,1]-\tilde{R}_k[n-1,K]|=0,~\lim_{n\to\infty}|\tilde{R}_k[n,i]-\tilde{R}_k[n,i-1]|=0~\forall{i,k},\label{eq:R_converge_decentral}\\
&\lim_{n\to\infty}|\hat{R}_k[n,i]-\tilde{R}_k[n,i]|=0~\forall
i,k.\label{eq:R_converge_decentral2}
\end{align}
\end{subequations}
\end{Claim}
%=============================================================================

%=============================================================================
\begin{Claim}\label{claim: bounded dec}
For each $i$, $\hat \ub[n,i]$ generated by Algorithm 2 is bounded for all $n$.
\end{Claim}
%=============================================================================
The proof of Claim 3 is presented in Appendix \ref{Appendix: proof
claim3}. Similar to Claim 1, \eqref{eq:x_converge_decentral} to
\eqref{eq:y2_converge_decentral} imply that the restrictive
approximations in \eqref{UMX_subprob_aprx_e} and
\eqref{UMX_subprob_aprx_f} are asymptotically tight as $n\rightarrow
\infty$. Since problem \eqref{UMX_subprob_aprx} satisfies the
Slater's condition, the dual variable vector ${\bm{\lambda}}[n,i]$
is bounded \cite{BK:Bertsekas2003_analysis}. Moreover, $\hat
\ub[n,i]$ is also bounded by Claim 4. Now let us consider the
primal-dual solution pair $(\hat \ub[n,i],{\bm{\lambda}}[n,i])$ for
all $i=1,\ldots,K$. Since they are all bounded, there exists a
subsequence $\{n_1,\dots,n_\ell,\dots\}\subseteq\{1,\dots,n,\dots\}$
and limit points
$\hat{\ub}^\star[i]\triangleq(\hat{\Wb}_i^\star,\{\hat{R}_k^\star[i]\},\{\hat{x}_{ik}^\star\}_k,\{\hat{y}_k^\star[i]\},\{\hat{z}_k^\star[i]\})$
and
${\bm{\lambda}}^\star[i]\triangleq({\lambda}^{\mathrm{b}\star}_i[i],
\{{\lambda}_k^{\mathrm{b}\star}[i]\}_{k\ne{i}},{\lambda}^{\mathrm{d}\star}[i],
\{{\lambda}_k^{\mathrm{e}\star}[i]\}_{k\ne{i}},\{{\lambda}_k^{\mathrm{f}\star}[i]\}_k,
{\lambda}^{\mathrm{g}\star}_i[i]$,
$\{{\lambda}_k^{\mathrm{g}\star}[i]\}_{k\ne{i}},{\lambda}^{P\star}[i],
\{{\lambda}^{\delta\star}_k[i]\}_k)\succeq\mathbf{0}$ for all $i$,
such that
\begin{align}\label{limit point dec}
\lim_{\ell\to\infty}\hat{\ub}[n_\ell,i] &=\hat
\ub^\star[i],~\lim_{\ell\to\infty}{\bm{\lambda}}[n_\ell,i]={\bm{\lambda}}^\star[i]
\end{align}
for $i=1,\ldots,K$. By \eqref{eq:R_converge_decentral} and
\eqref{eq:R_converge_decentral2}, we see that both
$\hat{R}_k[n_\ell,i]$ and $\tilde{R}_k[n_\ell,i]$  converge to the
same limit point, and they are the same for all $i$, i.e.,
\begin{equation}\label{eq:same_accumulation_R}
\hat{R}_k^\star[1]=\hat{R}_k^\star[2]=\cdots=\hat{R}_k^\star[K]\triangleq
\tilde{R}_k^\star,~k=1,\dots,K.
\end{equation}
Analogously, by \eqref{eq:x_converge_decentral} to \eqref{eq:y2_converge_decentral}, we have that
\begin{align}\label{eq:same_accumulation_y}
\hat{y}_k^\star[1]=\hat{y}_k^\star[2]=\cdots=\hat{y}_k^\star[K]\triangleq
\hat{y}_k^\star,~k=1,\dots,K, \\
\hat{z}_k^\star[1]=\hat{z}_k^\star[2]=\cdots=\hat{z}_k^\star[K]\triangleq
\hat{z}_k^\star,~k=1,\dots,K.\label{eq:same_accumulation_z}
\end{align}

Then, it follows from the inner approximation properties in
\eqref{eq:inner_approximation_b} to
\eqref{eq:inner_approximation_h}, \eqref{eq:x_converge_decentral},
\eqref{eq:y_converge_decentral}, and \eqref{limit point dec} to
\eqref{eq:same_accumulation_z} that the KKT conditions in
\eqref{UMX_subprob_aprx_KKT_1} and \eqref{UMX_subprob_aprx_KKT_2}
converge along the subsequence $\{n_1,\dots,n_\ell,\dots\}$ to
{\small\begin{subequations}\label{2UMX_subprob_aprx_KKT_1}
\begin{align}
\!\!\!\!\!\!\!\!\!\!\frac{\partial\mathcal{L}^{[i]}(\hat{\ub}^\star[i],
{\bm{\lambda}}^\star[i])}{\partial\Wb_i}&={\lambda}^{P\star}[i]\mathbf{I}_{N_t}-({\lambda}^{\mathrm{d}\star}[i]+{\lambda}_i^{\delta\star}[i])\Qb_{ii}\!+\!{\displaystyle\sum_{k\ne{i}}}
\left({\lambda}_k^{\mathrm{e}\star}[i]\frac{\partial{\Psi}_{ik}(\hat{\Wb}_i^\star,\hat{x}_{ik}^\star)}{\partial\Wb_i}
-{\lambda}_k^{\delta\star}[i]\Qb_{ik}\right)
\succeq\mathbf{0},\label{2UMX_subprob_aprx_KKT_1_a}
\\
\frac{\partial\mathcal{L}^{[i]}(\hat{\ub}^\star[i],{\bm{\lambda}}^\star[i])}{\partial{R_j}}
&=-\frac{\partial{U}(\tilde{R}_1^\star,\ldots,\tilde{R}_K^\star)}
{\partial{R_j}}+{\lambda}_{\mathrm{j}}^{\mathrm{f}\star}
[i]
\frac{\partial{\Phi}_{j}(\tilde{R}_j^\star,\hat{y}_j^\star)}
{\partial{R_j}}\ge0~\forall j,\label{2UMX_subprob_aprx_KKT_1_b}
\\
\frac{\partial\mathcal{L}^{[i]}(\hat{\ub}^\star[i],{\bm{\lambda}}^\star[i])}{\partial{x_{ii}}}&=
{\lambda}^{\mathrm{b}\star}_i[i]\frac{\partial \Theta_i^{[i]}(\hat{x}_{ii}^\star,\hat{y}_i^\star,\hat{z}_i^\star,\{\hat x_{ki}^\star\}_{k\neq i})  }{\partial{x_{ii}}}+{\lambda}^{\mathrm{d}\star}[i]e^{\hat{x}_{ii}^\star}-
{\lambda}^{\mathrm{g}\star}_i[i]e^{\hat{y}_i^\star-\hat{x}_{ii}^\star}=0,
\label{2UMX_subprob_aprx_KKT_1_c}
\\
\frac{\partial\mathcal{L}^{[i]}(\hat{\ub}^\star[i],{\bm{\lambda}}^\star[i])}{\partial{x_{ij}}}
&={\lambda}_j^{\mathrm{b}\star}[i]\frac{\partial \Theta_j^{[i]}(\hat{x}_{ij}^\star,\hat{y}_j^\star,\hat{z}_j^\star,\{\hat{ x}_{kj}^\star\}_{k\neq i})}
{\partial{x_{ij}}}+{\lambda}_j^{\mathrm{e}\star}[i]\frac{\partial{\Psi}_{ij}(\hat{\Wb}_i^\star,
   \hat{x}_{ij}^\star)}{\partial x_{ij}}=0~\forall j\in \mathcal{K}_i^c,\label{2UMX_subprob_aprx_KKT_1_d}
   \\
%\frac{\partial\mathcal{L}^{[i]}(\hat{\ub}^\star[i],{\bm{\lambda}}^\star[i])}{\partial{y_i}}
%&={\lambda}^{\mathrm{b}\star}_i[i]\frac{\partial \Theta_i^{[i]}(\hat{x}_{ii}^\star,\hat{y}_i^\star,\hat{z}_i^\star,\{\hat{x}_{ki}^\star\}_{k\neq i})  }{\partial{y_i}}+{\lambda}_i^{\mathrm{f}\star}[i]
%\frac{\partial{\Phi}_{i}(\tilde{R}_i^\star,\hat{y}_i^\star)}
%{\partial{y_i}}
%+{\lambda}^{\mathrm{g}\star}_i[i]e^{\hat{y}_i^\star-\hat{x}_{ii}^\star}=0,
%\label{2UMX_subprob_aprx_KKT_1_e}
%\\
\frac{\partial\mathcal{L}^{[i]}(\hat{\ub}^\star[i],{\bm{\lambda}}^\star[i])}{\partial{y_j}}
&={\lambda}_j^{\mathrm{b}\star}[i]\frac{\partial \Theta_j^{[i]}(\hat{x}_{ij}^\star,\hat{y}_j^\star,\hat{z}_j^\star,\{\hat{x}_{kj}^\star\}_{k\neq i})}
{\partial{y_j}}+{\lambda}_j^{\mathrm{f}\star}[i] \frac{\partial{\Phi}_{j}(\tilde{R}_j^\star,\hat{y}_j^\star)}{\partial{y_j}}+
{\lambda}_j^{\mathrm{g}\star}[i]e^{\hat{y}_j^\star-\hat{x}_{jj}^\star}=0~\forall j,
\label{2UMX_subprob_aprx_KKT_1_e}
\\
%\frac{\partial\mathcal{L}^{(i)}({\ub}[i],{\bm{\lambda}}^\star[i])}{\partial{z_i}}
%&={\lambda}^{\rm b\star}[i]\frac{\partial \Theta_i^{[i]}(\hat{x}_{ii}^\star,\hat{y}_i^\star,\hat{z}_i^\star,\{\hat{x}_{ki}^\star\}_{k\neq i})}{\partial{z_i}}-{\lambda}^{\rm g\star}[i]=0,\label{2UMX_subprob_aprx_KKT_1_g}
%\\
\frac{\partial\mathcal{L}^{[i]}(\hat{\ub}^\star[i],{\bm{\lambda}}^\star[i])}{\partial{z_j}}
&={\lambda}_j^{\mathrm{b}\star}[i]\frac{\partial \Theta_j^{[i]}(\hat{x}_{ij}^\star,\hat{y}_j^\star,\hat{z}_j^\star,\{\hat{x}_{kj}^\star\}_{k\neq i})}{\partial{z_j}}-{\lambda}_j^{\mathrm{g}\star}[i]=0~\forall j, \label{2UMX_subprob_aprx_KKT_1_f}
\end{align}
\end{subequations}}
and {\small
\begin{subequations}\label{2UMX_subprob_aprx_KKT_2}
\begin{align}
{\lambda}^{\mathrm{i}\star}[i]\cdot(\tr(\hat{\Wb}_i^\star)-P_i)&=0,~
\frac{\partial\mathcal{L}^{[i]}(\hat{\ub}^\star[i],{\bm{\lambda}}^\star[i])}
{\partial{R_j}}\tilde{R}_j^\star=0~\forall{j},\label{2UMX_subprob_aprx_KKT_2_b}\\
\frac{\partial\mathcal{L}^{[i]}(\hat{\ub}^\star[i],
{\bm{\lambda}}^\star[i])}{\partial\Wb_i}\cdot
\hat{\Wb}_i^\star&=\zerob,~
{\lambda}_j^{\delta\star}[i]\cdot(\delta-\tr(\hat{\Wb}_i^\star\Qb_{ik}))
=0~\forall{j}.\label{2UMX_subprob_aprx_KKT_2_d}
\end{align}
\end{subequations}}
\hspace{-0.1cm}It can be observed from \eqref{bar R3} that, for
$\rho_i<1$, $\tilde{R}_j[n,i]$ is strictly greater than zero for all
$i,j,n$; therefore, $\tilde{R}_j^\star>0$ for all $j$, which
indicates that
$\frac{\partial\mathcal{L}^{[i]}(\hat{\ub}^\star,\hat{\bm{\lambda}}^\star[i])}{\partial{R}_j}=0$
for all $i,j$ by \eqref{2UMX_subprob_aprx_KKT_2_b}.
Substituting this into \eqref{2UMX_subprob_aprx_KKT_1_b} for all $i=1,\ldots,K$, gives rise to
\begin{align}\label{lambda equal 1}
{\lambda}_j^{\mathrm{f}\star}[1]=\cdots={\lambda}_j^{\mathrm{f}\star}[K]=
\frac{\partial{U}(\tilde{R}_{1}^\star,\ldots,\tilde{R}_{K}^\star)}{\partial{R}_j}\left(\frac{\partial{\Phi}_{j}(\tilde{R}_j^\star,\hat{y}_j^\star)}
{\partial{R_j}}\right)^{-1}\triangleq{\lambda}_j^{\mathrm{f}\star},
~j=1,\dots,K.
\end{align}
In addition, one can verify that
\begin{align}
&\frac{\partial\Theta_j^{[1]}(\hat{x}_{1j}^\star,\hat{y}_j^\star,\hat{z}_j^\star,\{\hat{x}_{kj}^\star\}_{k\neq1})}{\partial{y_j}}=\cdots=\frac{\partial\Theta_j^{[K]}(\hat{x}_{Kj}^\star,\hat{y}_j^\star,\hat{z}_j^\star,\{\hat{x}_{kj}^\star\}_{k\neq{K}})}{\partial{y_j}},\notag\\
&\frac{\partial\Theta_j^{[1]}(\hat{x}_{1j}^\star,\hat{y}_j^\star,\hat{z}_j^\star,\{\hat{x}_{kj}^\star\}_{k\neq1})}{\partial{z_j}}=\cdots=\frac{\partial\Theta_j^{[K]}(\hat{x}_{Kj}^\star,\hat{y}_j^\star,\hat{z}_j^\star,\{\hat{x}_{kj}^\star\}_{k\neq{K}})}{\partial{z_j}},\notag
\end{align}
which, together with \eqref{2UMX_subprob_aprx_KKT_1_e}
\eqref{2UMX_subprob_aprx_KKT_1_f} and \eqref{lambda equal 1}, lead to
\begin{align}\label{lambda equal 2}
{\lambda}^{\mathrm{b}\star}_j[1]=\cdots={\lambda}^{\mathrm{b}\star}_j[K]\triangleq{\lambda}^{\mathrm{b}\star}_j,~
{\lambda}^{\mathrm{g}\star}_j[1]=\cdots={\lambda}^{\mathrm{g}\star}_j[K]\triangleq{\lambda}^{\mathrm{g}\star}_j~\forall j.
\end{align}
Finally, by \eqref{2UMX_subprob_aprx_KKT_1},
\eqref{2UMX_subprob_aprx_KKT_2}, \eqref{lambda equal 1} and
\eqref{lambda equal 2}, we conclude that
$(\{\hat{\Wb}_i^\star\},\{\tilde{R}_k^\star\},\{\hat{x}_{ik}^\star\}_k,
\{\hat{y}_k^\star\},\{\hat{z}_k^\star\})$ and
$(\{{\lambda}^{\mathrm{b}\star}_i\},\{{\lambda}^{\mathrm{d}\star}[i]\},
\{\{{\lambda}_k^{\mathrm{e}\star}[i]\}_{k\ne{i}}\}_i,\{{\lambda}_k^{\mathrm{f}\star}\},
\{{\lambda}^{\mathrm{g}\star}_i\},\{{\lambda}^{P\star}[i]\},
\{{\lambda}^{\delta\star}_k[i]\})$ satisfy the KKT conditions of
problem \eqref{UMX_ChVar}. The proof is completed.
\hfill{$\blacksquare$}

%%%%%%%%%%%%%%%%%%%%%%%%%%%%%%%%%%%%%%%%%%%%%%%%%%%%%%%%%%%%%%%%%
\section{Proof of Claim
\ref{claim:xyR_converge_decentral}}\label{Appendix: proof claim3}
%%%%%%%%%%%%%%%%%%%%%%%%%%%%%%%%%%%%%%%%%%%%%%%%%%%%%%%%%%%%%%%%%

%Similar to Claim \ref{claim:x_converge}, it can be proved that
%\begin{align*}
%&\lim_{n\to\infty}|\hat{x}_{ik}[n]-\bar{x}_{ik}[n-1]=0\forall{i},k\in\mathcal{K}_i^c,\\
%&\lim_{n\to\infty}|\hat{y}_i[n,1]-\bar{y}_k[n-1,K]|=0,~\lim_{n\to\infty}|\hat{y}_k[n,i]-\bar{y}_k[n,i-1]|=0\forall{i,k},\\
%&\lim_{n\to\infty}|\tilde{R}_k[n,i]-\hat{R}_k[n,i]|=0\forall{i,k}.
%\end{align*}

%Similar to Claim \ref{claim:x_converge}, equations
%\eqref{eq:x_converge_decentral}, \eqref{eq:y_converge_decentral} and
%\eqref{eq:R_converge_decentral2} can be proved as follows.
The ideas of the proof are similar to that of Claim 1. Because
constraints \eqref{UMX_subprob_aprx_e} and
\eqref{UMX_subprob_aprx_f} hold with equality at the optimum, we have
\begin{align}\label{eq:appdx_clm3_2}
e^{\bar{x}_{ik}[n]}\le{e^{\hat{x}_{ik}[n]}},~~\hat{R}_k[n,i]\le\frac{1}{\ln2}\ln(1+e^{\hat{y}_k[n,i]})
\end{align}
for all $k\in\mathcal{K}_i^c$, $i=1,\dots,K$. Also by
\eqref{UMX_subprob_aprx_c}, \eqref{UMX_subprob_aprx_h} and
\eqref{bar R3}, we have
\begin{align*}
1&=\rho_j\exp(\sigma_j^2e^{\hat{y}_j[n,i]-\bar{x}_{jj}[n-u_{ji}]})
\left(1+e^{-\bar{x}_{jj}[n-u_{ji}]+\hat{x}_{ij}[n]+\hat{y}_j[n,i]}\right)
\prod_{\substack{k\ne{j}\\k\ne{i}}}
\left(1+e^{-\bar{x}_{jj}[n-u_{ji}]+\bar{x}_{kj}
[n-u_{ki}]+\hat{y}_j[n,i]}\right) \notag \\
&=\rho_j\exp(\sigma_j^2e^{\bar{y}_i[n,i]-\bar{x}_{ii}[n-u_{ji}]})
\left(1+e^{-\bar{x}_{jj}[n-u_{ji}]+\bar{x}_{ij}[n]+\bar{y}_j[n,i]}\right)
\prod_{\substack{k\ne{j}\\k\ne{i}}}
\left(1+e^{-\bar{x}_{jj}[n-u_{ji}]+\bar{x}_{kj}
[n-u_{ki}]+\bar{y}_j[n,i]}\right),
\end{align*}
for all $j\in\mathcal{K}_i^c$. Using the above equation and
\eqref{eq:appdx_clm3_2} and the monotonicity of exponential
function, we obtain $\hat{y}_j[n,i]\le\bar{y}_j[n,i]$. Thus,
\begin{align}
%\hat{R}_i[n,i]&\le\frac{1}{\ln2}\ln(1+e^{\hat{y}_i[n,i]})
%=\frac{1}{\ln2}\ln(1+e^{\bar{y}_i[n,i]})=\tilde{R}_i[n,i],\label{eq:appdx_clm3_3}\\
\hat{R}_j[n,i]&\le\frac{1}{\ln2}\ln(1+e^{\hat{y}_j[n,i]})
\le\frac{1}{\ln2}\ln(1+e^{\bar{y}_j[n,i]})=\tilde{R}_j[n,i]~\forall j\in \mathcal{K}_i^c,\label{eq:appdx_clm3_30}
\end{align}
Similarly, by \eqref{UMX_subprob_aprx_b}, \eqref{UMX_subprob_aprx_h} and \eqref{bar R3}, we have
\begin{align}
\hat{R}_i[n,i]&\le\frac{1}{\ln2}\ln(1+e^{\hat{y}_i[n,i]})
=\frac{1}{\ln2}\ln(1+e^{\bar{y}_i[n,i]})=\tilde{R}_i[n,i].\label{eq:appdx_clm3_3}
\end{align}
Using the same arguments as in obtaining \eqref{eq equality} to
\eqref{stop 2} in Appendix A, we can show that
\eqref{eq:R_converge_decentral2}, \eqref{eq:x2_converge_decentral},
\eqref{eq:y2_converge_decentral}, \eqref{eq:y_converge_decentral}
and
\begin{align}
%&\lim_{n\to\infty}(\tilde{R}_k[n,i]-\hat{R}_k[n,i])=0~\forall{i}.\label{eq:appdx_clm3_5}\\
%&\lim_{n\to\infty}|\hat{x}_{ik}[n]-\hat{x}_{ik}[n]|=0~\forall{i},k,,\\
&\lim_{n\to\infty}|\hat{x}_{ik}[n]-\bar{x}_{ik}[n-1]|=0~\forall{i},k\in\mathcal{K}_i^c,
%&\lim_{n\to\infty}|\hat{y}_k[n,1]-\bar{y}_k[n-1,K]|=0,~\lim_{n\to\infty}|\hat{y}_k[n,i]-\bar{y}_k[n,i-1]|=0~\forall{i,k}.
\end{align}
which is \eqref{eq:x_converge_decentral} for $k\ne{i}$, are true.
%
%\begin{align*}
%1&=\rho_i\exp(\sigma_i^2e^{\bar{y}_i[n,i-1]-\bar{x}_{ii}[n-1]})\prod_{k\ne{i}}(1+e^{-\bar{x}_{ii}[n-1]+\bar{x}_{ki}[n-u_{ki}]+\bar{y}_i[n,i-1]})\\
%&=\rho_i\exp(\sigma_i^2e^{\hat{y}_i[n,i]-\hat{x}_{ii}[n]})\prod_{k\ne{i}}(1+e^{-\hat{x}_{ii}[n]+\bar{x}_{ki}[n-u_{ki}]+\hat{y}_i[n,i]})~\forall{i}.
%\end{align*}
%
%
%\begin{align}
%\hat{R}_i[n,i]&\le\frac{1}{\ln2}\ln(1+e^{\hat{y}_i[n,i]})
%=\frac{1}{\ln2}\ln(1+e^{\bar{y}_i[n,i]})=\tilde{R}_i[n,i],\label{eq:appdx_clm3_3}\\
%\hat{R}_j[n,i]&\le\frac{1}{\ln2}\ln(1+e^{\hat{y}_j[n,i]})
%\le\frac{1}{\ln2}\ln(1+e^{\bar{y}_j[n,i]})=\tilde{R}_j[n,i]~\forall j\in \mathcal{K}_i,\label{eq:appdx_clm3_30}
%\end{align}
%and
%\begin{align}
%&\lim_{n\to\infty}(e^{\hat{x}_{ik}[n]-\bar{x}_{ik}[n]})=0~\forall{i,k},\label{eq:appdx_clm3_4}\\
%&\lim_{n\to\infty}(\tilde{R}_k[n,i]-\hat{R}_k[n,i])=0~\forall{i}.\label{eq:appdx_clm3_5}
%\end{align}
%Then, following the proof in Claim 1 using the
%Taylor series expansion of $e^{\hat{x}_{ik}[n]}$,
%$\log_2(1+e^{\hat{y}_k[n,i]})$, we can obtain
%\begin{align*}
%&\lim_{n\to\infty}|\hat{x}_{ik}[n]-\bar{x}_{ik}[n-1]|=0~\forall{i},k\in\mathcal{K}_i^c,\\
%&\lim_{n\to\infty}|\hat{y}_k[n,1]-\bar{y}_k[n-1,K]|=0,~\lim_{n\to\infty}|\hat{y}_k[n,i]-\bar{y}_k[n,i-1]|=0~\forall{i,k}.
%\end{align*}
%Equations \eqref{eq:appdx_clm3_3} and \eqref{eq:appdx_clm3_5}
%also imply that \eqref{eq:y2_converge_decentral} is true.
What remains is to prove \eqref{eq:R_converge_decentral} and
$\displaystyle{\lim_{n\to\infty}}|\hat{x}_{ii}[n]-\bar{x}_{ii}[n-1]|=0~\forall{i}$.

It follows from \eqref{eq:y_converge_decentral},
\eqref{eq:y2_converge_decentral} and the triangle inequality that
\[
\lim_{n\to\infty}|\bar{y}_k[n,1]-\bar{y}_k[n-1,K]|=0,~\lim_{n\to\infty}|\bar{y}_k[n,i]-\bar{y}_k[n,i-1]|=0~\forall{i,k},
\]
which, by the definition in \eqref{distributed_ybar}, is equivalent to
\eqref{eq:R_converge_decentral}. By considering \eqref{bar R3} for transmitter $i-1$, and the fact that
\eqref{UMX_subprob_aprx_b} holds with equality at the optimal
point for transmitter $i$, we can obtain
\begin{align*}
1&=\rho_i\exp(\sigma_i^2e^{\bar{y}_i[n,i-1]-\bar{x}_{ii}[n-1]})\prod_{k\ne{i}}(1+e^{-\bar{x}_{ii}[n-1]+\bar{x}_{ki}[n-u_{ki}]+\bar{y}_i[n,i-1]})\\
&=\rho_i\exp(\sigma_i^2e^{\hat{y}_i[n,i]-\hat{x}_{ii}[n]})\prod_{k\ne{i}}(1+e^{-\hat{x}_{ii}[n]+\bar{x}_{ki}[n-u_{ki}]+\hat{y}_i[n,i]})~\forall{i}.
\end{align*}
Since both $\{\hat{y}_i[n,i]\}_{n=1}^\infty$ and
$\{\bar{y}_i[n,i-1]\}_{n=1}^\infty$ are bounded, and by
\eqref{eq:y_converge_decentral}, we obtain from
the above equation that
\[
\lim_{n\to\infty}|\hat{x}_{ii}[n]-\bar{x}_{ii}[n-1]|=0~\forall{i}.
\]
Thus the proof of Claim \ref{claim:xyR_converge_decentral} has been
completed.\hfill{$\blacksquare$}
%%%%%%%%%%%%%%%%%%%%%%%%%%%%%%%%%%%%%%%%%%%%%%%%%%%%%%%%%%%%%%%%%%%%%%%%%%%%%%%%%%%%%%%%%%%%

%%%%%%%%%%%%%%%%%%%%%%%%%%%%%%%%%%%%%%%%%%%%%%%%%%%%%%%%%%%%%%%%%%%%%%%%%%%%%%%%%%%%%%%%%%%%
\bibliographystyle{IEEEtran}
\bibliography{IFC_references}

\end{document}